\documentclass[12pt]{elsarticle}

\usepackage{amsfonts}
\usepackage{amsmath}
\usepackage{epstopdf}
\usepackage{geometry}
\usepackage{graphicx}
\usepackage{setspace}
\usepackage{tikz}
\usepackage{tabularx}
\usepackage{url}

\newenvironment{definition}[1][Definition:]{\begin{trivlist}
\item[\hskip \labelsep {\bfseries #1}]}{\end{trivlist}}

\DeclareMathOperator{\Pers}{Pers}

\begin{document}

\title{Coordinate-Free Quantification of Coverage in Dynamic Sensor Networks}
\author[ncsu]{Jennifer Gamble\corref{cor}}
\ead{jpgamble@ncsu.edu}

\author[uiuc]{Harish Chintakunta}
\ead{hkchinta@illinois.edu}

\author[ncsu]{Hamid Krim}
\ead{ahk@ncsu.edu}

\cortext[cor]{Corresponding author}
\address[ncsu]{Electrical and Computer Engineering, North Carolina State University}
\address[uiuc]{Coordinated Science Laboratory, University of Illinois Urbana Champaign}

\begin{abstract}
We present a novel set of methods for analyzing coverage properties in dynamic sensor networks. The dynamic sensor network under consideration is studied through a series of snapshots, and is represented by a sequence of simplicial complexes, built from the communication graph at each time point. A method from computational topology called zigzag persistent homology takes this sequence of simplicial complexes as input, and returns a `barcode' containing the birth and death times of homological features in this sequence. We derive useful statistics from this output for analyzing time-varying coverage properties. \\
Further, we propose a method which returns specific representative cycles for these homological features, at each point along the birth-death intervals. These representative cycles are then used to track coverage holes in the network, and obtain size estimates for individual holes at each time point. A weighted barcode, incorporating the size information, is then used as a visual and quantitative descriptor of the dynamic network coverage.
\end{abstract}


\maketitle

\section{Introduction}

Wireless sensor networks began to gain attention and popularity when technological advances allowed for the development of small, low-cost wireless sensors. These simple devices could be distributed over a region, with each sensor (or `node') gathering data about its local environment for purposes of monitoring, detecting or reporting. In subsequent years, the study of wireless sensor networks exploded, with research into methodologies for the different layers of the sensor network protocol stack (physical, data link, network, transport and application layers), each developing into their own sub-field. Areas of application include military, industrial, and environmental monitoring and tracking. See \cite{akyildiz2002} and \cite{yick2008} for surveys of the field.

One of the issues in sensor networks that quickly gained interest is the so-called `coverage problem' \cite{huang2005}. Given a set of (typically homogeneous) sensors, each with the ability to sense some region of immediate proximity to it, one wishes to make statements about the sensing ability of the entire network, taken as a whole. An initial question is whether every point in a region of interest is covered by at least one sensor. As sensor networks developed, it was no longer realistic to assume a static network, and node mobility became a factor in network analysis and design. It became clear that mobility of nodes could be considered not only for initial deployment \cite{poduri2004} \cite{howard2002}, but to improve coverage over time \cite{liu2005}. Thus, the development of methods to study dynamic, or time-varying sensor networks has become increasingly important.

A number of methods for determining area coverage were developed, as well as methods for efficient node deployment to provide complete or optimal coverage, see \cite{wang2011} for a survey. Such methods require geometric information about the locations of the sensors, or their distances from each other, in addition to information about the geometry of the coverage area for each sensor. Methods from computational and stochastic geometry have been used to study the coverage properties of dynamic sensor networks when complete geometric information is available \cite{peres2011}. The coverage is described using statistics such as the proportion of uncovered area at each time point, or the proportion uncovered over a time interval (where a point is considered covered if it is covered at any time during the interval). These descriptors have been used to analyze and compare various mobility models for dynamic networks, to determine advantages and disadvantages of each, as well as optimal strategies for intruder detection \cite{liu2013}.

It is often desirable to avoid assuming the availability of geometric information, such as global coordinates for the nodes, or distances between them. Instead, `coordinate-free' methods compute network properties using only local, binary information about which nodes are within communication range of each other. De Silva and Ghrist \cite{deSilva2007} were the first to propose a rigorous method for determining coverage which did not require location or distance information, but employed tools from simplicial homology theory (see Section \ref{StaticHom} for details). Such homological methods are able to give guarantees that a network is covered at a single time point, or over a time interval, using only coordinate-free data.

Other researchers have used coordinate-free data to study network coverage by detecting approximate boundaries of coverage holes in static networks. Some methods (such as in \cite{Kroller2006} or \cite{Li2006}) define interior nodes using specifically structured sub-graphs (`flowers' or `3MeSH rings', respectively), while another method defines boundary nodes by using breaks in iso-contours formed by hop distance from a base node \cite{Funke2005}. One method estimates the boundary by using a multi-step procedure built using the cuts in a shortest path tree which `forks' around coverage holes \cite{Wang2006}. All of these methods can obtain good experimental results, but are relatively dependent on the network having a high density, so the holes are large compared to the distances between neighboring sensors \cite{Khedr2009}.

In this paper, we consider the study of coverage properties of sensor networks which are both coordinate-free and time-varying. Information from the network is available as a series of discrete-time snapshots, where each node returns a list of the other nodes that are within its local area. Using this, we compute the number of coverage holes at each time point, as well as information about estimated hole sizes, and how the holes persist over time. This information is summarized in a `barcode' describing the birth and death times of homological features in the network over time, and we describe the relationship between these features and the coverage properties. The barcode is obtained by employing a method from the mathematical field of computational topology, called zigzag persistent homology (\cite{carlsson2010}, \cite{carlsson2009b}). We also propose an additional algorithm which returns specific cycles in the network characterizing the coverage holes over time, which aid in estimating the size of the holes.

The method we describe here is the only one currently available which can quantify the coverage dynamics in a coordinate-free network. We will also see that it correlates well with other coverage measures which utilize full geometric information. Further, the barcode includes information about how coverage holes form, merge, split and close in the time-varying network, which is not available using existing methods (whether geometric information is included or not). In the past, homological methods have been able to give guarantees that a network is covered at a single time point, or over a time interval, while geometric methods have been used to obtain summary statistics which describe the time-varying nature of the network coverage. Here, we use homological, coordinate-free methods to obtain a descriptor of the dynamic network coverage.


As our primary contributions, we propose how the `barcode' output from zigzag persistence can be used as a quantitative descriptor of time-varying coverage in a network, and moreover describe an algorithm we developed for choosing a specific geometrically-relevant cycle for each coverage hole in the network at each time point. The utility of the barcode is illustrated by using it to quantify and compare coverage dynamics for different models of sensor mobility. Our novel representative cycles are used in conjunction with a hop distance-based method to obtain size estimates for the holes, and this information is incorporated back into the barcodes, giving a visual and quantitative summary of the dynamic network coverage. Further examples demonstrate the effectiveness of this descriptor for tracking small coverage holes appearing in dense networks, identifying expanding failure regions, and monitoring the maintenance of a protective barrier of mobile sensors around a guarded region.

The organization of this paper is as follows: In Section \ref{StaticHom} we will first describe the basics of simplicial homology, and how it has been used effectively to give global coverage guarantees for both static and dynamic coordinate-free networks. In Section \ref{DynamicHom} we will outline our primary computational tool, zigzag persistent homology, and describe the additional types of coverage results it allows. Section \ref{HoleSizes} details the hop distance-based filtration, and its use in estimating hole sizes for a given simplicial complex. Section \ref{Tracking} gives our method for obtaining specific representative cycles, and how these cycles can be used with the hop distance filtration to enhance the barcode with estimated size information for each bar at each time point. This is followed by examples illustrating the utility of the method, and concluding remarks.

\section{Preliminaries}\label{StaticHom}
The sensor network coverage model we use assumes homogeneous, isotropic sensors with sensing radius $r$, so that each sensor is at the center of its associated coverage region, which is a disk of radius $r$. This `Boolean disk coverage model' is the most widely used sensor coverage model in the literature \cite{wang2011}. Throughout this paper, we will assume the network consists of $n$ sensors, indexed 1 through $n$. If sensor $i$ is located at $\mathbf{x}_i \in \mathbb{R}^2$, then denote the disk of radius $r$ centered at $\mathbf{x}_i$ as $B(\mathbf{x}_i,r)$. Then the coverage region $\mathcal{R}$, for the entire network, is the union of all such disks:

\begin{equation}\label{Rc}
\mathcal{R} = \bigcup_{i=1}^n B(\mathbf{x}_i,r)
\end{equation}

To study the coverage holes appearing in $\mathcal{R}$ two concepts are useful: the concept of \textit{homology}, and that of representing a sensor network with a \textit{simplicial complex}. Homology is a mathematical method which, intuitively, is used to define and categorize holes in spaces, (which are exactly the features of interest here, and are called \emph{topological features}). Thus, coverage analysis reduces to analysis of the topology of the space $\mathcal{R}$. The tools for this analysis come from the field of algebraic topology in mathematics, which quantifies the topology of a space by assigning algebraic invariants called homology groups. Representing a space as a simplicial complex (which can be done using local information only), provides a discrete combinatorial representation enabling computations of the homology groups. Thus in the sensor network setting, using reasonably coarse local information, specifically the assumptions that each node can sense in a local coverage disc of fixed radius, and that each sensor has a list of the other nodes within a known communication range, guarantees can be made about coverage for the entire network \cite{deSilva2007}.

\subsection{Simplicial homology}\label{SimpHom}

The theory of homology has a long and rich history, with results available in much greater generality than necessary for our purposes here (see \cite{hatcher2001} for a good introduction to algebraic topology, including homology theory). The situation we will be considering is when the spaces under analysis are simplicial complexes, which has the convenient byproduct that computing homology reduces to matrix calculations. First, we define a simplicial complex, and its homology.

\begin{definition}
A \textit{$k$-simplex} is a set of $k+1$ vertices, or singleton elements. Any subset of the $k+1$ vertices forming a simplex is called a \textit{face} of the simplex, where each face is also a simplex itself.

A \textit{simplicial complex}, $\mathcal{K}$, is a set of simplices such that any simplex in $\mathcal{K}$ also has all of its faces in $\mathcal{K}$.
\end{definition}

A simplicial complex can be thought of as a higher-dimensional analogue to a graph. Although simplicial complexes can be represented purely abstractly as a collection of sets of vertices (as above), they are also often defined or visualized as being embedded in Euclidean space. That setting is useful because vertices are represented as specific coordinate points, and simplices as their convex hull. In that case, a $0$-simplex is a vertex (also called a node), a $1$-simplex is an edge between two vertices, a $2$-simplex is a triangle, and higher dimensional simplices are defined analogously. For computational purposes the abstract combinatorial representation is used, because its discrete nature lends itself well to compact storage and calculations. In particular, this representation allows for straightforward computation of homology.

\begin{figure}[htp]
\begin{center}
\begin{tabular}{ll}
\includegraphics[scale=0.25]{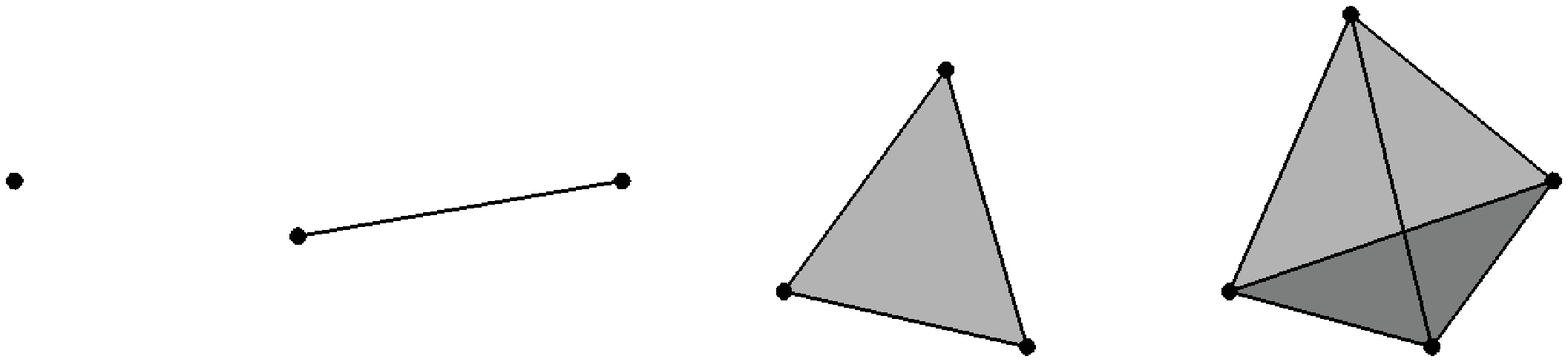} & \includegraphics[scale=0.4]{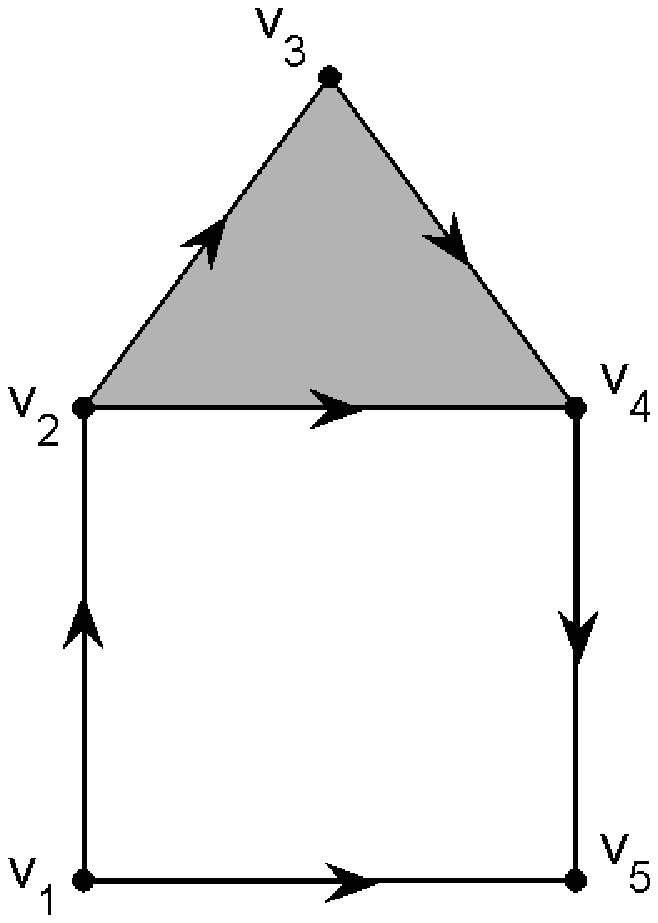} \\
\end{tabular}
\end{center}
\caption{(Left) 0-, 1-, 2-, and 3-dimensional simplices. (Right) An example of a small simplicial complex, with vertices labeled and orientations indicated on the edges. \label{SimpComp}}
\end{figure}

\begin{definition}
\textit{(Homology)} Given a simplicial complex $\mathcal{K}$ we build the \textit{chain spaces} $\mathcal{C}_0$, $\mathcal{C}_1$, $\mathcal{C}_2$, $\ldots$, where $\mathcal{C}_k$ is the vector space formed by using the $k$-simplices as basis elements. We then encode information about the specific structure of the simplicial complex in the \textit{boundary maps} $\partial_1$, $\partial_2$, $\ldots$, where \[ \partial_k: \mathcal{C}_k \rightarrow \mathcal{C}_{k-1} \]
describes explicitly how the $k$-simplices are connected to the $(k-1)$-simplices. For $k$-simplex $\sigma = [v_0, v_1, \ldots, v_k]$, the boundary map $\partial_k$ maps $\sigma$ onto the alternating sum of its faces:
\[ \partial_k \sigma = \sum_{i=0}^k (-1)^i[v_0, \ldots, \hat{v}_i, \ldots, v_k] \]
where $\hat{v}_i$ indicates vertex $v_i$ is removed. Note that the above definition of the boundary operator depends on the initial ordering of the simplex, which is referred to as \emph{orientation}. The simplices are assigned arbitrary orientations. Then the \textit{$k$-th homology group} is defined to be
\[ H_k(\mathcal{K}) = \ker(\partial_k)/\mbox{im}(\partial_{k+1}) \]
and the \textit{$k$-th Betti number} (denoted $\beta_k$) of the simplicial complex $\mathcal{K}$ is the rank of $H_k(\mathcal{K})$.
\end{definition}

To understand this definition, let us look at what $\ker(\partial_k)$ and $\mbox{im}(\partial_{k+1})$ mean individually. In general, $\partial_k$ maps a $k$-simplex $\sigma$ onto its boundary (which is made up of $(k-1)$-simplices), so if $\sigma = [v_i, v_j]$ is an edge, then $\partial_1 \sigma = v_j - v_i$ is the difference of $\sigma$'s vertices. Similarly, if $\sigma = [v_i, v_j, v_k]$ is a triangle (a $2$-simplex), then $\partial_2 \sigma = [v_j, v_k] - [v_i, v_k] + [v_i, v_j]$ is the alternating sum of its edges. An element $c$ in the chain space $\mathcal{C}_k$ is just a linear combination of $k$-simplices $\sigma_1, \ldots, \sigma_{n_k}$,
\[ c = \sum_{i=1}^{n_k} a_i \sigma_i \]
and can be written as a vector $c = [a_1, \ldots, a_{n_k}]$ of length $n_k =$ number of $k$-simplices in $\mathcal{K}$. The coefficients $a_i$ come from a field $\mathbb{F}$ (such as the real numbers), but we choose to perform our computations over the field $\mathbb{Z}_2 = \{0,1\}$ (in this case, the interpretation is that simplices with nonzero coefficients are the ones present in the chain $c$). The boundary operator $\partial_k$ is written as a $n_{k-1} \times n_k$ matrix, so the computation of the boundary for any chain reduces to the matrix multiplication $\partial_k c$. Any chain with boundary zero (i.e. any $c$ such that $\partial_k c = 0$) is called a \textit{cycle}, and so $\ker \partial_k$ is the set of all $k$-cycles. In particular, the boundary of a simplex will form a cycle, which implies that all boundaries are themselves cycles (i.e. $\mbox{im}\partial_{k+1} \subseteq \ker \partial_k$). This also implies the general property that $\partial_k \partial_{k+1} = 0$. We can now reinterpret the definition of homology as ``cycles which are not boundaries''.

\begin{definition}
Two cycles $c_1$ and $c_2$ are \textit{homologous} (written $c_1 \sim c_2$) if their difference is can be written as a linear combination of boundaries. The set of all cycles that are homologous to a given cycle (say $c$) is called a \textit{homology class} (denoted $[c]$). All cycles in the same homology class will surround exactly the same hole (or set of holes). When a specific cycle is chosen to represent an entire homology class, it is called a \textit{representative cycle}. The span of the homology classes defined by $k$-cycles form the $k$-th homology space.
\end{definition}

It is in this sense that the rank of the $k$-th homology group (the Betti number $\beta_k$) counts the number of $k$-dimensional `holes' in the simplicial complex. Intuitively, $\beta_0$ counts the number of connected components, $\beta_1$ counts the number of `holes' as we normally think of them (empty regions that one can form a loop around), $\beta_2$ counts the number of enclosed voids, and higher-dimensional homology is defined analogously.

In Figure \ref{SimpComp}, the cycle formed by edges $[v_2, v_3]$, $[v_3, v_4]$, and $[v_2, v_4]$ is the boundary of the triangle $[v_2, v_3, v_4]$, and thus is equivalent to zero (trivial) with respect to homology. The cycle formed by edges $[v_1, v_2]$, $[v_2, v_4]$, $[v_4, v_5]$, and $[v_1, v_5]$, which we denote by $c$, cannot be written as the boundary of triangles, and is thus non-trivial with respect to homology. Note also that the non-trivial cycle $c$, is homologous to the cycle formed by edges $[v_1, v_2]$, $[v_2, v_3]$, $[v_3, v_4]$, $[v_4, v_5]$, and $[v_1, v_5]$.

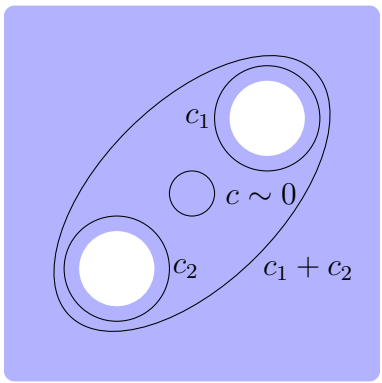
\begin{figure}
\centering
\begin{tikzpicture}
\fill[blue!30!white,rounded corners]  (-2.5,-2.5) rectangle (2.5,2.5);
\fill[white] (1,1) circle (0.5) (-1,-1) circle (0.5) ;
\draw (0,0) circle (0.3) (0.3,0) node[anchor=west]{$c\sim 0$};
\draw (1,1) circle (0.7) (0.4,1) node[anchor=east]{$c_1$};
\draw (-1,-1) circle (0.7) (-0.4,-1) node[anchor=west]{$c_2$};
\draw[rotate=45] (0,0) ellipse (2.3 and 1.2);
\draw (0.8,-1) node[anchor=west]{$c_1+c_2$}  ;
\end{tikzpicture}
\caption{A space with two holes, and three non-trivial homology classes (any two of which are linearly independent). \label{HomBasis}}
\end{figure}

A final concept we would like to emphasize is that of a \textit{homology basis}. As seen in the above definitions, given a simplicial complex $K$, the $k$-th homology group $H_k(K)$ is a vector space of dimension $\beta_k$, and therefore any linearly independent set of $\beta_k$ homology classes form a basis for $H_k(K)$. As an example, consider Figure \ref{HomBasis}
, which illustrates a space with two holes (so $\beta_1 = 2$). The cycle $c$ does not surround any holes, and is homologous to zero (i.e. it is trivial). The homology class $[c_1]$ contains to all cycles which surround only the righthand hole, and is represented by cycle $c_1$. Similarly, $c_2$ represents the homology class of cycles surrounding the lefthand hole. Note that the cycle $c_1 + c_2$ is homologous to the sum of the cycles $c_1$ and $c_2$. Thus, this space has three distinct, non-trivial homology classes: $[c_1]$, $[c_2]$, and $[c_1+c_2]$, any two of which form a basis for the first homology (eg. $\{[c_1],[c_2]\}$ form a basis, but so does $\{[c_1],[c_1+c_2]\}$). Given a compact region of the plane, such as the one shown, there exists a \textit{canonical basis} for its first homology, namely the basis with one homology class surrounding each of the holes ($[c_1]$ and $[c_2]$ in our example). This result is a specific case of the more general principle of Alexander Duality (see, for example Ch. 5 of \cite{Miller2005}). The concept of a canonical homology basis will be come relevant for us again in Section \ref{Tracking}.

\subsection{Simplicial complex representation of a sensor network}\label{SimpHomSensor}
For the purposes of analyzing the coverage region of a sensor network, we are interested in computing the homology of $\mathcal{R}$ (the coverage region for the network - defined in Equation \ref{Rc}). Specifically, we are interested in $\beta_1 = \mbox{rank}(H_1(\mathcal{R}))$, the rank of the first homology group, to determine how many holes are present in the network. Given a set of sensors, one can build a simplicial complex by using the sensors as vertices, and adding higher dimensional simplices (edges, triangles, etc) between them based on the distances between the sensors/vertices. Two common ways to build a simplicial complex from a set of points are using the \v{C}ech, and the Vietoris-Rips complexes. For the following definitions, assume vertex $v_i$ corresponds to sensor $i$, which has location $\mathbf{x}_i \in \mathbb{R}^2$, and the disk of radius $r$ centered at $\mathbf{x}_i$ is denoted $B(\mathbf{x}_i,r)$.

\begin{definition}
A \textit{\v{C}ech complex} contains the $k$-simplex formed by vertices \\ $\{v_0, v_1, \ldots, v_k\}$ whenever
\[ \bigcap_{i=0}^k B(\mathbf{x}_i,r) \neq \emptyset \]
\end{definition}

\begin{definition}
A \textit{Vietoris-Rips complex} (also referred to as a \textit{Rips complex}) includes the $k$-simplex formed by vertices $\{v_0, v_1, \ldots, v_k\}$ whenever
\[ B(\mathbf{x}_i,r) \cap B(\mathbf{x}_j,r) \neq \emptyset \mbox{ for all } 0 \leq i < j \leq k. \]
\end{definition}

\begin{figure}[htbp]
\begin{center}
\begin{tabular}{ll}
\includegraphics[scale=0.4]{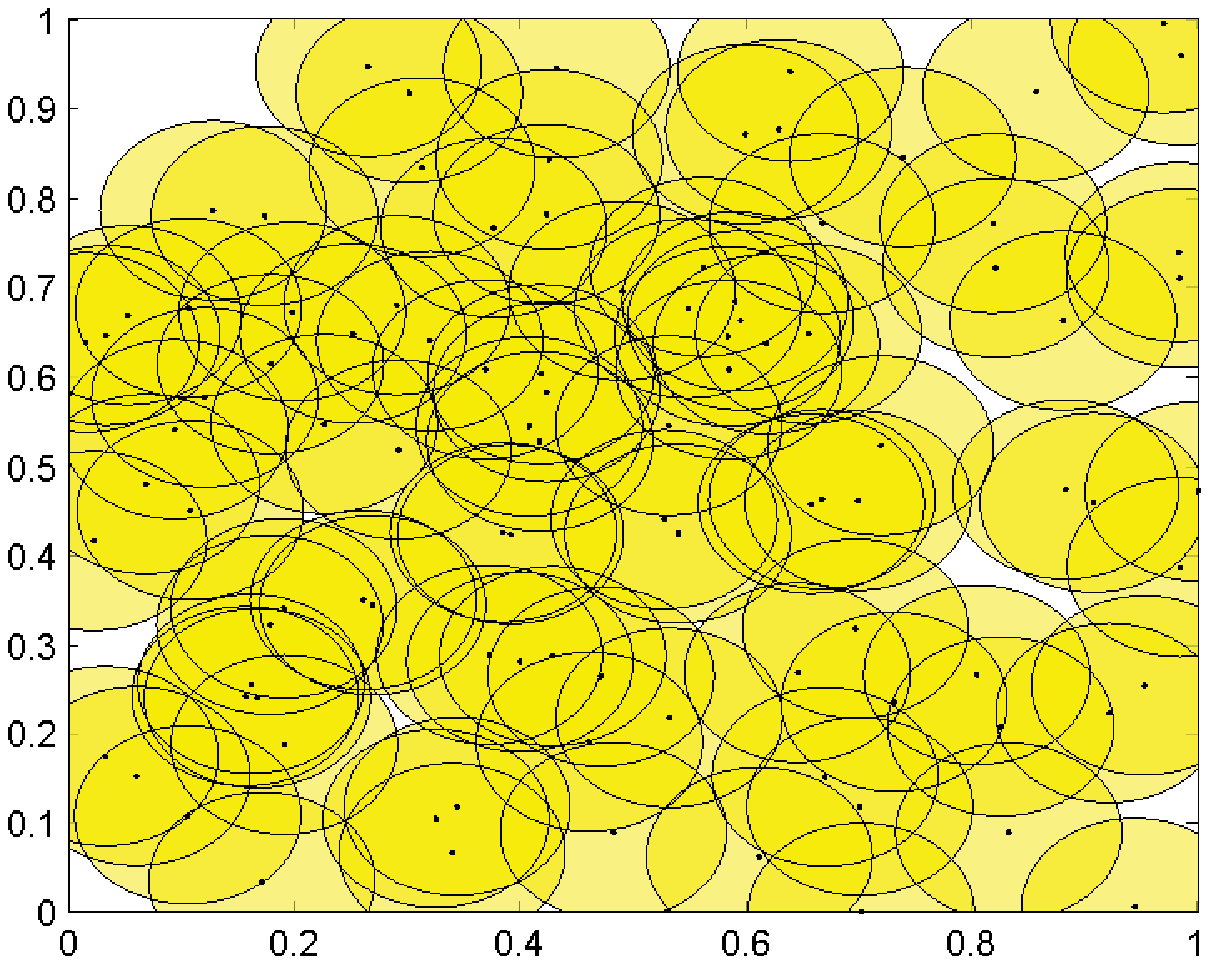} & \includegraphics[scale=0.4]{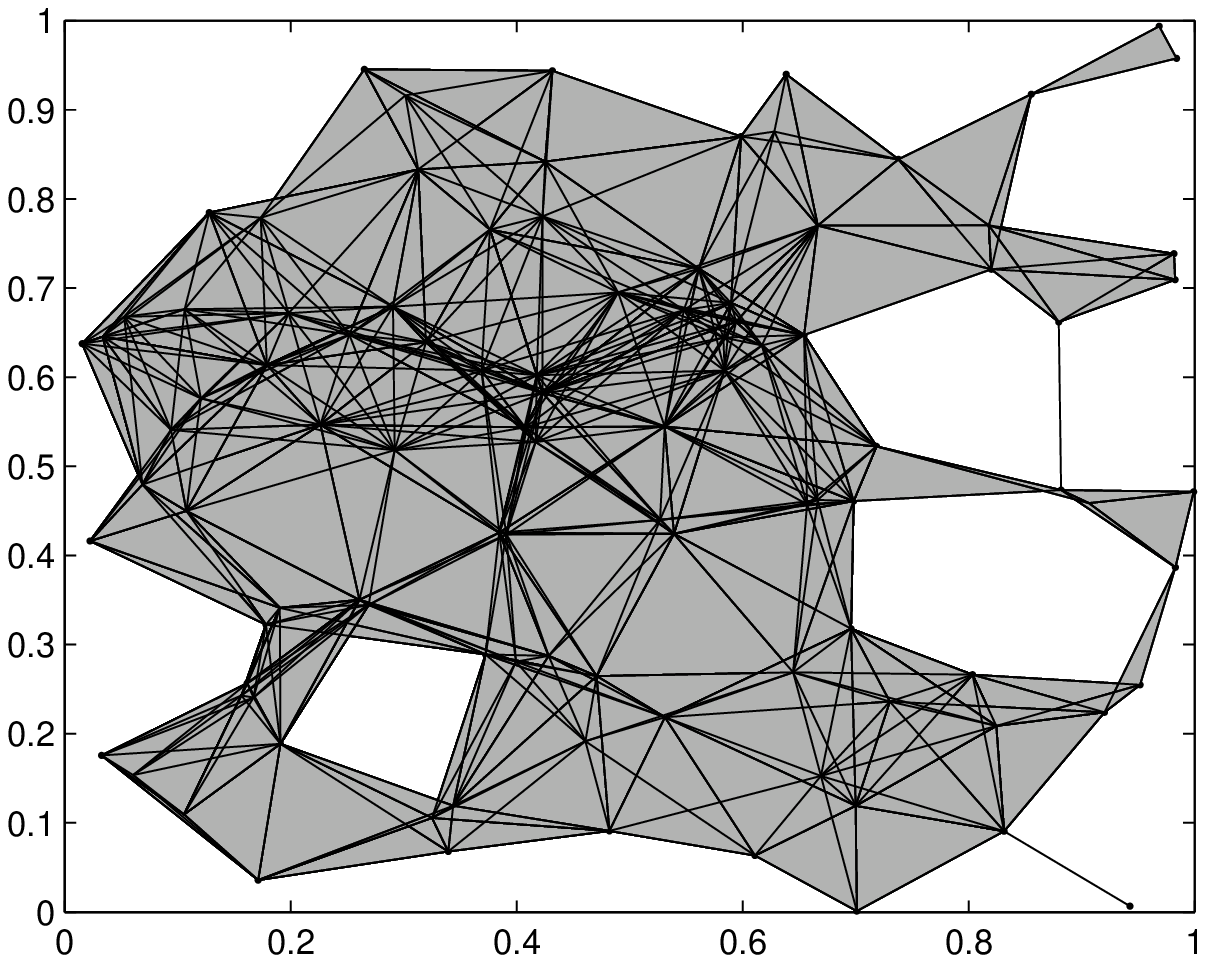} \\
\end{tabular}
\end{center}
\caption{(Left) The coverage region $\mathcal{R}$ for a sensor network. (Right) The associated Rips complex. \label{CoverageArea}}
\end{figure}

\begin{figure}[htbp!]
\begin{center}
\begin{tabular}{c}
\includegraphics[scale=0.4]{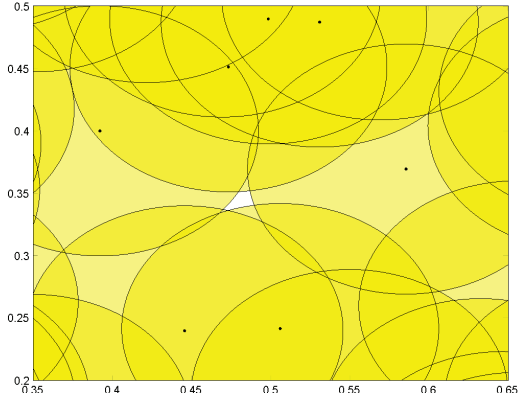} \\
\end{tabular}
\end{center}
\caption{A situation where three coverage disks each intersect pairwise, but there is no triplet=wise intersection. The associated triangle will be `filled-in' in the Rips complex, but not the \v{C}ech complex (thus, the \v{C}ech complex reflects the true homology of the coverage region). \label{CechvsRips}}
\end{figure}

In other words, the \v{C}ech complex contains the higher-dimensional simplex formed by a group of sensors whenever all the coverage disks of those sensors have nonempty intersection, and the Rips complex contains the higher-dimensional simplex whenever the coverage disks of a group of sensors all intersect pairwise. The coverage region formed by the union of coverage disks for a sensor network is shown in Figure \ref{CoverageArea} (left), with the associated Rips complex (right). Note that computation of the $(k+1)$-wise intersection of disks in the \v{C}ech definition requires precise geometric information about the relative locations $\mathbf{x}_i$ of the sensors. For the Rips complex, on the other hand, once edges are formed between all sensors of distance less than $2r$, the information about which higher-dimensional simplices to include follows directly. This is equivalent to requiring only the binary information contained in the adjacency matrix for the communication graph (where sensors can communicate whenever they are within distance $2r$ from each other). Both the \v{C}ech and Rips complexes depend on choice of parameter $r$, and for a given value of $r$, the two complexes will differ precisely when a set of sensors are all pairwise within $2r$, but do not all intersect at any point. A 2D example of when the \v{C}ech and Rips complexes will differ is shown in Figure \ref{CechvsRips}. Since the three coverage disks intersect pairwise, but have no triplet-wise intersection (leaving a small area uncovered), the associated triangle will be in the Rips complex, but not the \v{C}ech.

The configuration displayed in Figure \ref{CechvsRips} also illustrates one of the properties of the \v{C}ech complex: it has the exact same homology (number of holes) as the coverage region $\mathcal{R}$, while the Rips complex can `miss' small coverage holes like this. The worst-case scenario in terms of missed area is when the three nodes form an equilateral triangle with edge lengths $2r$, which can be achieved network-wide when the sensors lie on a hexagonal lattice. In this case the holes account for $\sim7\%$ of the total area, but are not detected by the Rips complex. In practice, when the nodes are distributed randomly uniformly, the holes missed by the Rips complex amount to $\ll 1\%$ of the total area.

The results by De Silva and Ghrist \cite{deSilva2007} use this simplicial complex representation of a sensor network, and describe a precise relationship between the sensing radius and the communication radius of each node which allows coverage guarantees to be made. The sensing radius defines the coverage region, and the communication radius is used to build the Rips complex used for computing the homology, so their results allow very coarse binary information about pairwise communication to infer whether global coverage is achieved. They additionally consider a problem in dynamic networks: whether an \textit{evasion path} exists which would allow an intruder to remain undetected over a time interval. Their results give conditions which will guarantee that no such evasion path exists.

For our purposes, we will understand that although the holes detected by the first homology of the Rips complex do differ from the holes in the coverage region (in exactly the way described above), the holes which are missed are extremely small relative to the size of the network. We will therefore use the homology computed using the Rips complex as a sufficient approximation. This is a particularly safe assumption in the time-varying case, because a very small hole that remains very small over time is justifiably ignored. Thus, when we discuss `network coverage', we are referring to the coverage as characterized by the Rips complex.

An additional note on the use of the Rips complex to characterize the network: the only assumption that is really required is that whenever three sensors can communicate pairwise, then the entire triangle that they define is considered to be covered. Thus, the assumption that the coverage region is the union of identical coverage disks centered at each node, is somewhat stricter than necessary.

We now consider a time-varying network, which again has only pairwise communication information at each time point. We present a method here which, in addition to detecting global coverage, will track homological features over time, and give us information about the number and duration of coverage holes.

\section{Coverage properties of dynamic networks}\label{DynamicHom}

\subsection{Zigzag persistent homology}\label{ZigZag}
Zigzag persistent homology is a recently developed computational method to track homological features (such as those described in Section \ref{StaticHom}) through a sequence of spaces. In our setting, where sensor networks are represented by simplicial complexes and the first homology detects coverage holes, we employ this method to tell us about coverage holes in a time-varying sensor network. While we give a brief summary here, see \cite{carlsson2010} and \cite{carlsson2009b} for complete mathematical and algorithmic details (respectively) of zigzag persistence.

We use zigzag persistent homology to study a sequence of simplicial complexes

\[ K_1 \leftrightarrow K_2 \leftrightarrow \ldots \leftrightarrow K_n\]

Call this sequence $\mathcal{K}$, and assume each map `$\leftrightarrow$' is an inclusion: either `forward' as $K_i \rightarrow K_{i+1}$ or `backward' as $K_i \leftarrow K_{i+1}$. This sequence is studied by computing the associated homology spaces to obtain the \textit{zigzag persistence module}

\begin{equation}\label{ZZmod}
H_p(\mathcal{K}) = H_p(K_1) \leftrightarrow H_p(K_2) \leftrightarrow \ldots \leftrightarrow H_p(K_n)
\end{equation}

\begin{figure}[tbp]
\begin{center}
\begin{tabular}{ll}
\includegraphics[scale=0.4]{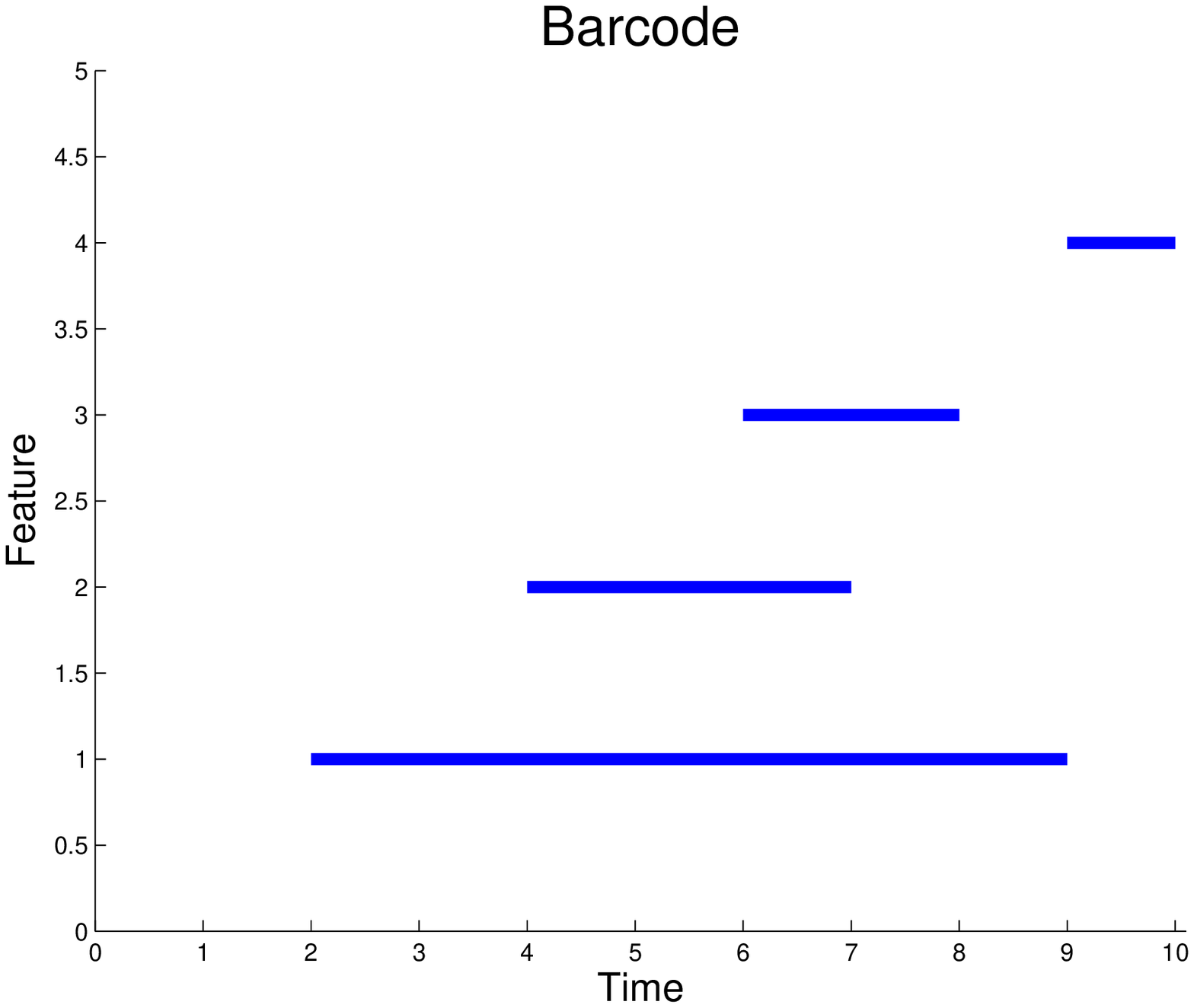} & \includegraphics[scale=0.4]{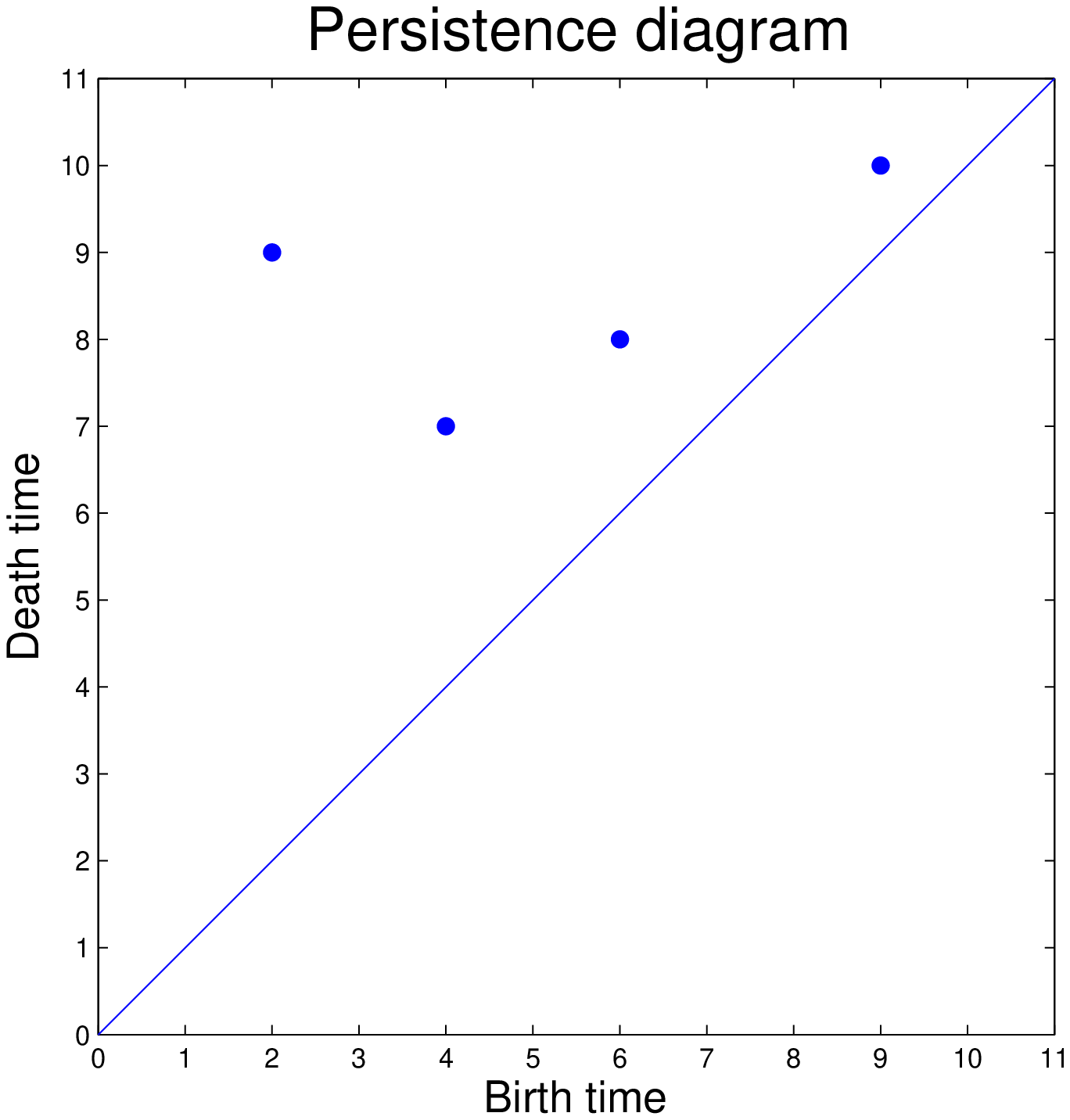} \\
\end{tabular}
\end{center}
\caption{The barcode (left) and persistence diagram (right) corresponding to the set of birth-death intervals  $\{[2,9], [4,7], [6,8], [9,10] \}$. \label{EgBarcode}}
\end{figure}

One of the main theorems in the theory of zigzag persistent homology, is that such a module can be uniquely decomposed. Each $H_p(K_i)$ is a vector space, and the module in Equation \ref{ZZmod} can be decomposed into a set of `interval modules', each consisting of one-dimensional vector spaces, for some range $[b,d]$, where $1 \leq b \leq d \leq n$, and zeros outside of this range (see \cite{carlsson2010} for details). The intervals in this decomposition are interpreted as the lifetimes of individual homological features in the sequence, which are summarized by their birth and death times ($b$ and $d$). In the sensor network setting, the decomposition of the zigzag persistence module for the first homology gives a list of birth and death times of the one-dimensional homological features in the sequence. These homological features describe the time-varying coverage the network, in a way described precisely in Section \ref{BarcodeDescriptor}. The multi-set of birth and death times
\[ \Pers(\mathcal{K}) = \{[b_j,d_j] \} \]
is the zigzag persistence of our sequence of spaces, and is represented pictorially in two common ways. The first is a \textit{barcode} where the $x$-axis represents time $t$, the $y$-axis represents individual homological features, and each feature is depicted as a horizontal line from its birth time ($b_i$) to death time ($d_i$). The second visual representation is a \textit{persistence diagram}, which plots the points $(b_i,d_i)$ on two-dimensional coordinate axes. Thus, all points lie above the diagonal (death occurs after birth), and points further from the diagonal indicate longer lifetimes. As an example, the barcode and persistence diagram corresponding to $\Pers(\mathcal{K}) = \{[2,9], [4,7], [6,8], [9,10] \}$ are shown in Figure \ref{EgBarcode}.

This output of a discrete set of birth and death times for homological features can be used to quantify the time-varying coverage for a given dynamic sensor network, as described in the following section.

\subsection{Barcodes as descriptors of coverage}\label{BarcodeDescriptor}

Here we describe how the framework of zigzag persistent homology can be used to describe information about time-varying coverage in a dynamic sensor network. Section \ref{SimpHomSensor} described how a sensor network is represented as a simplicial complex derived from the communication graph, and the first homology of this complex is used to determine coverage of the network. We now consider a sensor network that is varying with time, whose communication graph (and thus its associated simplicial complex) is available at a sequence of discrete time points. It is assumed that each sensor has a unique node identification number in $\{ 1,\ldots,n \}$, and so a correspondence can be made between the simplicial complex at one time point and the next.

Given the simplicial complexes at two consecutive time points $t_i$ and $t_{i+1}$, we do not have a direct inclusion map $K_{t_i} \rightarrow K_{t_{i+1}}$ or $K_{t_i} \leftarrow K_{t_{i+1}}$, because there may be a number of simplices that are present in $K_{t_i}$ but not in $K_{t_{i+1}}$, and vice versa. To employ the machinery of zigzag persistent homology, we require inclusion maps (either forward or backward) between consecutive spaces. To achieve this we map through the union space $K_{t_i} \cup K_{t_{i+1}}$, with each of the simplicial complexes $K_{t_i}$ and $K_{t_{i+1}}$ mapping by inclusion into $K_{t_i} \cup K_{t_{i+1}}$, as shown in Equation \ref{zzComplexes}. For a set of $T$ snapshots at time points $t_1, t_2, \ldots, t_T$, we thus obtain the sequence of simplicial complexes:

\begin{equation}\label{zzComplexes}
\begin{array}{rlrclrccclrl}
& \multicolumn{2}{c}{(K_{t_1} \cup K_{t_2})} & & \multicolumn{2}{c}{(K_{t_2} \cup K_{t_3})} & & & & \multicolumn{2}{c}{(K_{t_{T-1}} \cup K_{t_T})} & \\
& \nearrow & \nwarrow & & \nearrow & \nwarrow & & & & \nearrow & \nwarrow & \\
K_{t_1} & & & K_{t_2} & & & \mbox{ } \cdot \mbox{ } & \cdot & \mbox{ } \cdot \mbox{ } & & & K_{t_T} \\
\end{array}
\end{equation}

and the associated zigzag persistence module:

\vspace{-3mm}

\begin{small}
\[ \begin{array}{rlrclrccclrl}
& \multicolumn{2}{c}{H_1(K_{t_1} \cup K_{t_2})} & & \multicolumn{2}{c}{H_1(K_{t_2} \cup K_{t_3})} & & & & \multicolumn{2}{c}{H_1(K_{t_{T-1}} \cup K_{t_T})} & \\
& \nearrow & \nwarrow & & \nearrow & \nwarrow & & & & \nearrow & \nwarrow & \\
H_1(K_{t_1}) & & & H_1(K_{t_2}) & & & \mbox{ } \cdot \mbox{ } & \cdot & \mbox{ } \cdot \mbox{ } & & & H_1(K_{t_T}) \\
\end{array} \]
\end{small}

\begin{figure}[htp]
\begin{center}
\begin{tabular}{l}
\includegraphics[width=140mm,height=80mm]{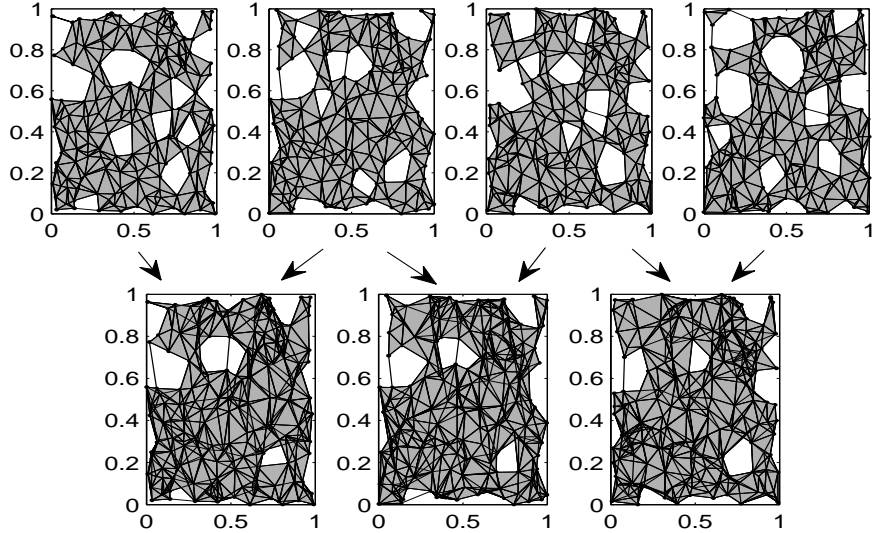} \\
\end{tabular}
\end{center}
\caption{Rips complexes for the coverage region at four time points (top row), as well as the union complexes that the zigzag algorithm maps through (bottom row) \label{zzrips}}
\end{figure}

See Figure \ref{zzrips} for Rips complexes of four time points in a dynamic network, with the union complexes used for mapping through.
From the zigzag persistence module above, the $\Pers(\mathcal{K}) = \{[b_j,d_j]\}$ containing the lifetimes of the homological features can be computed. At this stage it is worth noting the distinction between homology classes and coverage holes, as well as the lack of a straightforward definition of what a `time-varying coverage hole' is.

One characterization of a time-varying coverage hole is known as an `evasion path', which means that there exists a spatiotemporal path which remains uncovered. This can be thought of as a path that an intruder could travel in order to avoid detection. Homological methods have been used \cite{deSilva2006} to give necessary conditions for such an evasion path to exist, using the same coordinate-free setting assumed here. More recently, it has been further shown \cite{Adams2013} that the opposite implication does not hold. Specifically, the presence of interval $[b,d]$ in the zigzag persistence output, does not imply that there exists an evasion path over that same interval. Because of this, there is not a one-to-one relationship between birth-death intervals and evasion paths. Further, when a hole opens, travels around in space, and eventually closes, it is clear what is meant by `time-varying coverage hole', but in some cases a single hole may split into two, or two holes may merge into one. Because of the ambiguity introduced about `which hole' is obliterated or preserved during these processes, it is unclear what constitutes a single coverage hole over time. The tracking of lifetimes of homological features in zigzag persistence can, however, be done unambiguously, and although these are not interpreted as individual time-varying coverage holes, they are related to the coverage region in the following ways:

\begin{enumerate}
\item If a coverage hole appears at time $b$ and remains isolated (does not split or merge with any other holes) until it disappears at time $d$, then the exact interval $[b,d]$ will be present in $\Pers(\mathcal{K})$. This means that in the case where a time-varying coverage hole is well defined, its lifetime is exactly represented in the barcode.
\item If an evasion path exists over interval $[b,d]$, then there exists an interval in $\Pers(\mathcal{K})$ containing $[b,d]$. This means that no evasion paths will be missed.
\item If $\Pers(\mathcal{K}) = \{[b_j,d_j] \mbox{ } | \mbox{ } j \in 1,\ldots,m\}$ are the intervals output from zigzag persistence, then define $\Lambda_i = \{ j \in 1,\ldots,m\ \mbox{ } | \mbox{ } b_j \leq i \leq d_j\}$ to index the set of intervals which are `alive' at time $i$. Then
    \[ |\Lambda_i| = \beta_1(K_i) \]
    (the number of intervals alive at any time point is equal to the number of holes in the simplicial complex at that time).
\end{enumerate}

In light of this, we propose the use of the barcode/persistence diagram from zigzag persistent homology as a descriptor of the coverage of a network over time. In general, more bars and longer bars correspond to worse coverage. Since the computation only requires the Rips complexes (i.e. adjacency matrices of the communication graph) at each time point, this measure can be computed without requiring coordinates or distances between the sensors. In particular, summary statistics such as maximum and mean lifetimes of homological features can be computed, in addition to analysis of the barcode/persistence diagram as a whole. Metrics (such as the Wasserstein or bottleneck distances on persistence diagrams - see \cite{cohen2010lipschitz}, \cite{cohen2007stability}) have also been developed to compute pairwise distances between two persistence diagrams, which allows for quantification of differences between the coverage patterns of multiple time-varying networks.

At present, the only methods \cite{liu2013} available for analyzing coverage in dynamic sensor networks are to measure the coverage directly (using geometric information), and compute the proportion of uncovered area at each time point, or the proportion uncovered over a time interval (including a point as covered if it has been covered at any time during the interval).

In the following sections we describe how $\Pers(\mathcal{K})$ can be used effectively to quantify coverage in mobile sensor networks, and we illustrate its use in comparing mobility models. We further propose a method which is used in conjunction with the current zigzag persistence algorithm, and obtains specific representative cycles for each bar, which are adaptively tracked over time, and can then be used to obtain coarse size information about the holes present in the network.

\subsection{Comparing mobility models}\label{MobilityModels}

We present here some results on how the output from zigzag persistence can be used to characterize the coverage obtained by different mobility models for dynamic sensor networks. The analysis of coverage properties of mobility models previously used geometric descriptors to derive analytical results about the network, such as the limiting distribution of the nodes, the expected time-until-coverage for uncovered points, or expected proportion of uncovered area \cite{liu2013}. Ours is the first method which can additionally describe the dynamics of the coverage, in terms of the formation, duration, and behavior of coverage holes over time.

The two models we discuss are based on Brownian motion, and straight-line motion. For each of these, it is assumed that the nodes move independently from one another.

\subsubsection{Mobility patterns: Discrete Brownian and Straight Line}\label{MPsetup}

\emph{Discrete Brownian:} One model used to approximate the random movement of nodes in a large scale sensor network assumes each node moves independently and identically distributed (i.i.d.) according to a Brownian motion (eg. \cite{peres2011}). This is modeled in discrete-time by allowing each sensor to move according to a $2$-dimensional Gaussian distribution at each time step (with variance proportional to the time increment).

\emph{Straight Line:} A second commonly-used i.i.d. mobility model has each node choose an initial random direction and velocity, and then proceed (indefinitely) along this course (\cite{liu2005}, \cite{liu2013}). In this setting, at $t=0$ each node randomly chooses a direction $\theta \in [0, 2\pi)$ according to some distribution described by $f_\Theta(\theta)$, and randomly chooses a speed $v \in [v_{min}, v_{max}]$ according to a distribution described by $f_V(v)$. Typically $f_\Theta(\theta)$ and $f_V(v)$ are uniform distributions over their respective intervals, but other distributions are also possible. To make our simulations using the Discrete Brownian and Straight Line mobilty models directly comparable, we will choose the initial vector describing the velocity and direction for the Straight Line model from the same 2-dimensional Gaussian distribution used for each time step in the Discrete Brownian.

\subsubsection{Simulations}

Simulations were performed in a bounded region $[0,1]^2$, and for both mobility patterns the initial positions of the nodes were drawn from a uniform distribution over the region. When the movement of a sensor causes it to reach the boundary of the region, it bounces off with elastic (billiard-like) collisions, which will cause a change in the direction but not the speed.

\begin{figure}[htp]
\begin{center}
\begin{tabular}{cc}
\includegraphics[scale=0.35]{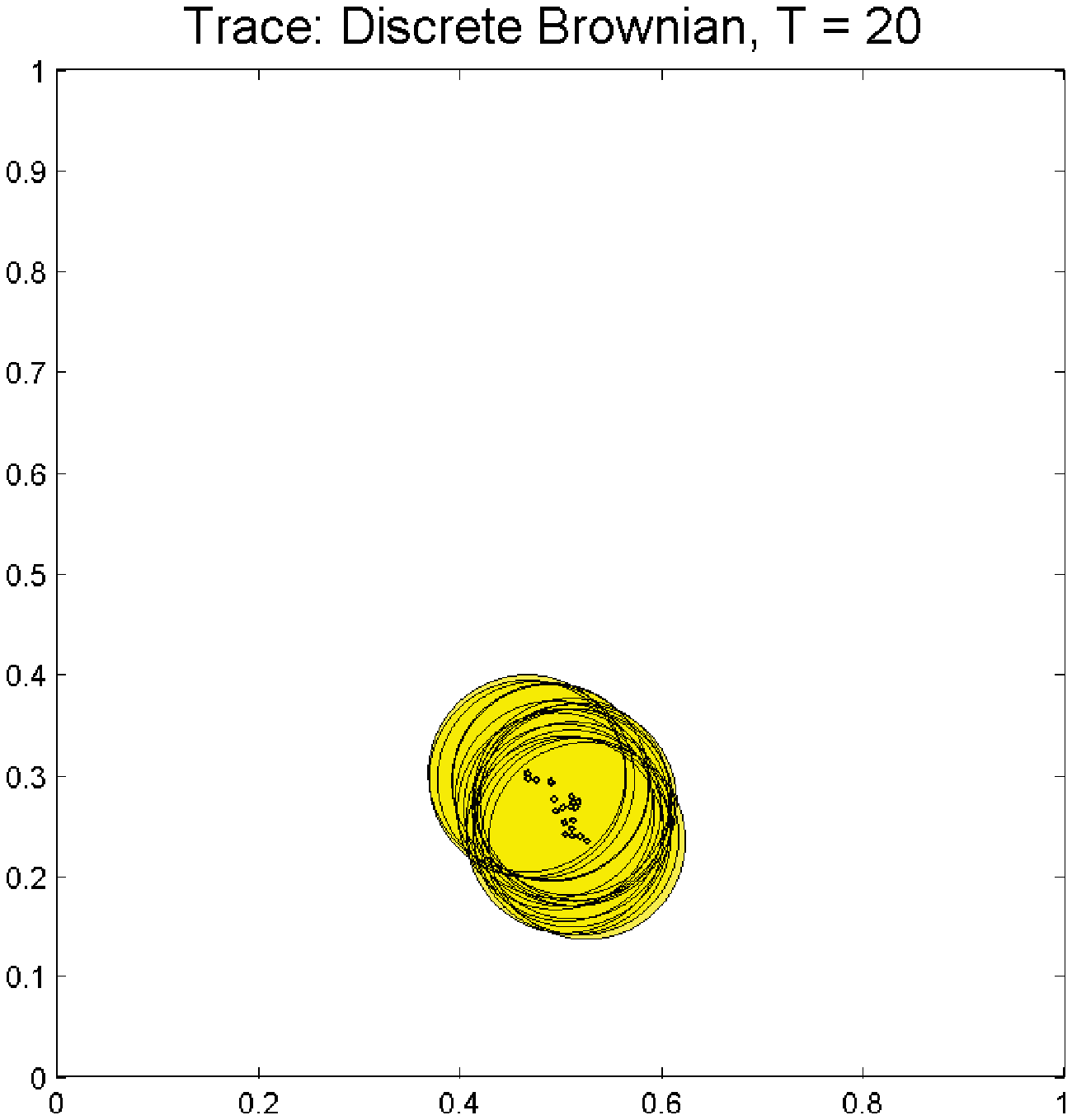} & \includegraphics[scale=0.35]{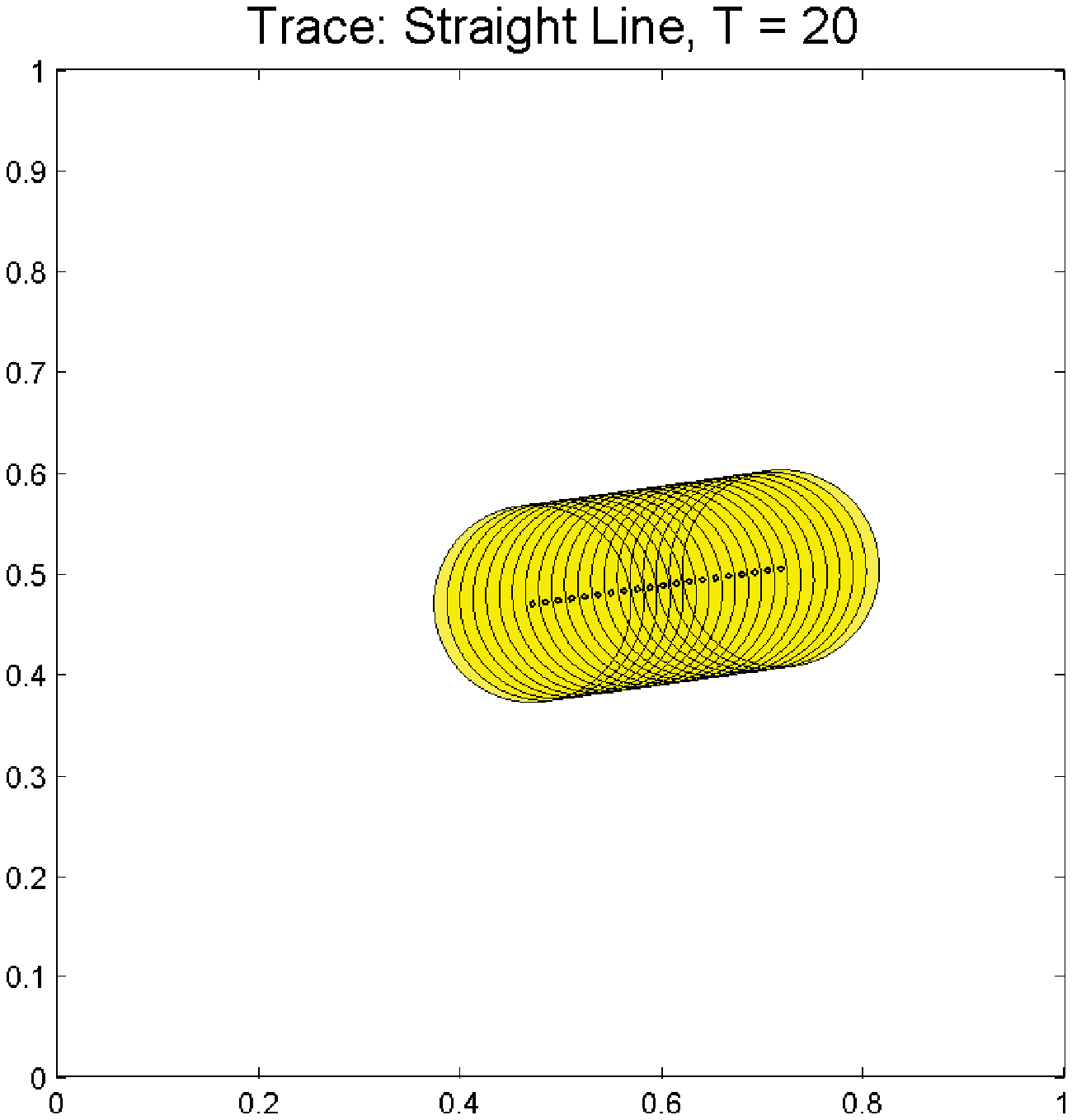} \\
\includegraphics[scale=0.35]{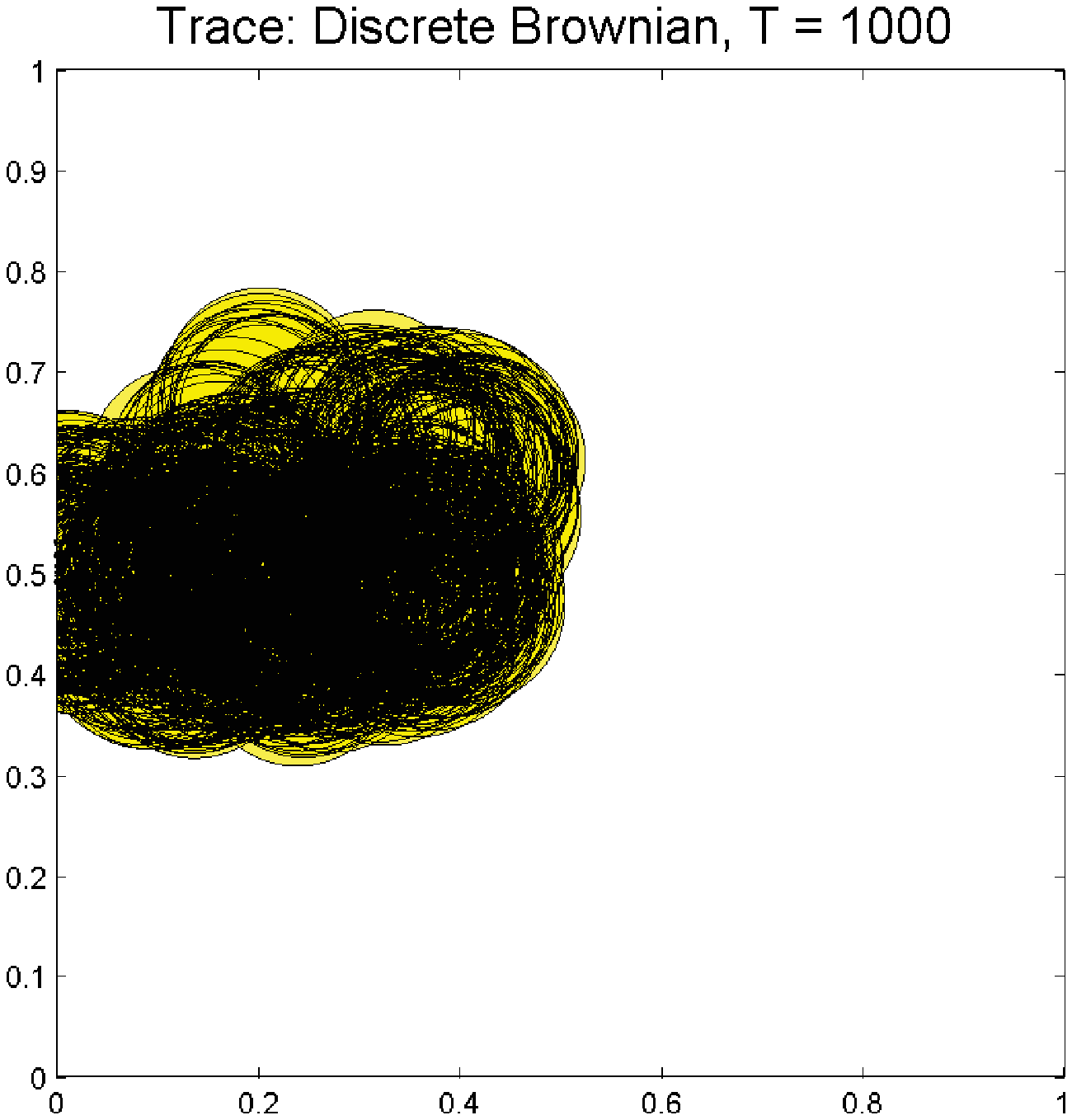} & \includegraphics[scale=0.35]{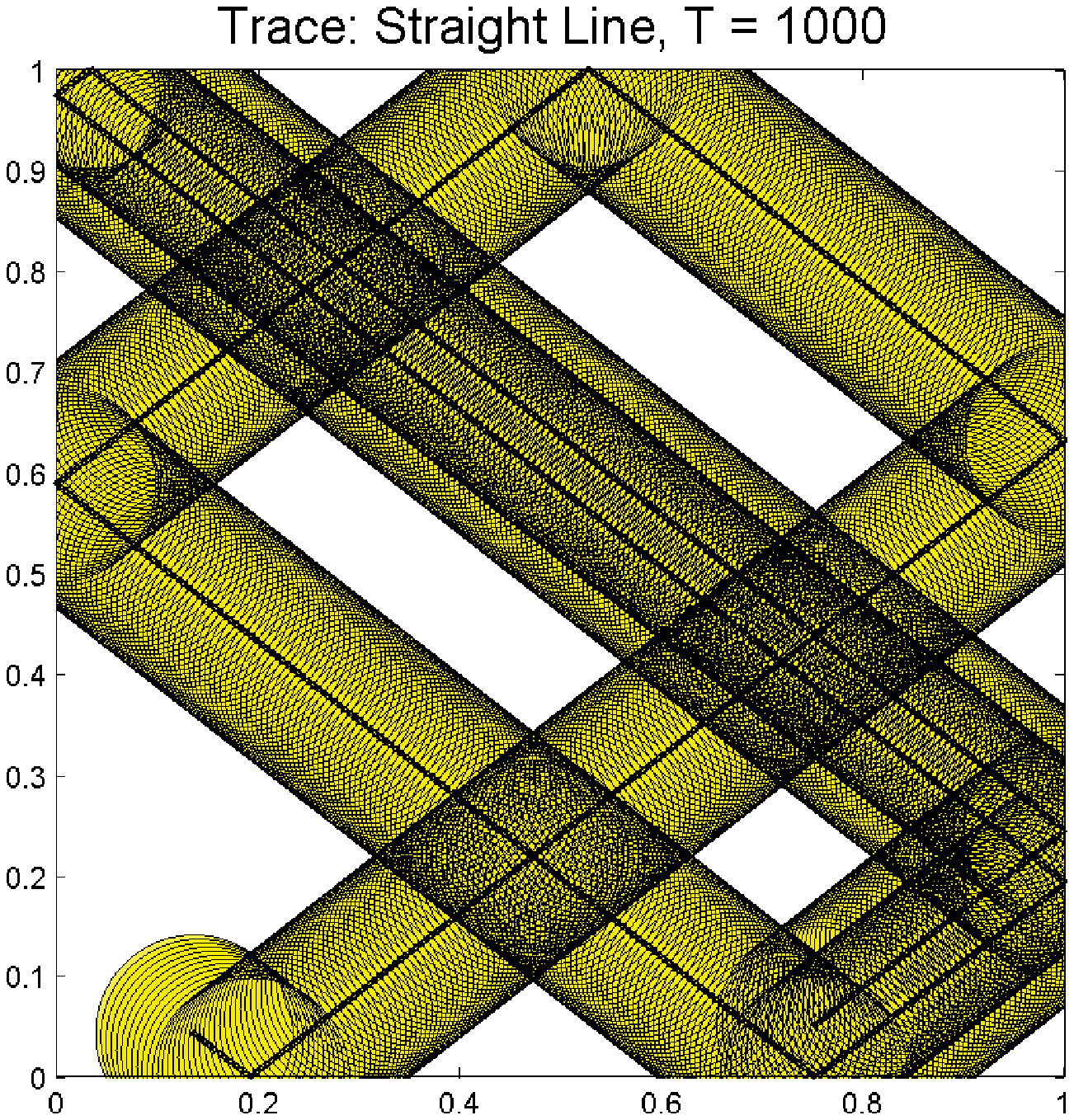} \\
\end{tabular}
\end{center}
\caption{The path of a single sensor following the Discrete Brownian (left) or Straight Line (right) mobility pattern, described in Section \ref{MPsetup} for $T=20$ time points (top row) or $T=1000$ time points (bottom row). \label{MPtraces}}
\end{figure}

Using $n=100$ nodes, over an interval of $T=50$ time points, 50 replications were generated for each mobility pattern. The simulations were paired, in that the initial coordinates of the sensors were the same for the two patterns, and were generated independently for each replication. All pairings for computing differences between the patterns, and computing the Wilcoxon signed rank were done by pairing the two replications (one from each mobility pattern) with the same initial configuration of sensors. The 2-dimensional Gaussian distribution used to initialize the movement in Straight Line pattern, and at each time point for the Discrete Brownian, had mean zero and standard deviation $0.1r$ (where $r$ is the radius of the coverage disk for each sensor). A trace of one sensor following each of the mobility patterns for $T = 20$ and $1000$ time points is shown in Figure \ref{MPtraces}.

For each replication, the sequence of $T$ simplicial complexes $K_1,\ldots,K_T$ (representing the sensor network at time points $1,\ldots,T$) are used along with the union complexes $K_i \cup K_{i+1}$ for $i=1,\ldots,T-1$ to build the sequence $\mathcal{K}$, as in Equation \ref{zzComplexes}. This is used as input to compute the zigzag persistence birth-death intervals $\Pers(\mathcal{K}) = \{[b_j,d_j]\}$ and associated representative cycles. Statistical analysis was performed to test for differences in the coverage properties of the two patterns, using both traditional coverage measures and descriptors obtained from our homological methods. The variables used for analysis are described in Section \ref{MPstats}.

For both of these mobility patterns, the spatial distribution of the nodes is stationary in time, uniformly distributed in $[0,1]^2$ at each time point \cite{LeBoudec2005}. Because of this, all point-wise coverage statistics, such as average proportion of uncovered area or average number of coverage holes at any time point, should be the same for the two mobility patterns. What we expect might differ between the two groups is the way in which coverage holes form, merge, split, and close, which can be detected in differences in the distribution of lifetimes of homology classes (the number and lengths of the intervals in $\Pers(\mathcal{K})$. i.e. bars in the barcode).

\subsubsection{Coverage statistics}\label{MPstats}

The output of zigzag persistent homology on the sequence of simplicial complexes gives a set of birth-death intervals $\Pers(\mathcal{K}) = \{[b_j,d_j]\mbox{ } | \mbox{ } j=1,\ldots,m\}$ for each replication (each represented as a barcode). Statistical analysis is performed using 50 barcodes for the Discrete Browian simulation runs, and 50 barcodes for the Straight Line simulations. The summary statistics extracted from the barcodes are described below, with results of the statistical analysis on theses variables detailed in Section \ref{MPresults}.

\vspace{2mm}
\begin{tabularx}{\linewidth}{lX}
\textbf{Variable} & \textbf{Description} \\
\textit{barcode} & An $m$-by-2 matrix containing the set of birth-death pairs for a given simulation run (the number $m$ will vary, run to run). This is the main descriptor of the homological features tracked using zig-zag persistence. \\
\textit{LTcounts} & A $T$-length vector with the counts of how many bars have length (lifetime) $t$, for $t = 1,\ldots,T$ in a given simulation run. i.e.) \textit{LTcounts}(1) is the number of bars that persist for only a single timepoint, \textit{LTcounts}(2) is the number of bars with a lifetime of 2,$\ldots$, \textit{LTcounts}($T$) is the number of bars that persist over the entire simulation run. \\
\textit{\# of bars} & (scalar) The number $m$ of birth-death intervals $\{[b_j,d_j]\mbox{ } | \mbox{ } j=1,\ldots,m\}$ in a barcode for a given simulation run. \\
\textit{sum of bars} & (scalar) The sum $\sum_{j=1}^m (d_j - b_j)$ of all bar lengths (lifetimes) in a given simulation run. \\
\textit{interval coverage} & A $T$-length vector giving the proportion of the simulation region covered by time $t$, for $t=1,\ldots,T$ in a given simulation run. A point in the simulation region is considered covered by time $t$, if it is covered at any point in the interval $[0,t]$. \\
\end{tabularx}
\vspace{2mm}

\begin{figure}[htp]
\begin{center}
\begin{tabular}{cc}
\includegraphics[scale=0.35]{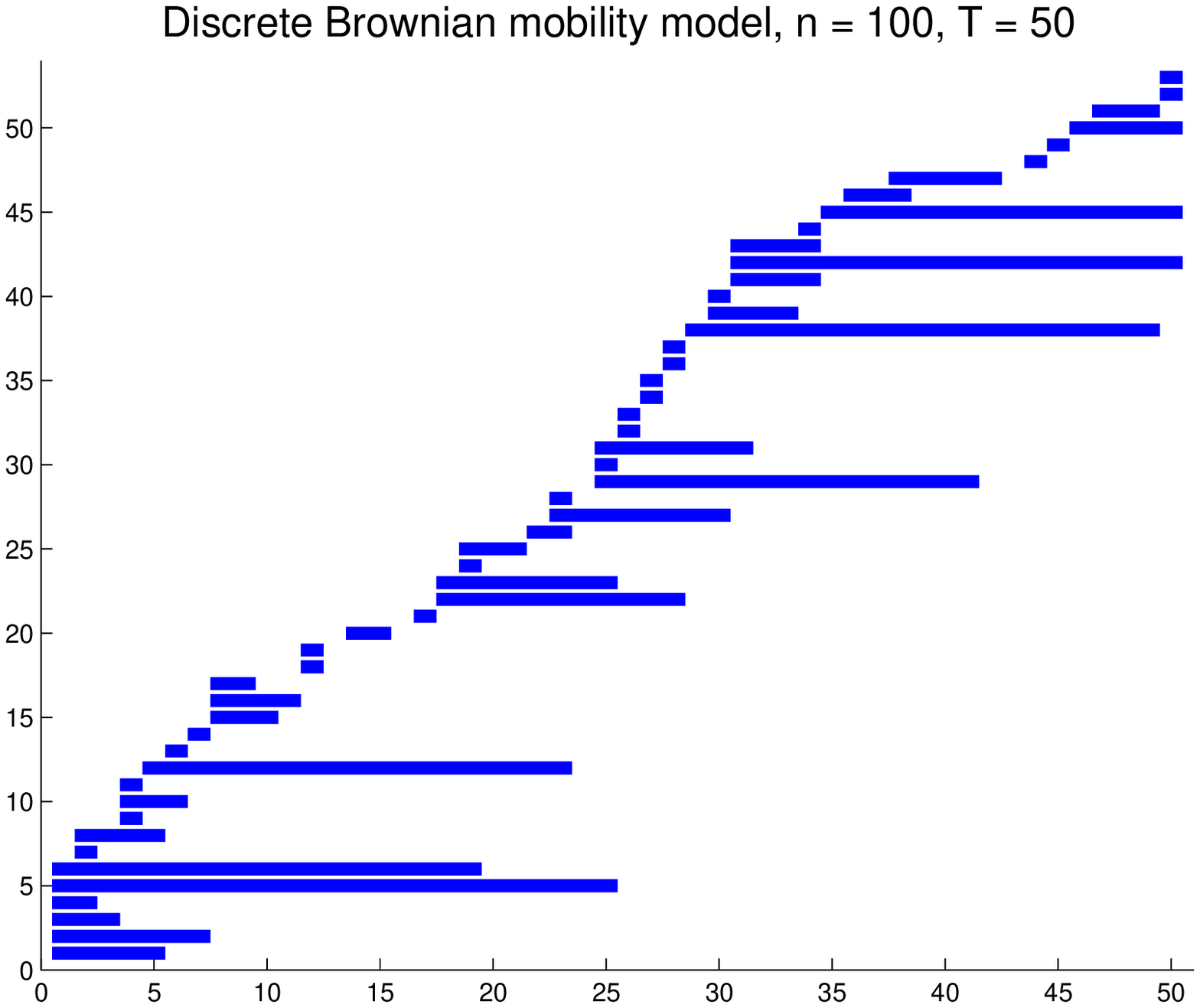} & \includegraphics[scale=0.35]{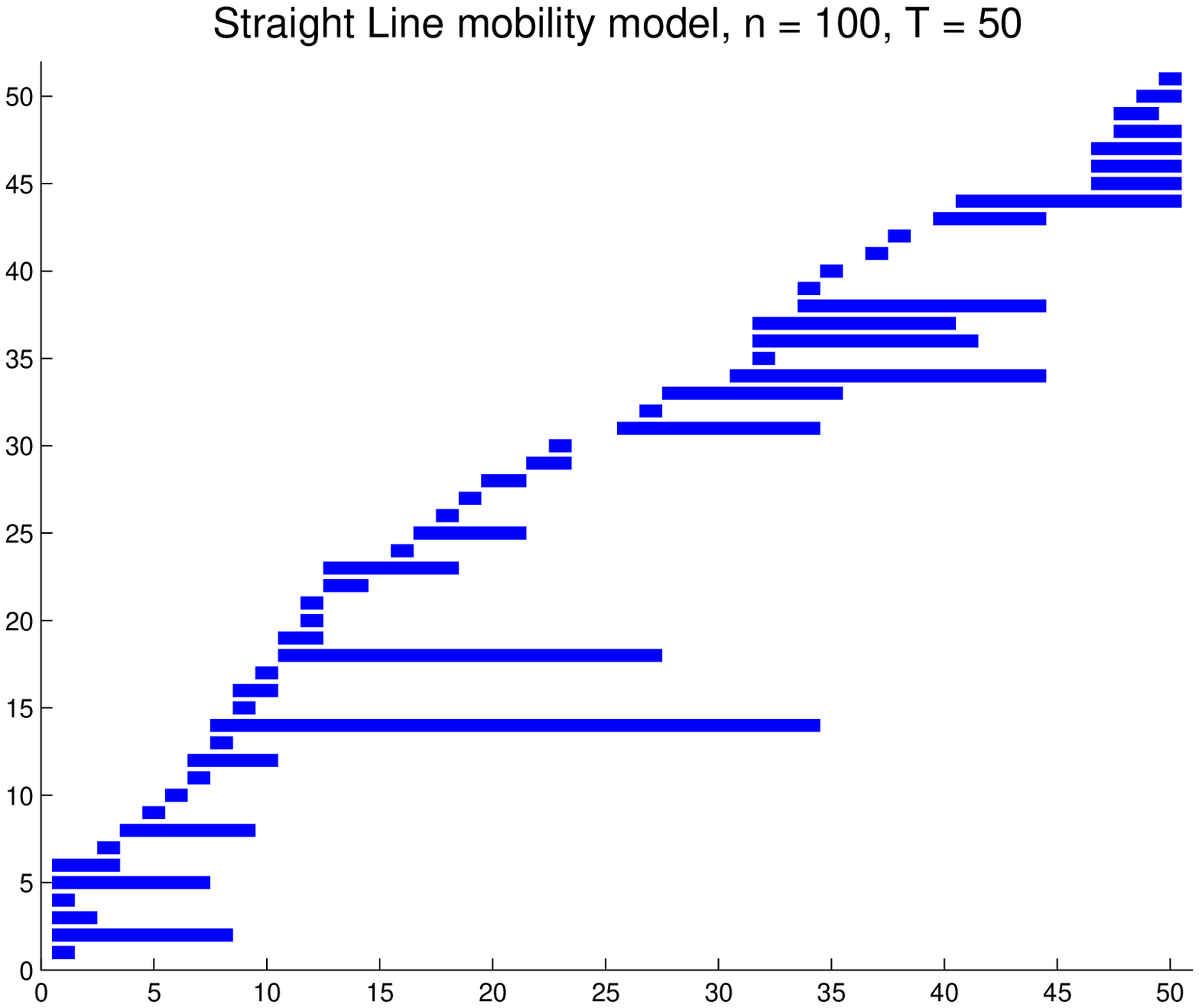} \\
\end{tabular}
\end{center}
\caption{Barcodes obtained for a single simulation run of the Discrete Brownian (left) and Straight Line (right) mobility patterns. \label{MPbarcodes}}
\end{figure}

Example barcodes from for one simulation run from each mobility pattern are shown in Figure \ref{MPbarcodes}. The colors of the bars will be used later when identifying bars with specific representative cycles, but are unimportant for now. Note that looking at a single pair of barcodes does not give a clear indication of whether there is a difference between the time-varying first homology of the two patterns, and we take a deeper look using statistical analysis. Since the mobility patterns are time-stationary, the variables involving the lifetimes of the homological features are of greatest interest, as opposed to those which depend on the specific birth or death time. This is why we do not analyze the barcodes directly for statistically significant differences, but instead look at lifetimes of the homological features.

\subsubsection{Results}\label{MPresults}

The results given in this section are each followed by a set of brackets containing the mean and standard deviation (in parentheses) for each group, followed by the $p$-value obtained from the nonparametric Wilcoxon signed rank test. This test was used because the assumption of normality necessary for a paired samples t-test was typically violated.

\begin{figure}[htp]
\begin{center}
\begin{tabular}{cc}
\includegraphics[scale=0.35]{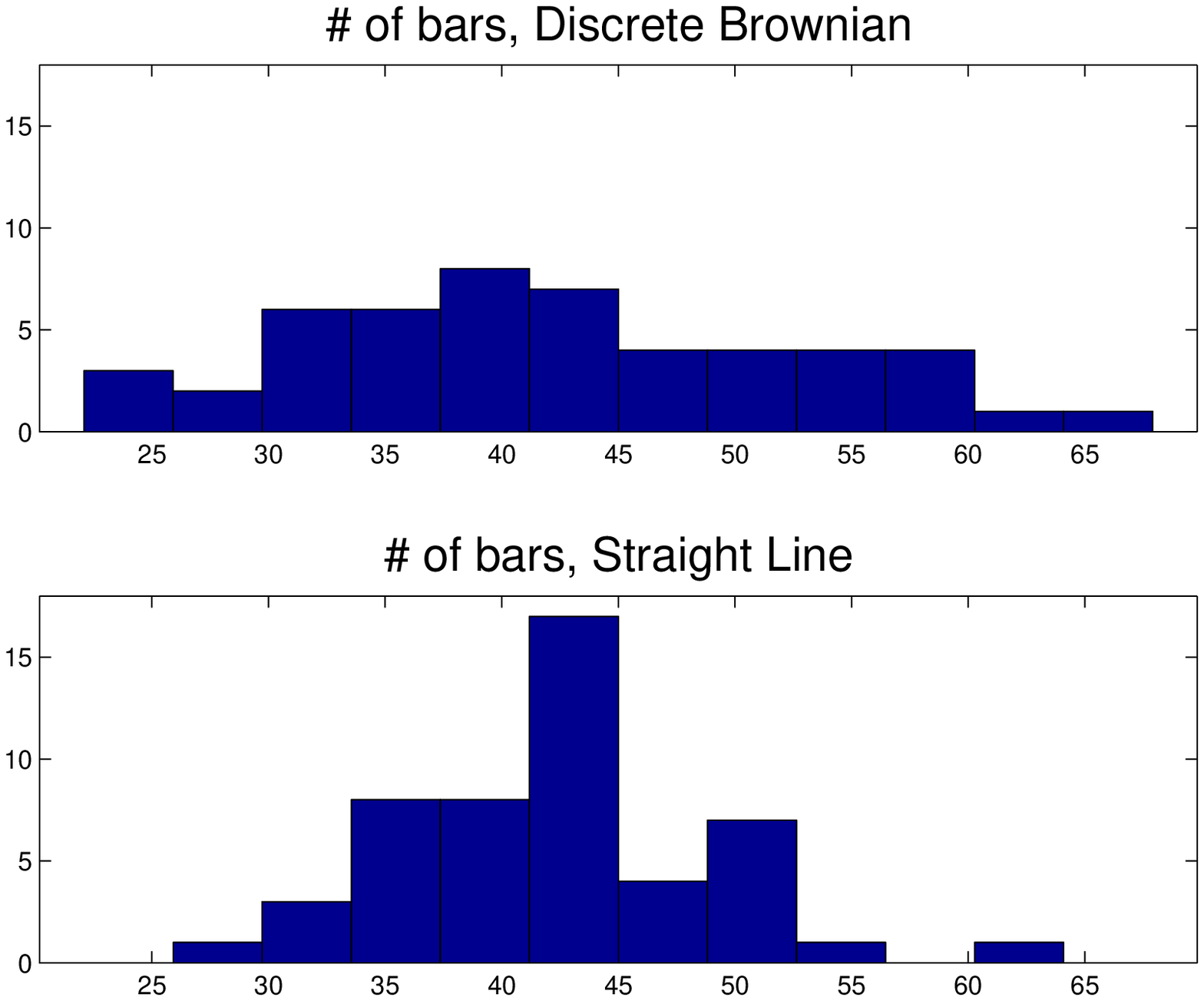} & \includegraphics[scale=0.35]{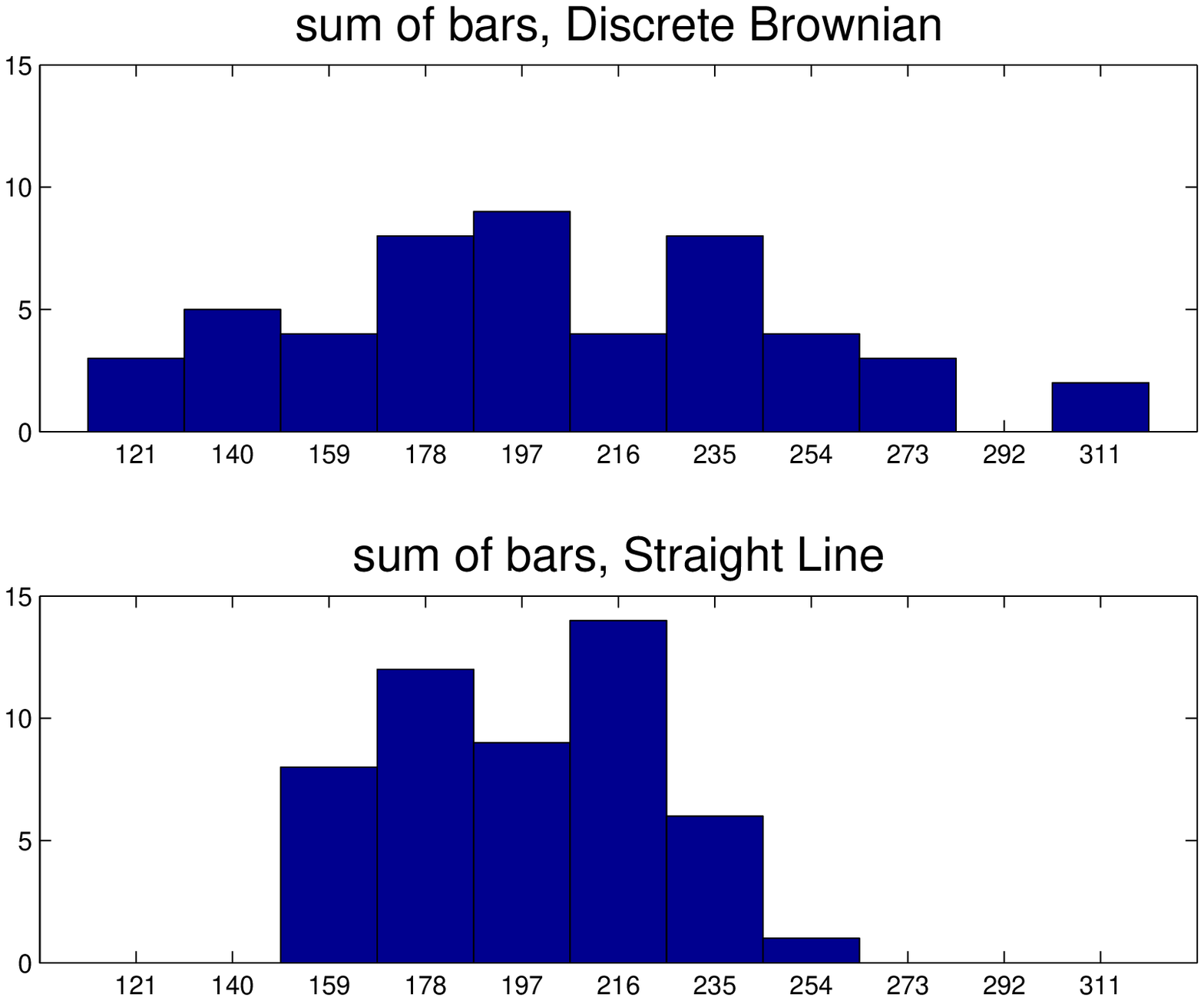} \\
\end{tabular}
\end{center}
\caption{Histograms of the counts for \emph{\# of bars} (left) and \emph{sum of bars} (right) for the Discrete Brownian (top row) and Straight Line (bottom row) mobility patterns. Note that the mean for the two patterns are similar for both variables, but the variance is greater in the Discrete Brownian pattern. \label{BarHists}}
\end{figure}

Comparing the Discrete Brownian and Straight Line mobility patterns, there is no statistically significant difference between the two groups for \emph{\# of bars} [DB=42.42(10.52), SL=42.6(6.42), $p=0.70$], or \emph{sum of bars} [DB=201.96(48.19), SL=197.58(25.50), $p=0.71$]. There is however a statistically significant difference in the variance of the two patterns for both \emph{\# of bars} ($p<0.001$) and \emph{sum of bars} ($p<0.0001$), with the Discrete Brownian pattern having larger variability than the Straight Line. Histograms for the distribution of \emph{\# of bars} and \emph{sum of bars} for the two patterns are shown in Figure \ref{BarHists}

\begin{figure}[htp]
\begin{center}
\begin{tabular}{cc}
\includegraphics[scale=0.35]{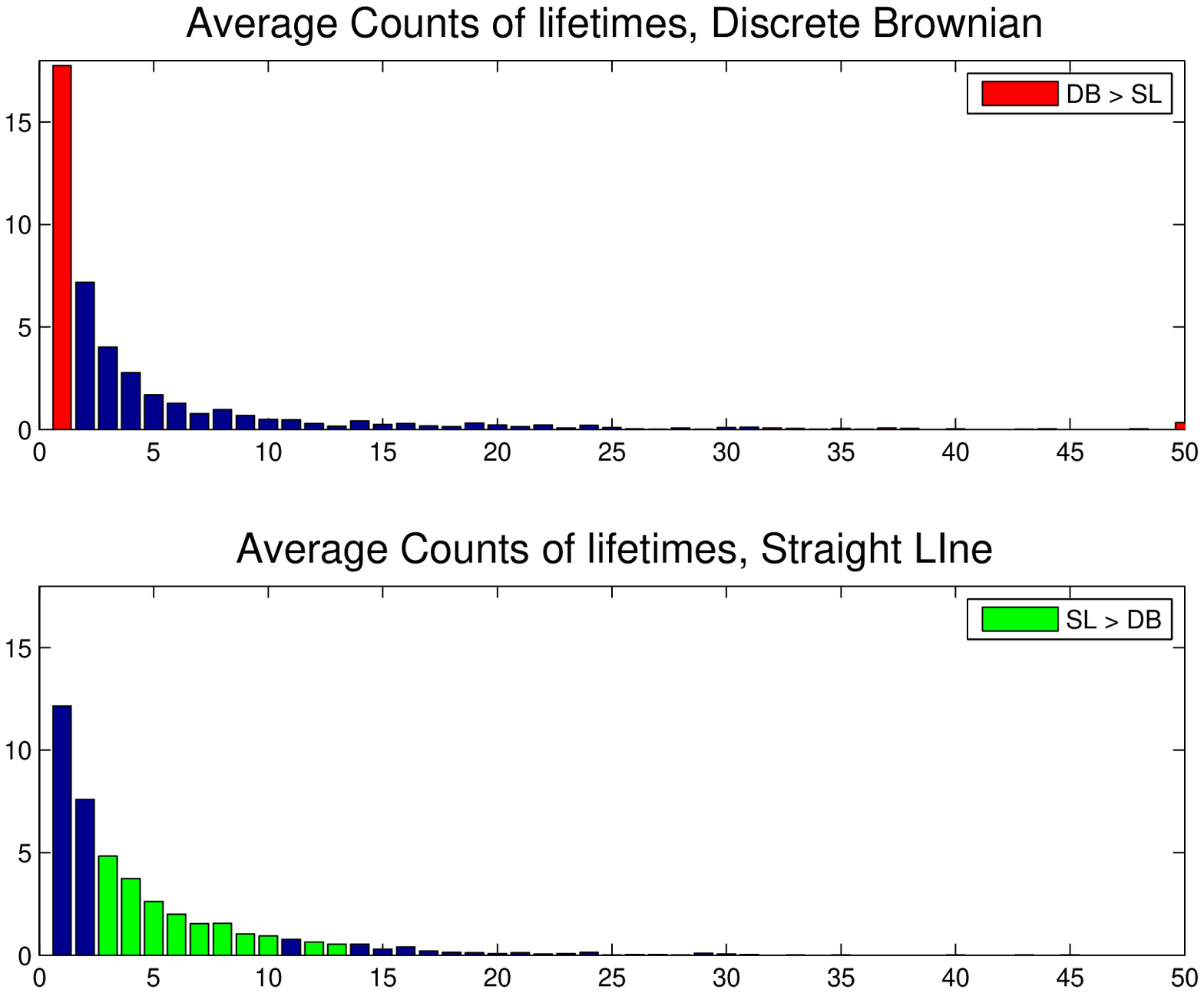} & \includegraphics[scale=0.35]{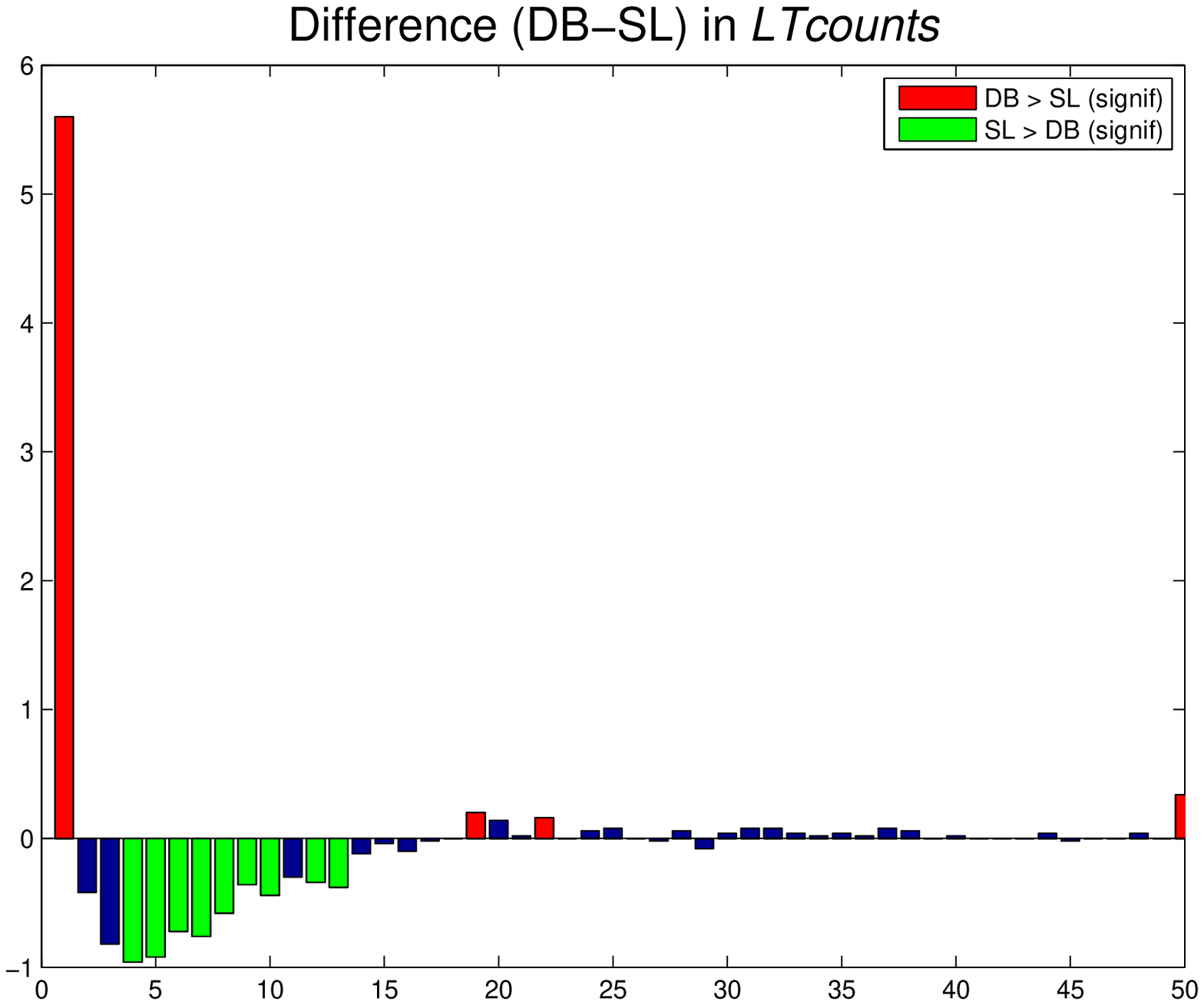} \\
\end{tabular}
\end{center}
\caption{\textit{LTcounts} for the Discrete Brownian (top left) and Straight Line (bottom left) mobility patterns, as well as the paired difference in \textit{LTcounts} (right), with the lifetimes whose frequency has a statistically significant difference between the groups highlighted. The lifetimes that occur more frequently in the Discrete Brownian pattern ($t=1,19,22$ and ,$50$) are highlighted in red in the top plot, and those that occur more frequently in the Straight Line pattern ($t=4,\ldots,13$) are highlighted in green in the bottom plot. \label{LTcounts}}
\end{figure}

The counts of lifetimes are distributed differently for the two groups. The Discrete Brownian mobility pattern has a significantly higher number of very short lifetimes (for $t=1$, $p<0.0001$), and very long lifetimes (for $t=50$, $p<0.001$). A few long lifetimes ($t=19$ and 22) also more frequent in the Discrete Brownian pattern, with moderate significance ($p<0.05$). The Straight Line mobility pattern has a significantly higher number of short-medium length lifetimes ($t=4,\ldots,13$, all have $0.001<p<0.05$). Even after a Bonferroni correction for multiple hypothesis testing, the differences for $t=1$ and $50$ are still statistically significant (at the level $p<0.001$). Histograms of \emph{LTcounts} for the two mobility patterns are shown in Figure \ref{LTcounts}, with the lifetimes that show statistically significant differences highlighted.

\begin{figure}[htp]
\begin{center}
\begin{tabular}{c}
\includegraphics[scale=0.5]{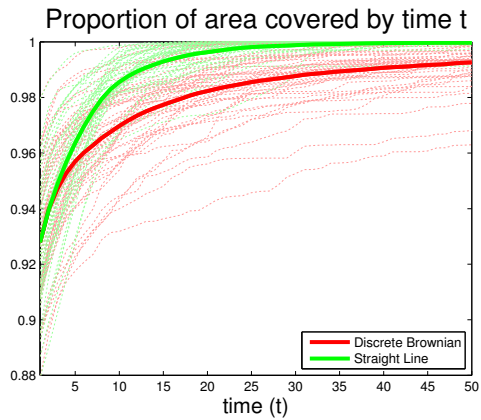} \\
\end{tabular}
\end{center}
\caption{The proportion of area covered over time interval $[0,t]$ for each of the two mobility patterns, with all simulation runs overlaid as dotted lines. Mean interval coverage is shown as a thick solid line for each mobility pattern (Discrete Brownian in red, Straight Line in green). \label{intervalCoverage}}
\end{figure}

\begin{figure}[htp]
\begin{center}
\begin{tabular}{c}
\includegraphics[scale=0.5]{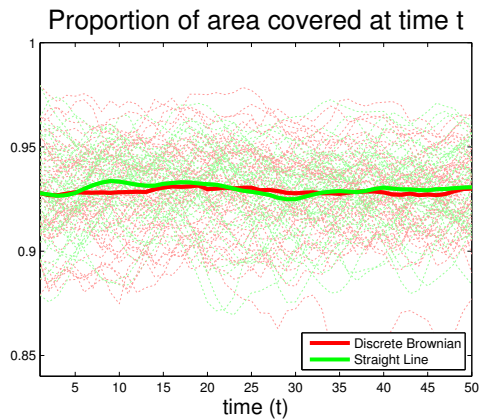} \\
\end{tabular}
\end{center}
\caption{The proportion of area covered at each time point for each of the two mobility patterns, with all simulation runs overlaid as dotted lines. Mean coverage is shown as a thick solid line for each mobility pattern (Discrete Brownian in red, Straight Line in green). \label{ProportionCovered}}
\end{figure}

The variable \textit{interval coverage} is a more traditional coverage measure, and we see that over time, the Straight Line mobility pattern will sweep out coverage of a greater proportion of the total area than the Discrete Brownian model. The interval coverage for all the replications are shown faintly in Figure \ref{intervalCoverage}, with the means for each pattern shown in bold. The difference between the two mobility patterns in \textit{interval coverage} is statistically significant for time points $t = 7,\ldots,T$ ($p< 0.001$). This is in agreement with previous work \cite{liu2005}, as well as the fact that a sensor traveling a path of fixed total length will cover the greatest area if it travels in a straight line. We additionally note that the time-point-wise coverage, measured by proportion of covered area, as expected shows no statistically significant difference between the patterns, see Figure \ref{ProportionCovered}.

\subsubsection{Discussion}
From the results presented here, it appears that although the two mobility patterns both have the same stationary distribution, and the same average energy expenditure at each time point, there is a difference in the time-varying coverage pattern displayed by the two models. For the Discrete Brownain mobility pattern, the erratic movement of the sensors results in many small holes appearing and disappearing quickly (usually present for only a single time-point). Additionally, this mobility pattern displays significantly more long-lasting coverage holes, which typically correspond to large holes that are present in the initial configuration (i.e. the mobility pattern does not fill in existing holes quickly). For the Straight Line mobility pattern, since the sensors are each following a smooth trajectory, the coverage holes seem to appear, grow, shrink and disappear smoothly, instead of appearing and disappearing immediately, or remaining uncovered for longer periods. In light of this, the Straight Line mobility model would be preferable in situations such as surveillance, or intruder detection, where it is important to quickly cover holes present in the initial deployment, and long-lasting coverage holes would prove costly. The Discrete Brownian model might be more desirable in circumstances where a thorough inspection takes precedence over time, such as in geographical surveying or environmental monitoring.

\section{Coordinate-free estimation of hole size}\label{HoleSizes}

The barcode obtained from zigzag persistence gives us a quantitative descriptor for the time-varying coverage of a network. Just as knowing the Betti number (number of holes) for a given simplicial complex tells us nothing about the hole sizes, the presence of a long bar in the barcode may or may not correspond to a large hole geometrically. Given that our network is described as a sequence of adjacency matrices (describing the simplicial complex at each snapshot, but without coordinate information), the best estimate available is the hop-length of the shortest cycle surrounding a hole. This can be obtained without having to compute the shortest cycle explicitly, by performing a hop-distance filtration on the simplicial complex (at each time point). For a simplicial complex $K$, the hop distance filtration is a nested sequence of simplicial complexes $K^1 \subseteq K^2 \subseteq \ldots$ defined as follows:

\begin{definition}
The \emph{hop distance filtration} on a simplicial complex $K$, performed up to a maximum hop distance of $m$, is a nested sequence of simplicial complexes $K^1 \subseteq K^2 \subseteq \ldots K^m$, defined inductively:
\begin{enumerate}
\item $K^1$ is the original complex $K$
\item $K^h$ contains all of the simplices of $K^{h-1}$, and adds edges between any nodes that were $h$ hops apart in $K$, as well as all possible higher-dimensional simplices (i.e. if three edges forming a triangle are present in $K^h$, the associated 2-simplex will be added to $K^h$ as well).
\end{enumerate}
\end{definition}

\begin{figure}[htp]
\begin{center}
\begin{tabular}{ccc}
\includegraphics[scale=0.35]{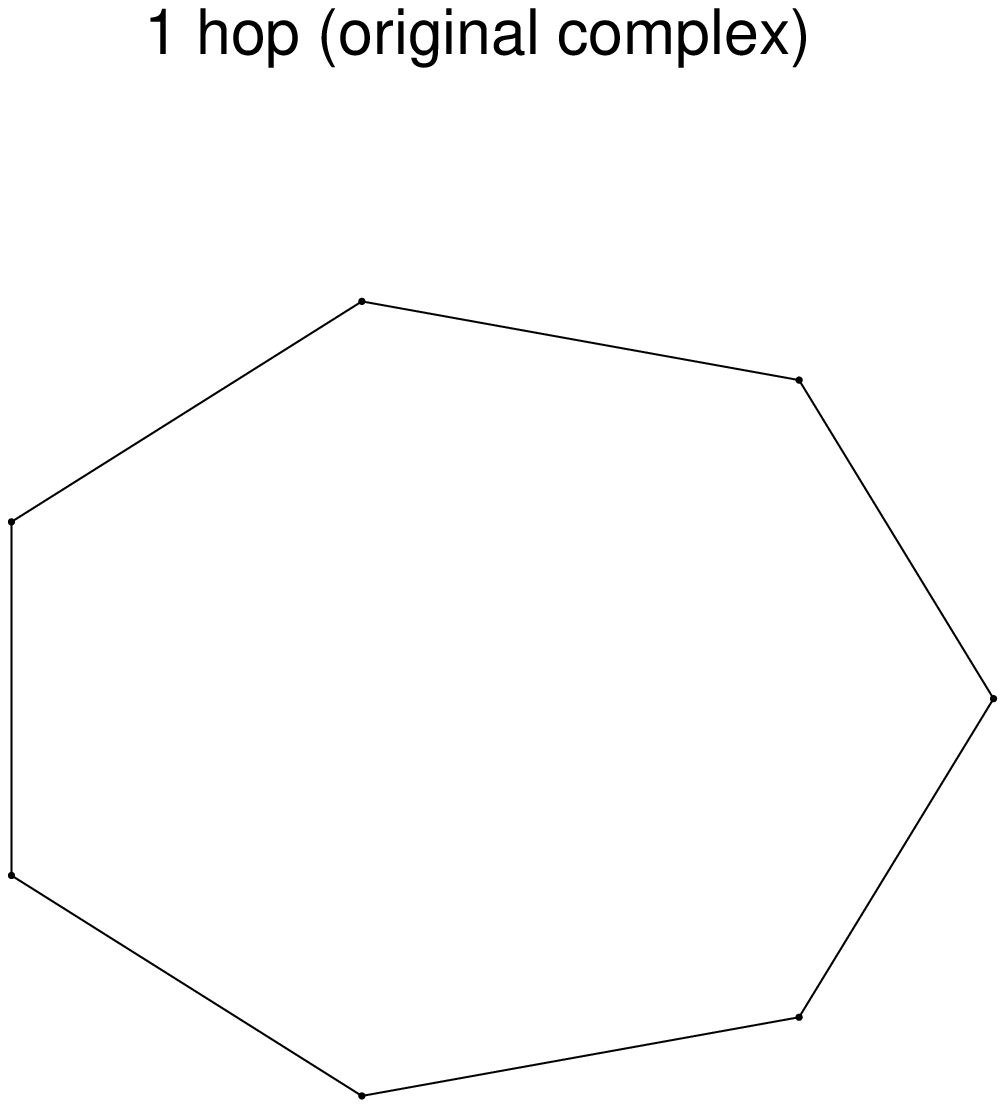} & \includegraphics[scale=0.35]{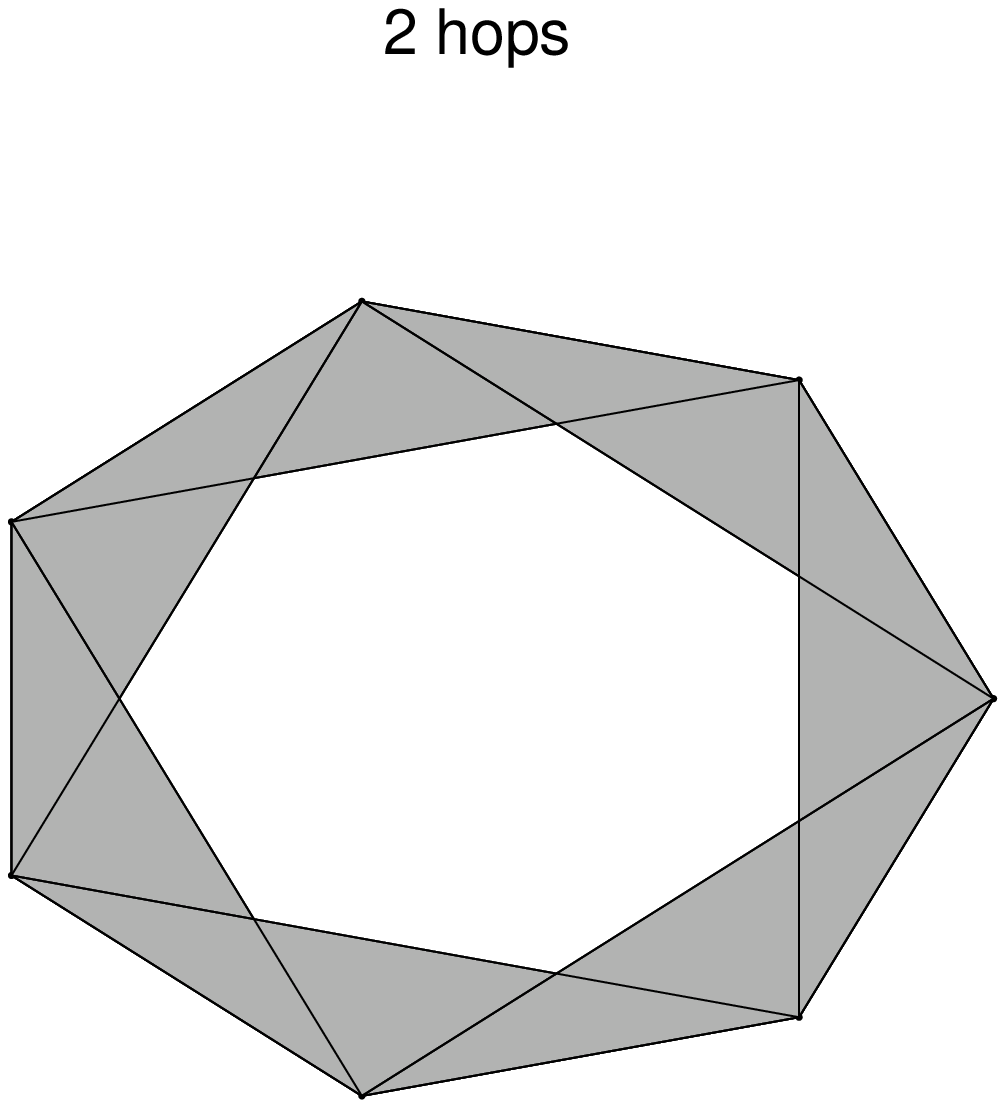} & \includegraphics[scale=0.35]{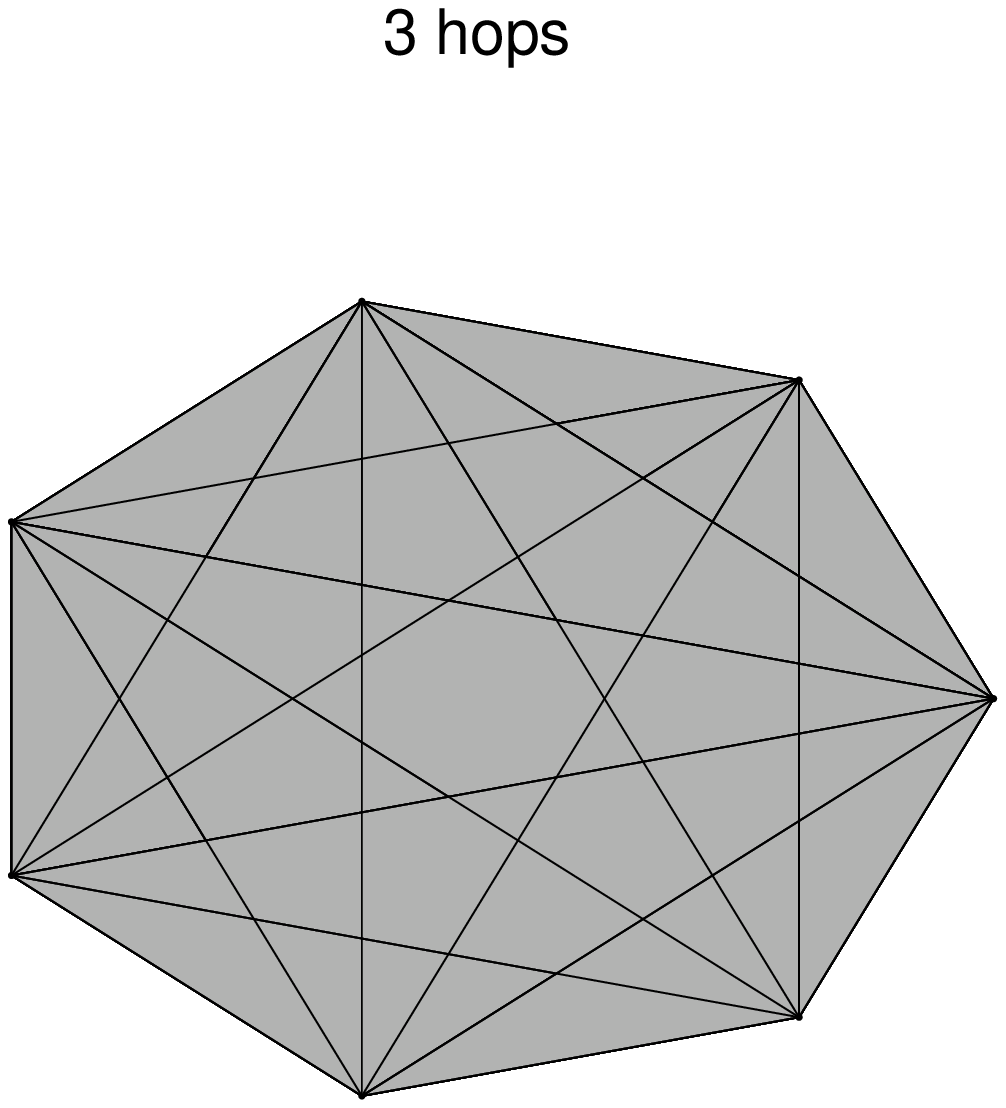} \\
\end{tabular}
\end{center}
\caption{Hop distance filtration on a simplicial complex $K$ consisting of a single loop (left), and the complexes $K^2$ (center) where there is still a non-trivial homology class, and $K^3$ (right) where the hole is completely filled in. \label{HopFilt}}
\end{figure}

A hop distance filtration for a simplicial complex consisting of a single loop is shown in Figure \ref{HopFilt}. Since the loop has a hop-length of 7 hops, it will become `filled in' by a triangle at a depth of 3 in the hop distance filtration (when edges are added between nodes that are three hops apart). Table \ref{HopSizes} gives the relationship between the hop-length of the shortest cycle surrounding a hole, and its persistence in the hop distance filtration. For a given simplicial complex, each of its holes will have a corresponding `depth' to which they persist in the hop distance filtration. The depths themselves can be taken as measures of the sizes of the holes, or alternatively the depths squared may be used (since the depth is a length-based measurement, its square will be proportional to area).

\begin{table}
\begin{center}
\begin{tabular}{|c|c|}
\hline
Hop-length of shortest & Persistence of hole in \\
cycle surrounding hole & hop distance filtration \\
\hline
4, 5, 6 & 1 \\
7, 8, 9 & 2 \\
10, 11, 12 & 3 \\
\vdots & \vdots \\
$3k + 1$, $3k+2$, $3k+3$ & k \\
\hline
\end{tabular}
\end{center}
\caption{The relationship between hole size (in terms of hop-length) and the depth it persists in the hop-distance filtration. \label{HopSizes}}
\end{table}

\begin{figure}[htp]
\begin{center}
\begin{tabular}{ccc}
\includegraphics[scale=0.24]{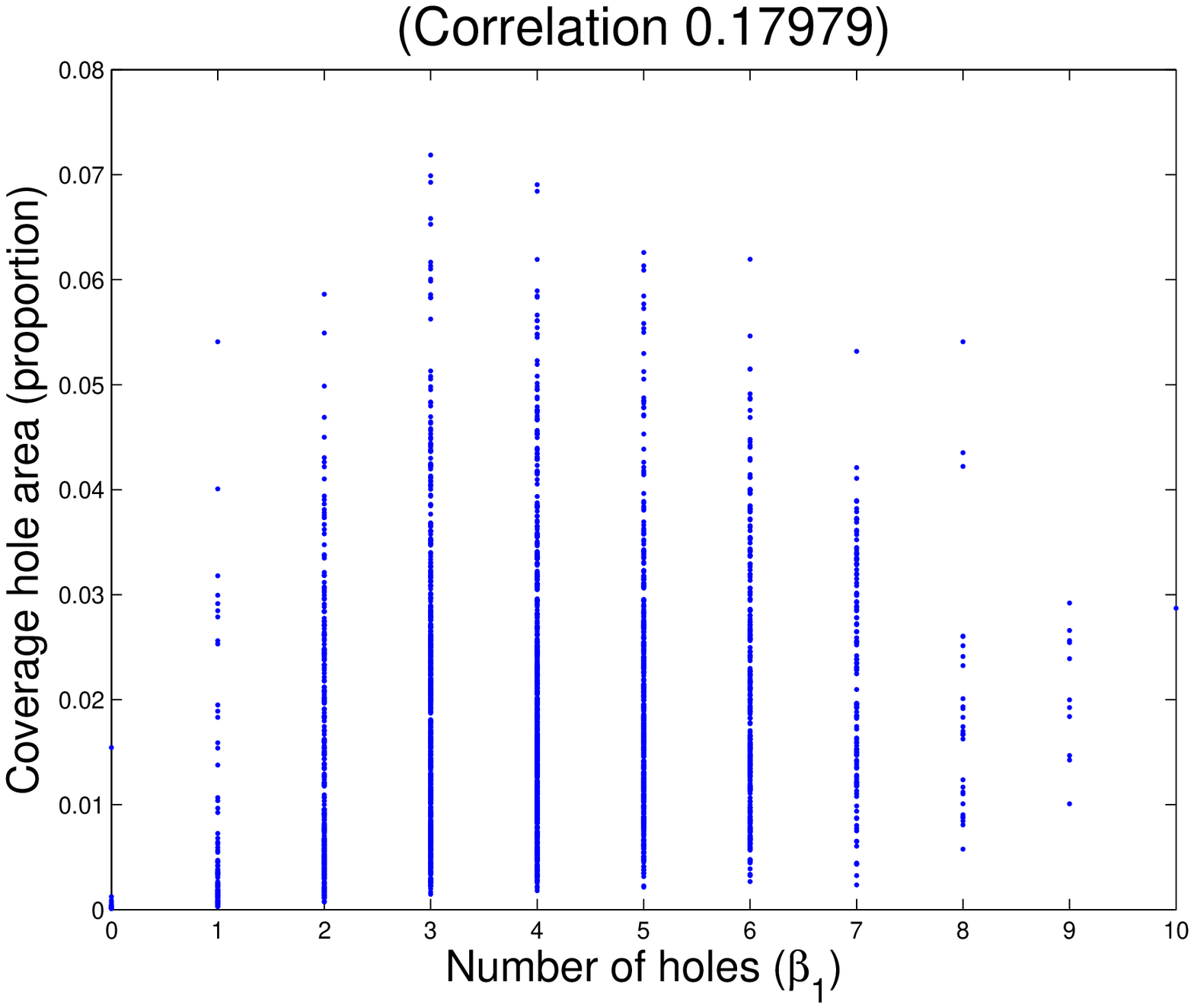} & \includegraphics[scale=0.24]{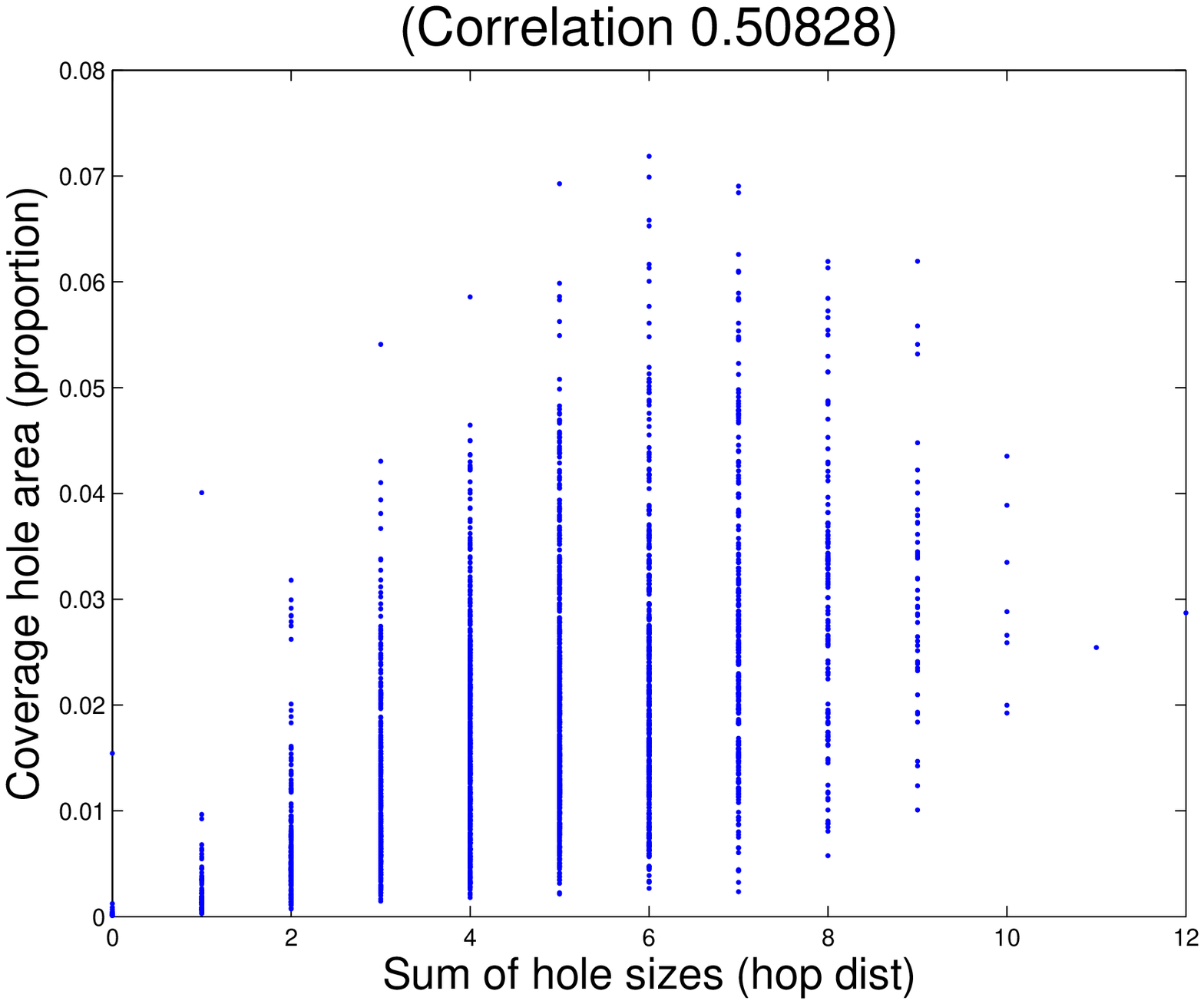} & \includegraphics[scale=0.24]{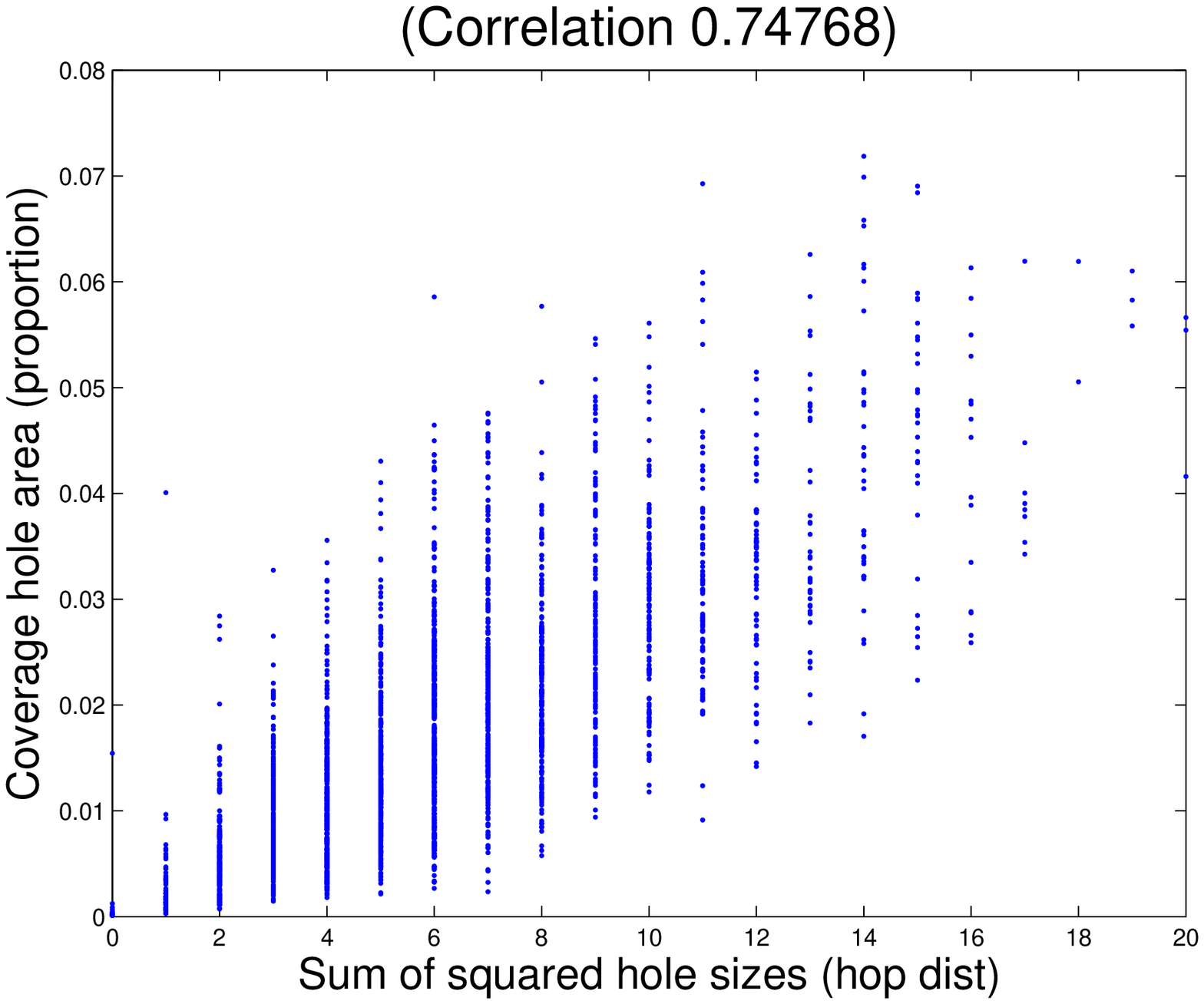} \\
\end{tabular}
\end{center}
\caption{Relationship between \emph{coverage hole area} (measured as proportion of total area) and various homological features. Left - first betti number $r = 0.176$), middle - sum of hole sizes (measured using depth in hop distance filtration, $r = 0.505$), right - sum of squared hole sizes (measured using depth in hop distance filtration, $r = 0.747$). \label{scatterplots}}
\end{figure}

To illustrate the benefit of incorporating hop-distance size estimates, we compare various possible homological descriptors of timepoint-wise coverage with the true geometric coverage information. This was done for each time point in all of the simulation runs for the Discrete Brownian model described in the previous section (for $50\times50 = 2500$ points), the results of which are shown in scatterplots in Figure \ref{scatterplots}. For a given sensor network we measure the geometric coverage by the proportion of total area contained inside the coverage holes (ignoring uncovered area along the boundary of the simulation region, which is undetectable by the simplicial complex), and refer to this measure as \emph{coverage hole area}. The homological coverage descriptors based on coordinate-free data only are:
\begin{enumerate}
\item The number of holes in the complex (first Betti number)
\item The sum of the hole sizes (measured by depth in the hop distance filtration)
\item The sum of the squared hole sizes (i.e. sum of squared depths)
\end{enumerate}
As mentioned above, the number of holes in a simplicial complex (the first Betti number) does not describe the hole sizes at all.
By combining information about the number of holes along with their estimated sizes, we are able to obtain a coordinate-free descriptor which correlates well with the true geometric information about the size of the coverage holes. This is rather surprising, since the coordinate-free information is very coarse relative to the geometric.

A note on the computational complexity of performing the hop distance filtration, since the simplicial complexes $K^h$ grow large quickly as $h$ increases. The filtration does not generate any new first homology, and all holes are present in the original complex, so only the death times need to be computed. This may be done by computing the first Betti number for each of the subsequent complexes, yielding the number of holes `killed' at each depth. A topology-preserving simplicial collapse \cite{Wilkerson2013} is performed on each of the complexes before computing the first homology, to increase efficiency of the computations. An alternative method to improve efficiency would be to compute persistent homology of the filtration using the Morse theoretic collapse algorithm presented in \cite{Mischaikow2013}.

In addition to using the hop-distance filtration as a measure of the sizes of the holes present in the network at each time point, we would like to link the hole sizes present at time $i$, with the bars (obtained from zigzag persistence) at time $i$, for each time point $i$. The hop depth information can be combined with the zigzag persistence, and represented together using a barcode that is enhanced with estimated size information for each bar at each time point. In order to do this, we need to make a choice for the homology class corresponding to each bar, as well as a specific representative cycle for that homology class. Then, observing when the inclusion of the cycle becomes trivial in the hop-distance filtration will tell us the size of the largest hole that cycle encircles. Unlike the set of birth-death intervals, the choice of homology class for each bar is not unique, so we would like our choice to be geometrically-motivated, and as close as possible to the `canonical basis' described at the end of Section \ref{SimpHom}, which would have each homology class surrounding exactly one hole. The method we propose to achieve this is described in the following section.

To obtain the set of depths for a given complex, it is not necessary to use persistent homology to compute explicitly the depth of each hole in the hop distance filtration. Since all the holes are present at $h=1$ (the original complex), simply computing the first Betti number for each of the subsequent complexes, until the first homology is trivial, will yield the number of holes that are `killed' at each depth. Since the sizes of the complexes grows large quickly as $h$ increases, this can be done efficiently by performing a topology-preserving simplicial collapse method \cite{Wilkerson2013} before computing the homology.

\section{Tracking representative cycles}\label{Tracking}

Given the set of intervals $\{[b_j,d_j]\}$ obtained from zigzag persistence, we want to have a choice of representative cycle for each interval, at each time point. The homology classes for this set of cycles should form a basis for the homology, and the choice of representative cycles over time should map into each other in a meaningful way. We present here a method, to be computed alongside the zigzag algorithm, which returns such representative cycles. For a more detailed mathematical and algorithmic description, see \cite{Gamble2014}.

Intuitively this method aims to compute a `canonical basis', where there is one representative cycle surrounding each hole. Given the Rips complex for a static sensor network, without an embedding or geometric information, such a canonical basis is impossible to obtain. In the time-varying setting however, a small amount of `canonical' information is available: when a coverage hole is first formed by the removal of a 2-simplex (triangle), the boundary of that triangle is known to surround exactly the hole of interest. The idea behind our method is then to use that boundary as the representative cycle for the homology class at its birth time, and propagate that information forward through the sequence of complexes as best as possible. The representative cycles we choose need to also be compatible with the interval decomposition in the zigzag algorithm, (the technical detail of this compatibility is described in \cite{Gamble2014}).

When applying this method alongside the zigzag algorithm, each bar in $\Pers(\mathcal{K}) = \{[b_j,d_j]\}$ is associated with a specific representative cycle at each time point. This associates each bar with a specific hole (or set of holes) that it surrounds, even though this information is not directly available to us. We can obtain size estimates for the hole(s) by including the representative cycle in the hop-distance filtration of the complex (at each time point), as described in Section \ref{HoleSizes}. If the representative cycles did indeed form a canonical basis, then the size information about each hole over time would be attached one-to-one with a corresponding bar.
Although guarantees of a true canonical basis are impossible, when implemented in practice the method gives representative cycles that are quite geometrically meaningful. Short-lived holes are typically surrounded by a tight cycle at their boundary, and holes that begin with the removal of a triangle and then grow in size are also well-tracked.

\subsection{Examples}
Here, we present a number of examples where the representative cycles and associated size estimates give useful and interesting results, unavailable through other methods. Recall that all of the results and computations discussed in this section are obtained using only the communication graph of the network at each time point, with no information about coordinates or distances between neighboring sensors.

\subsubsection{Tracking holes in a dense network}

\begin{figure}[htp]
\begin{center}
\begin{tabular}{ccc}
\includegraphics[scale=0.22]{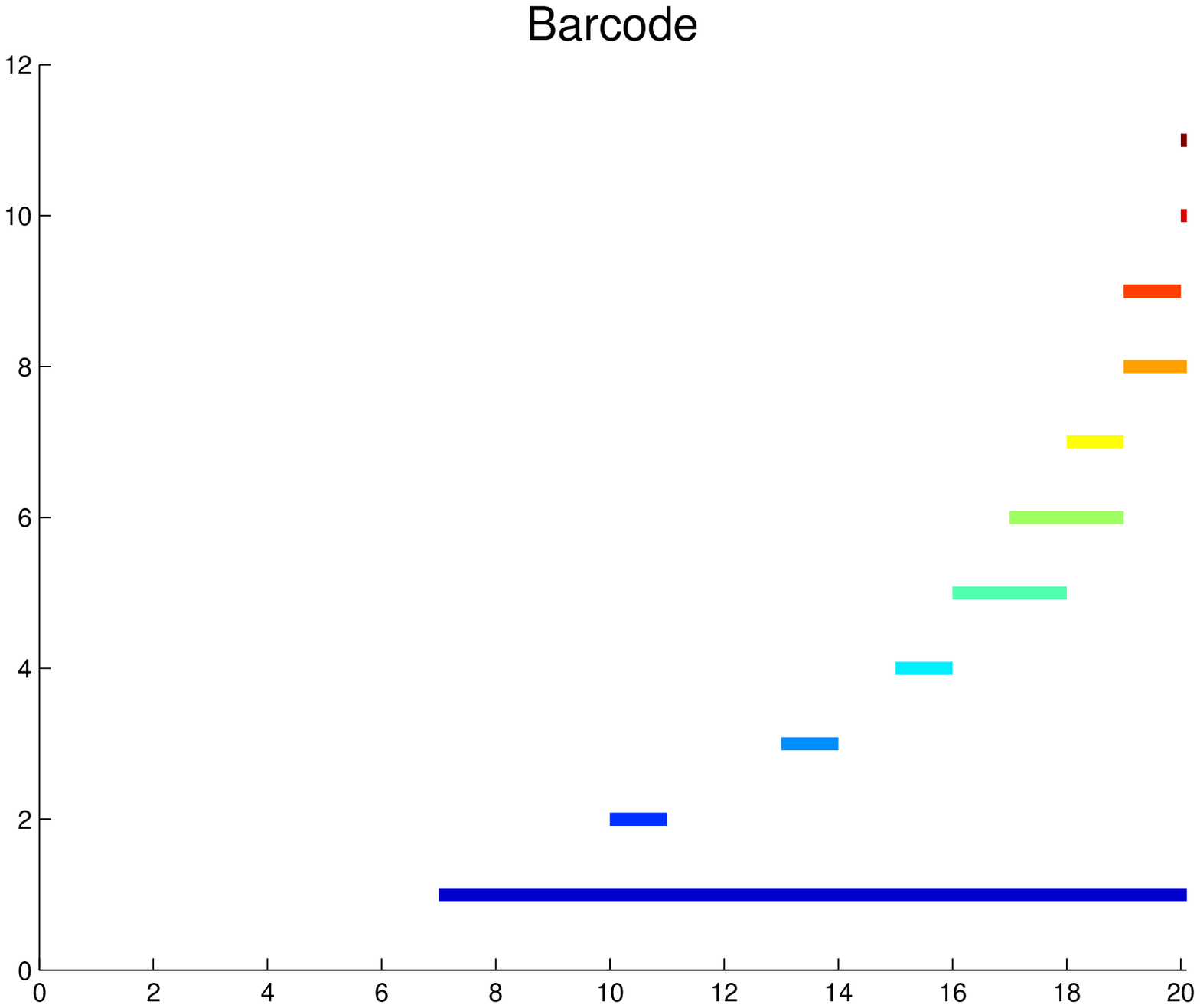} & \includegraphics[scale=0.22]{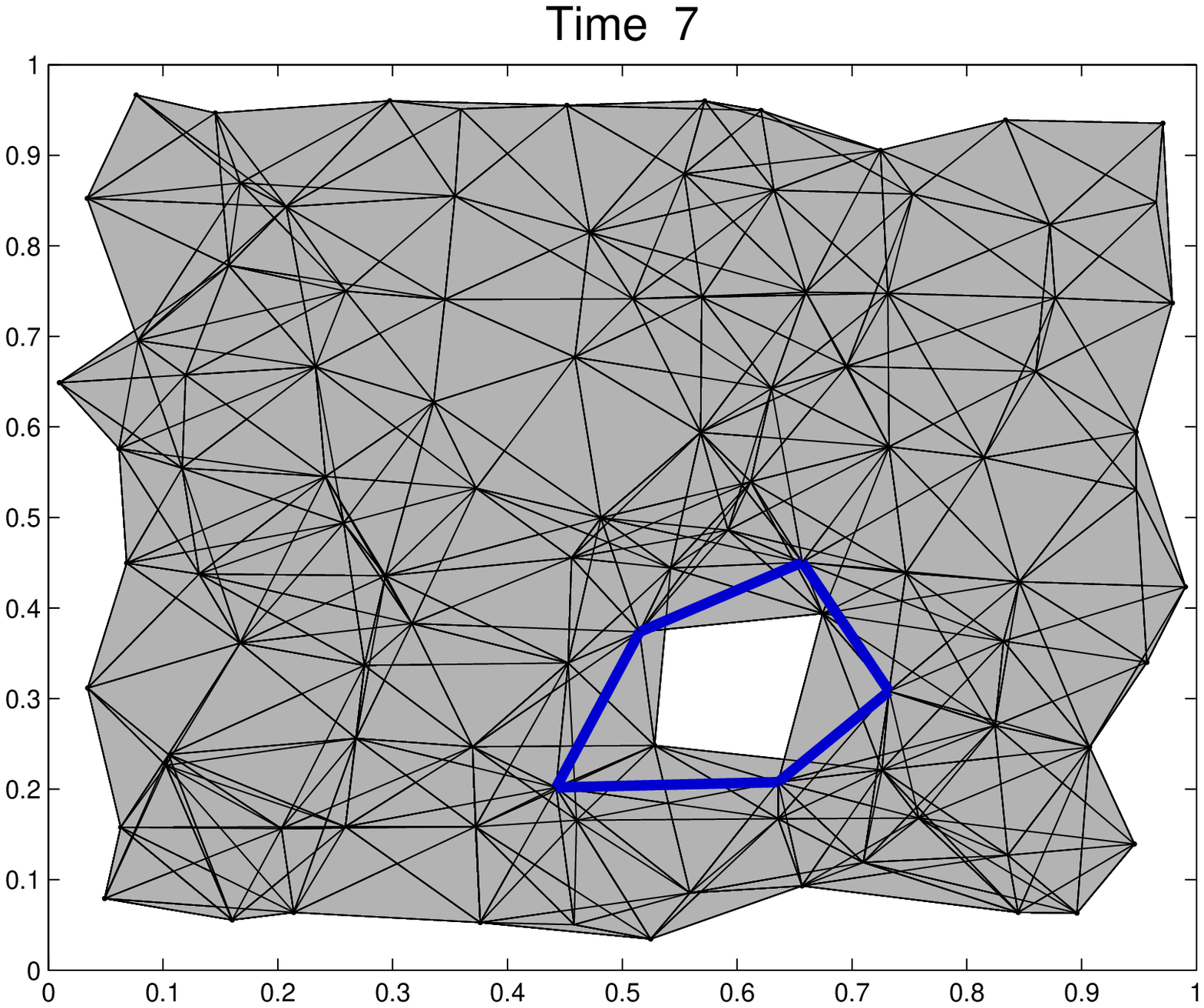} & \includegraphics[scale=0.22]{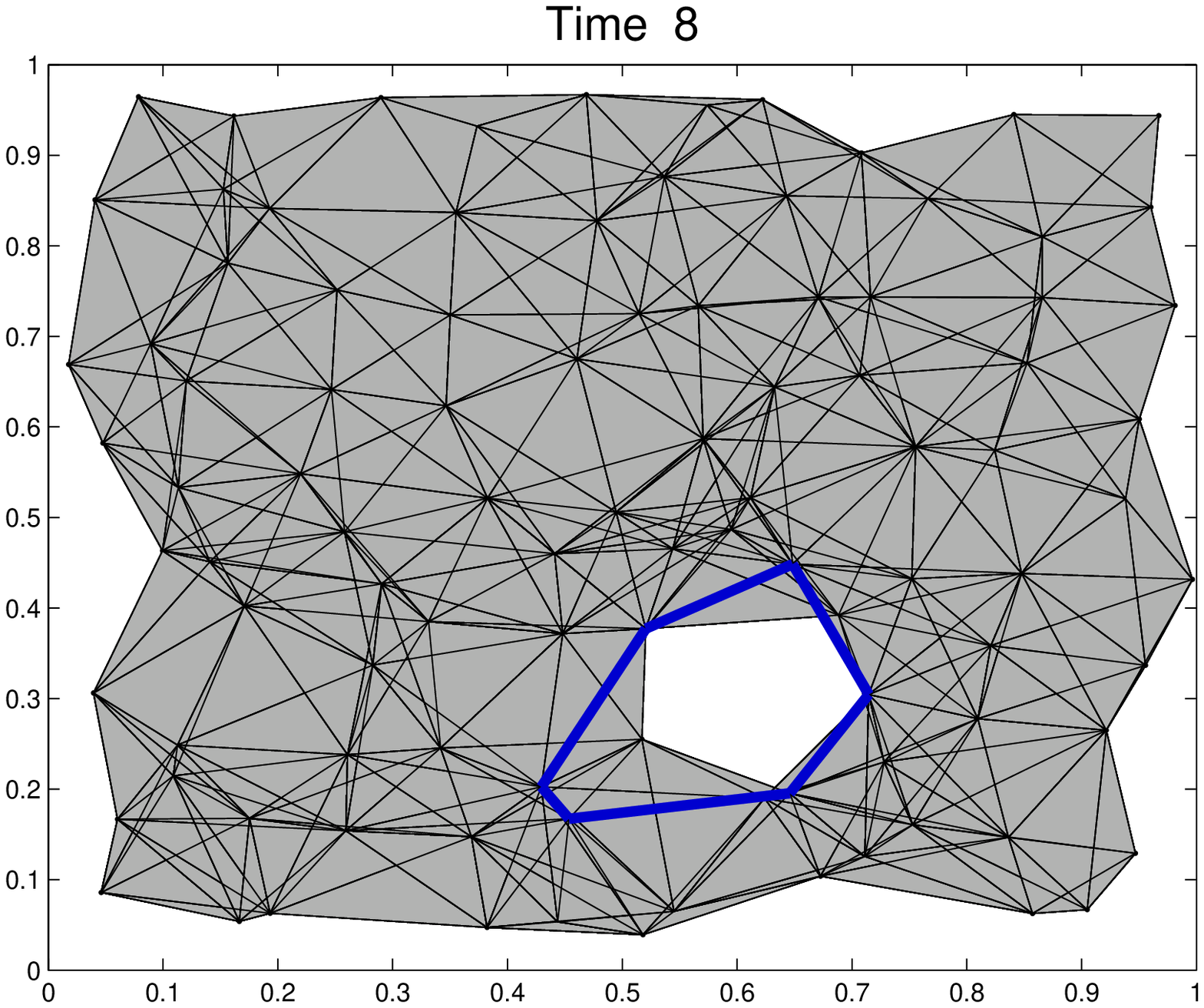} \\
\includegraphics[scale=0.22]{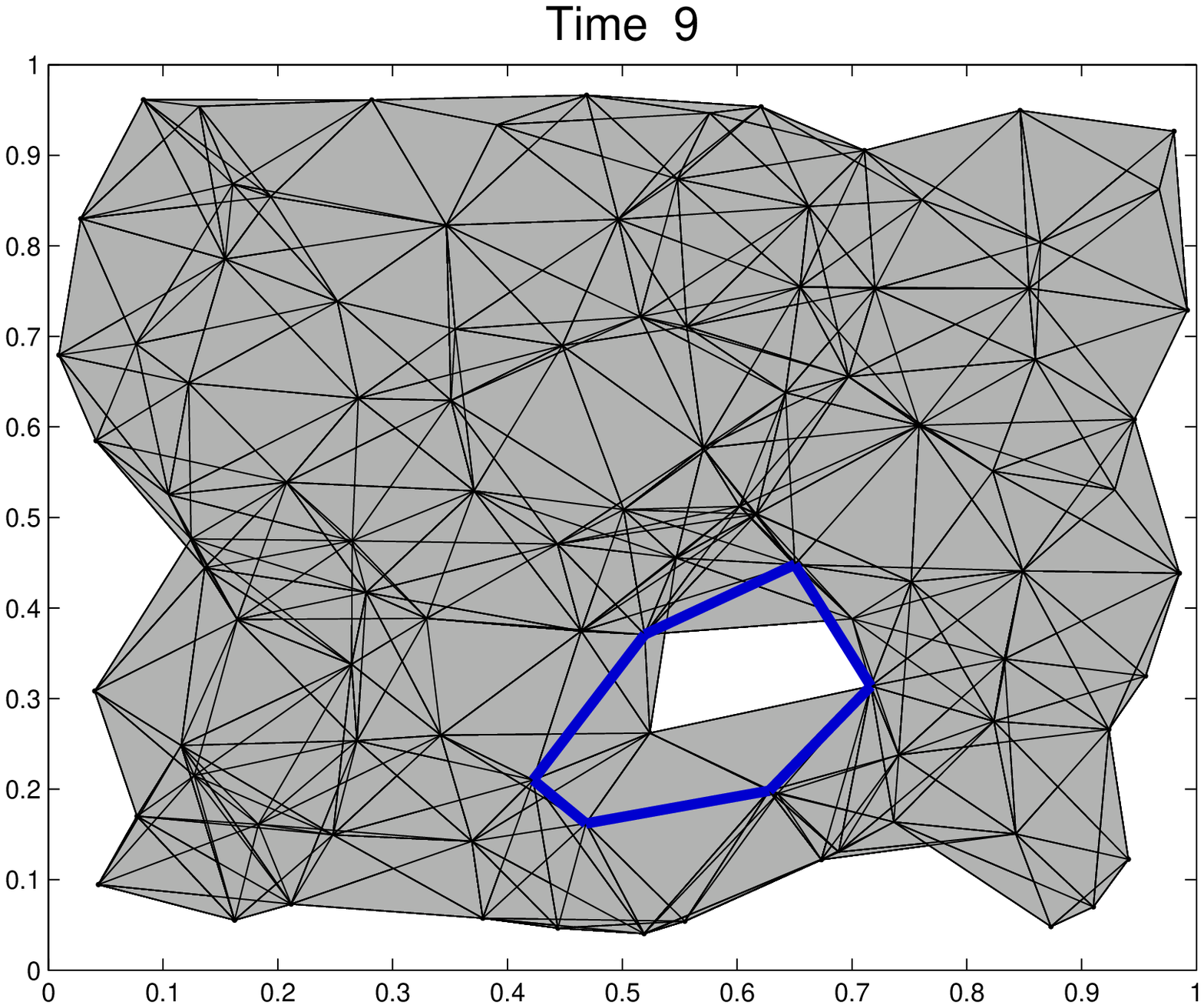} & \includegraphics[scale=0.22]{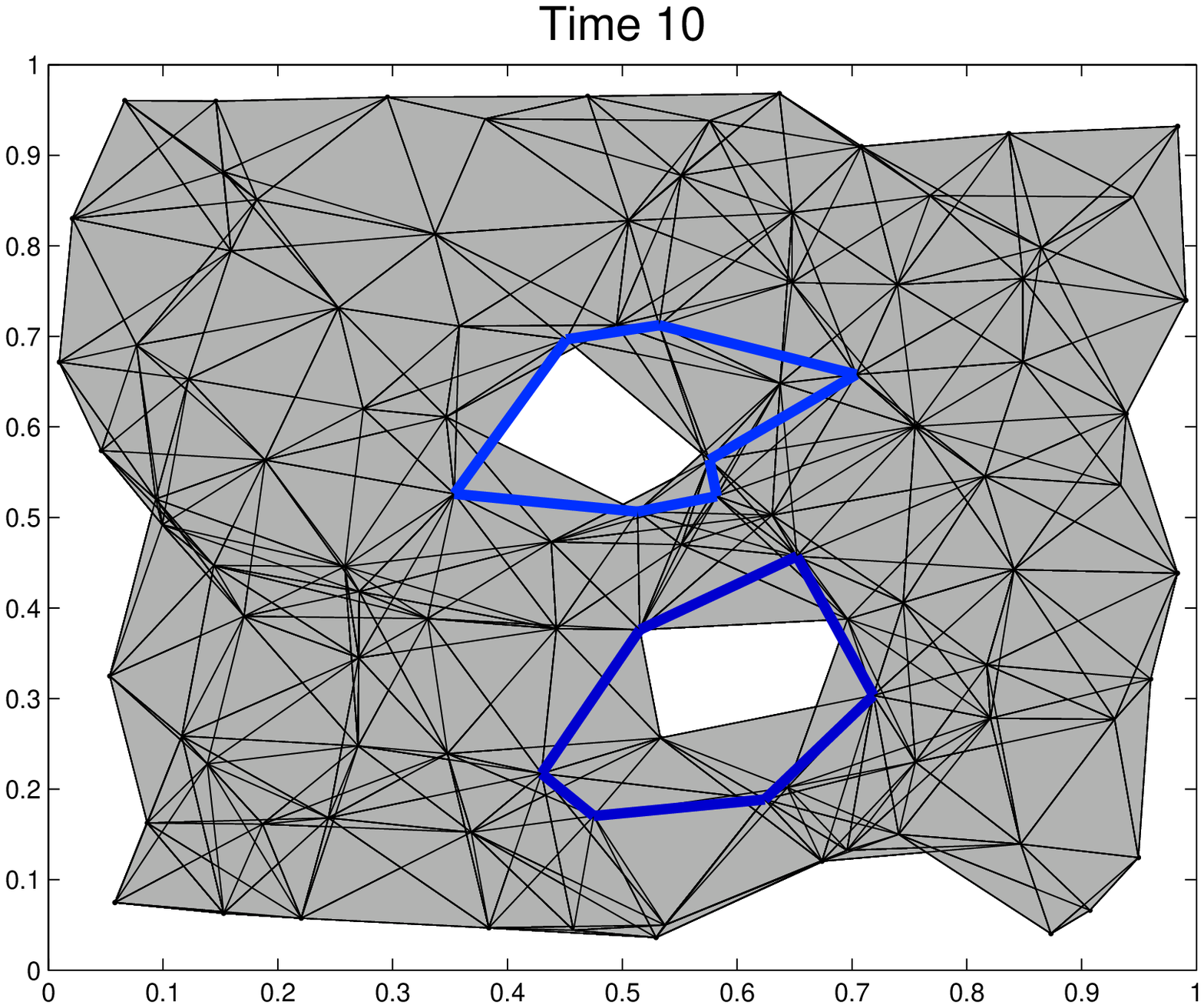} & \includegraphics[scale=0.22]{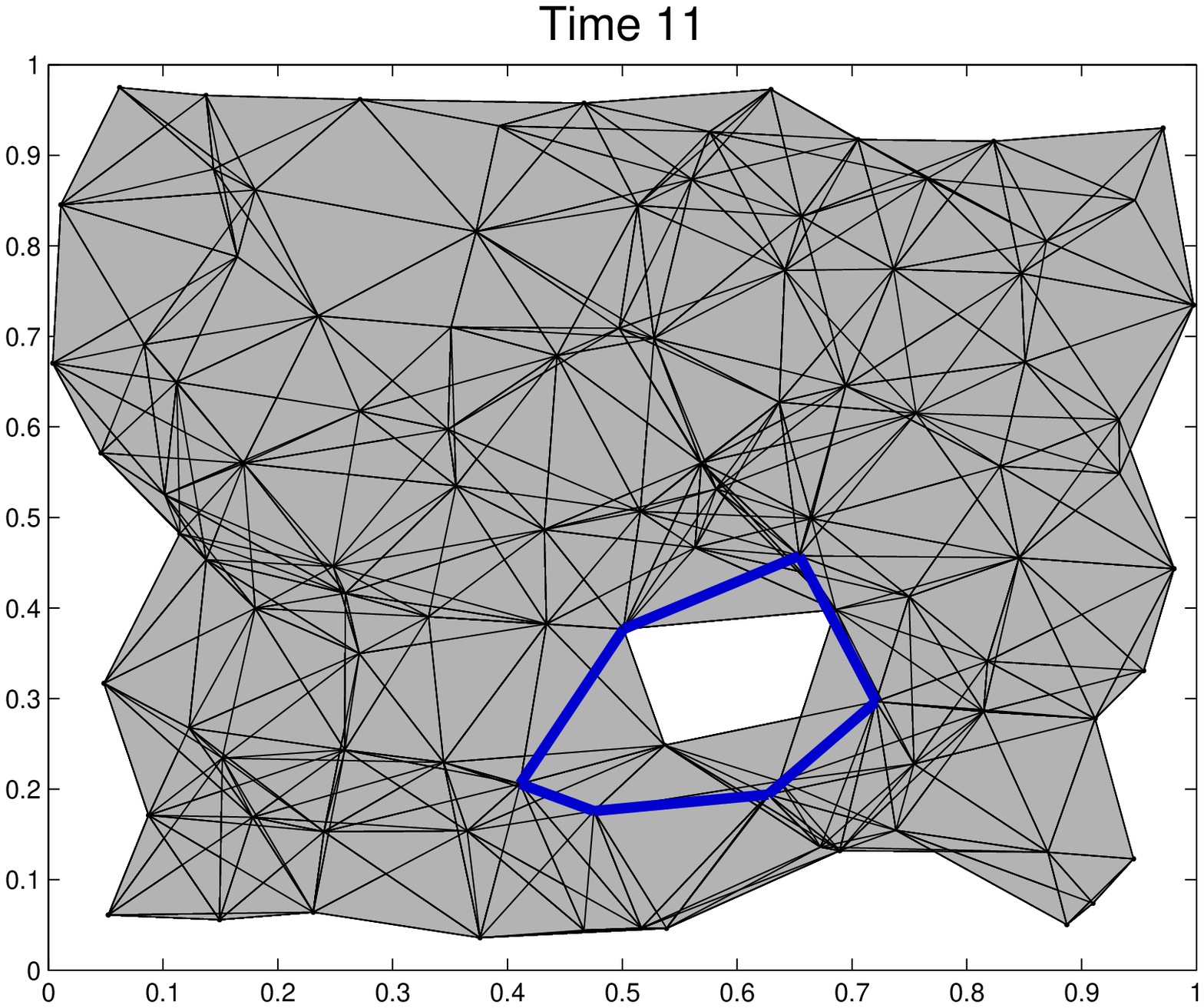}  \\
\includegraphics[scale=0.22]{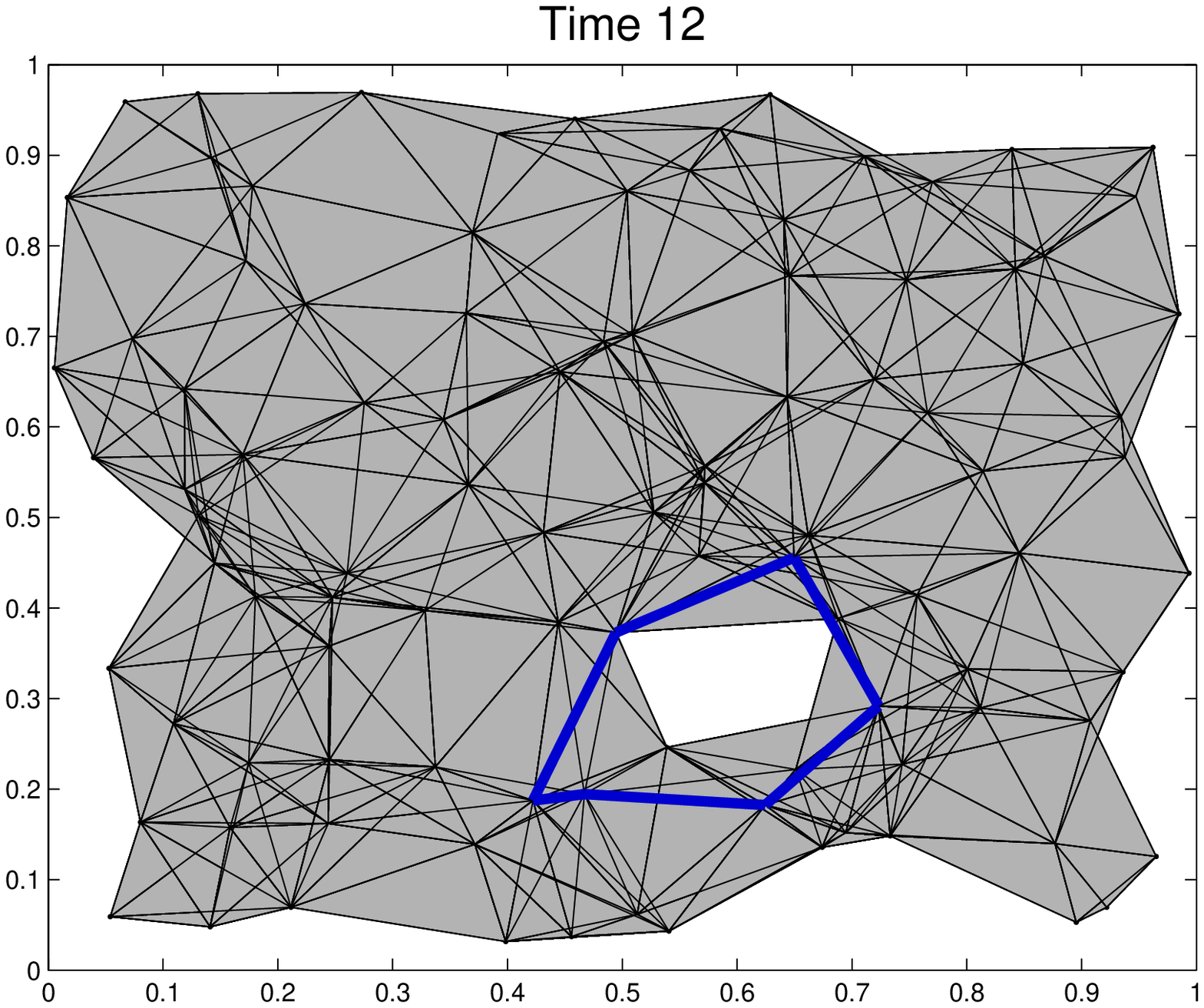} & \includegraphics[scale=0.22]{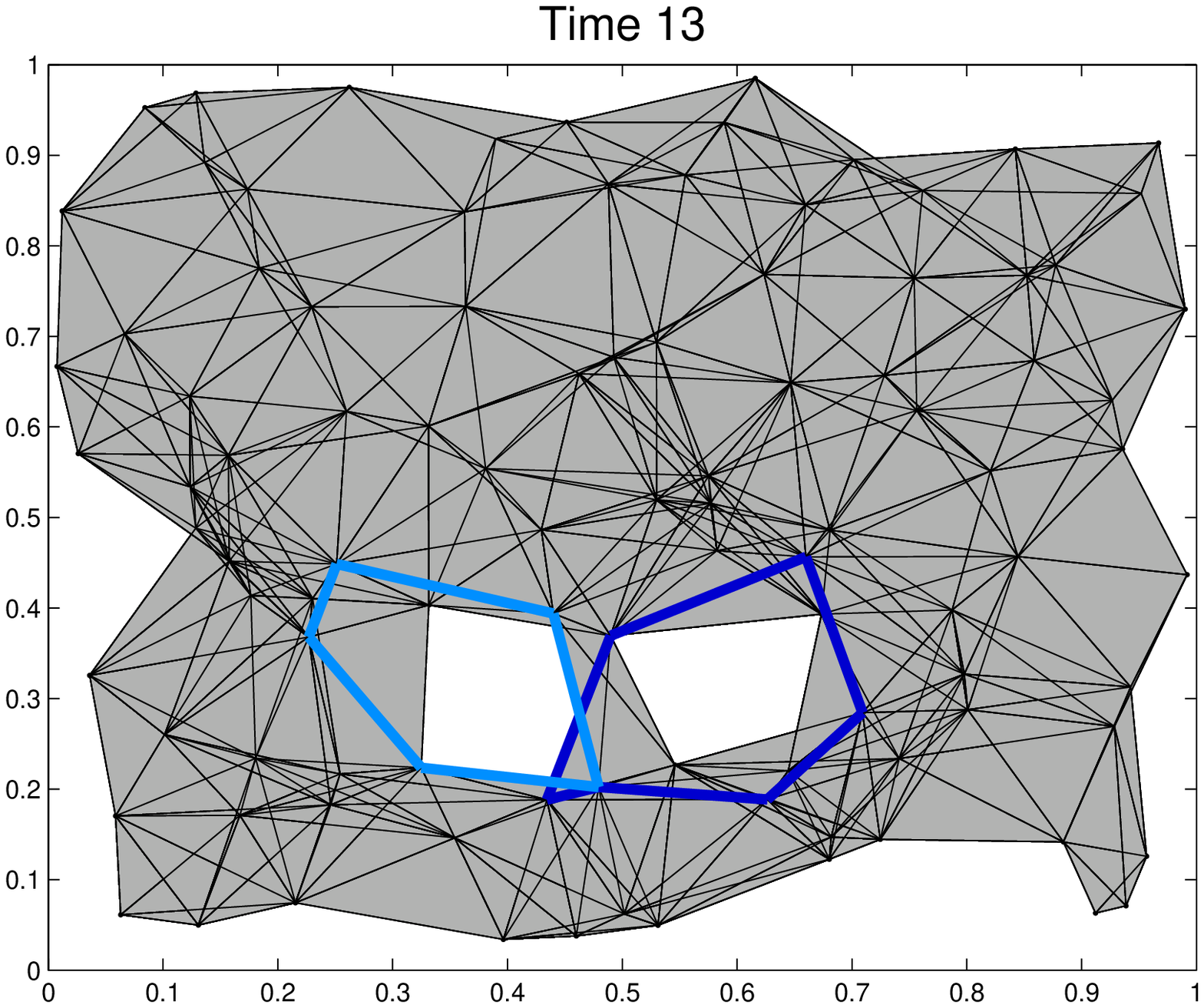} & \includegraphics[scale=0.22]{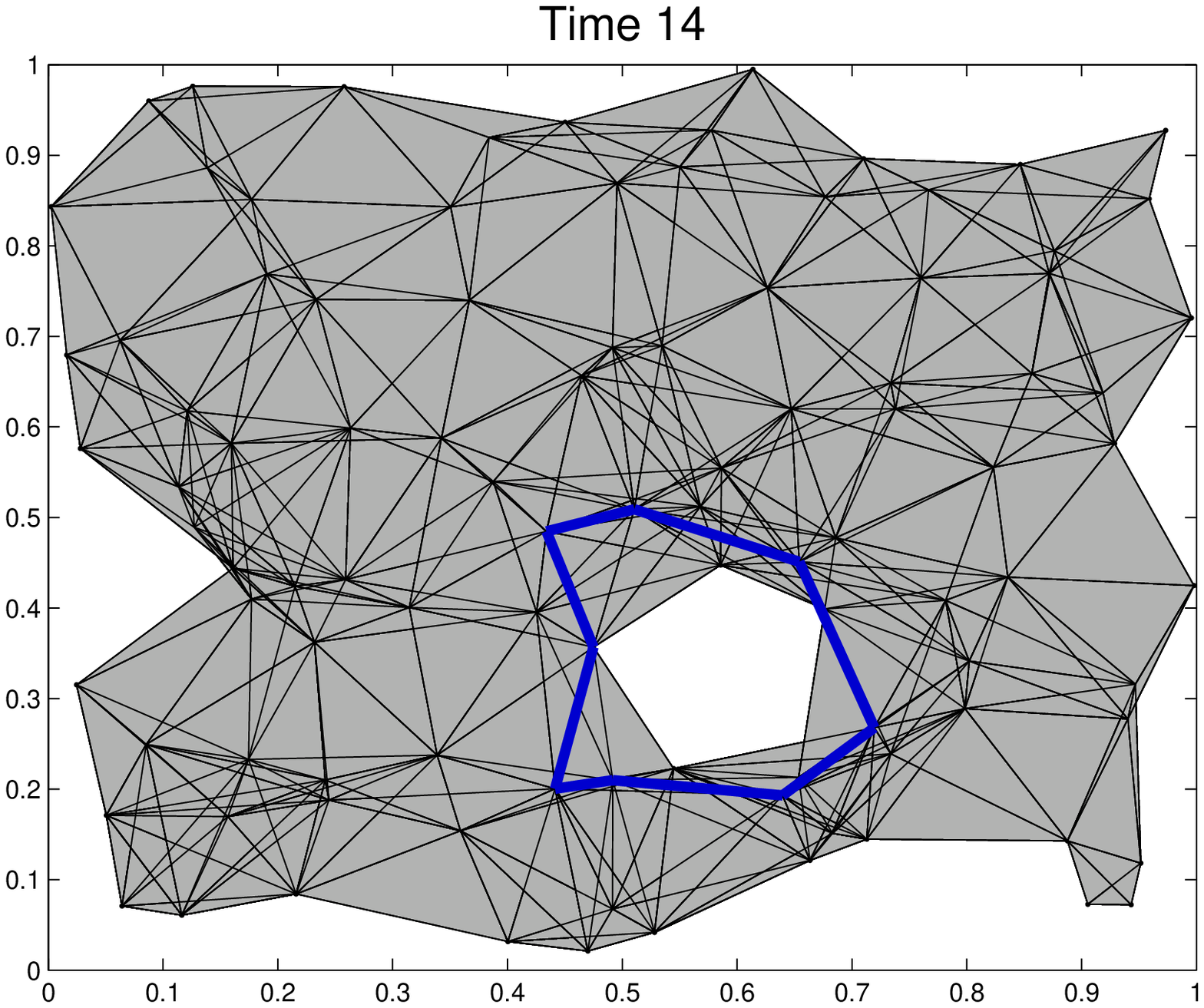} \\
\includegraphics[scale=0.22]{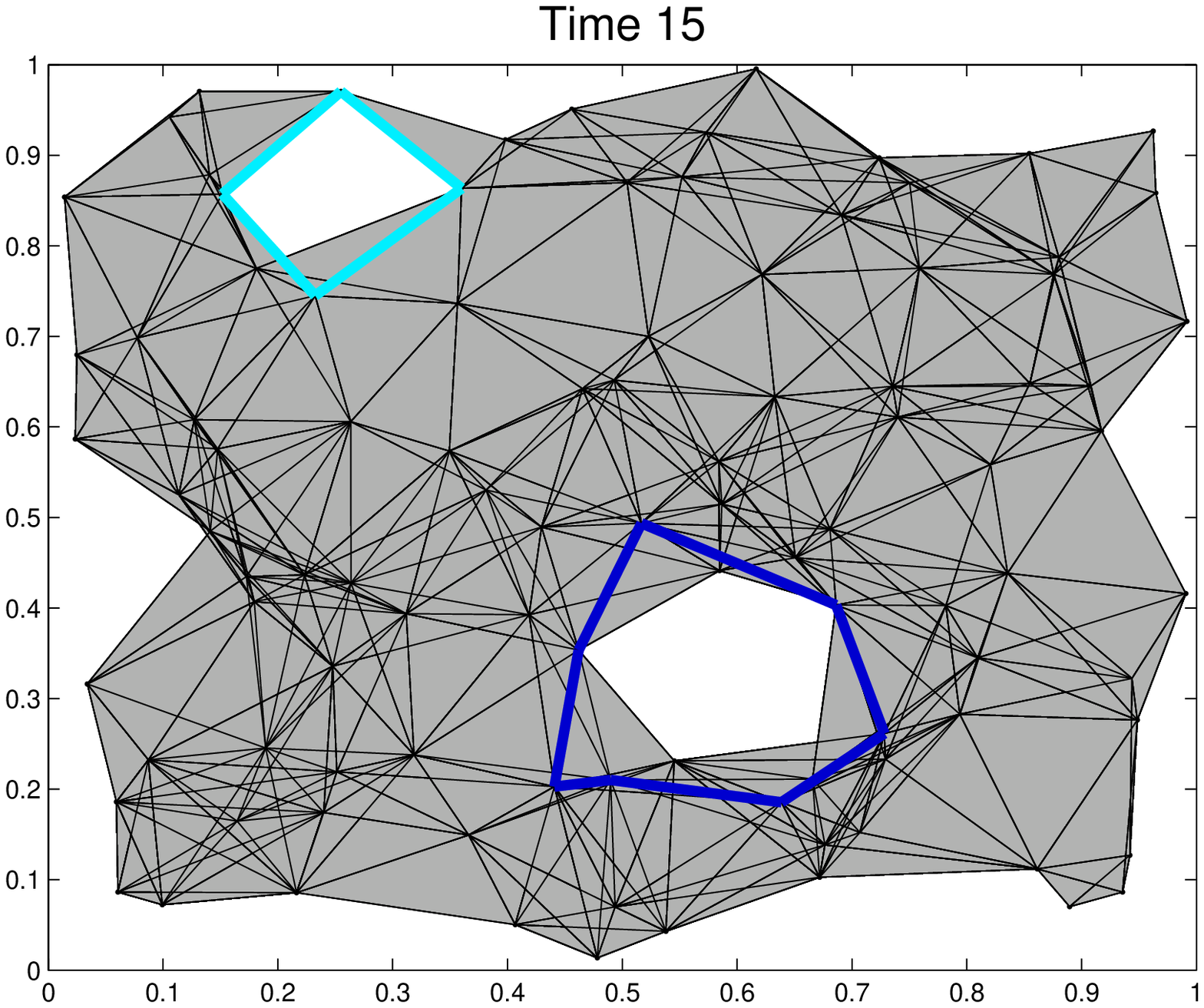} & \includegraphics[scale=0.22]{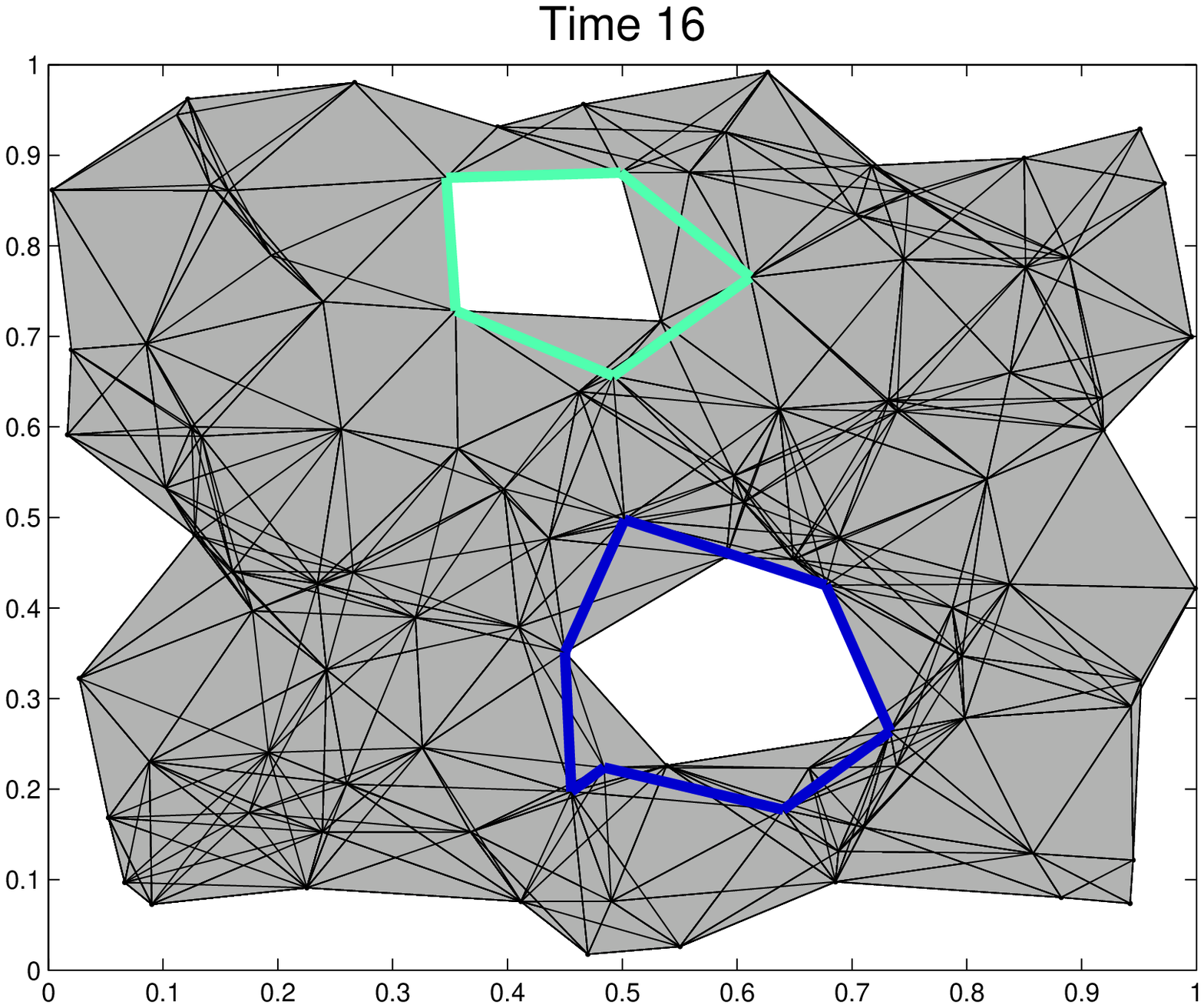} & \includegraphics[scale=0.22]{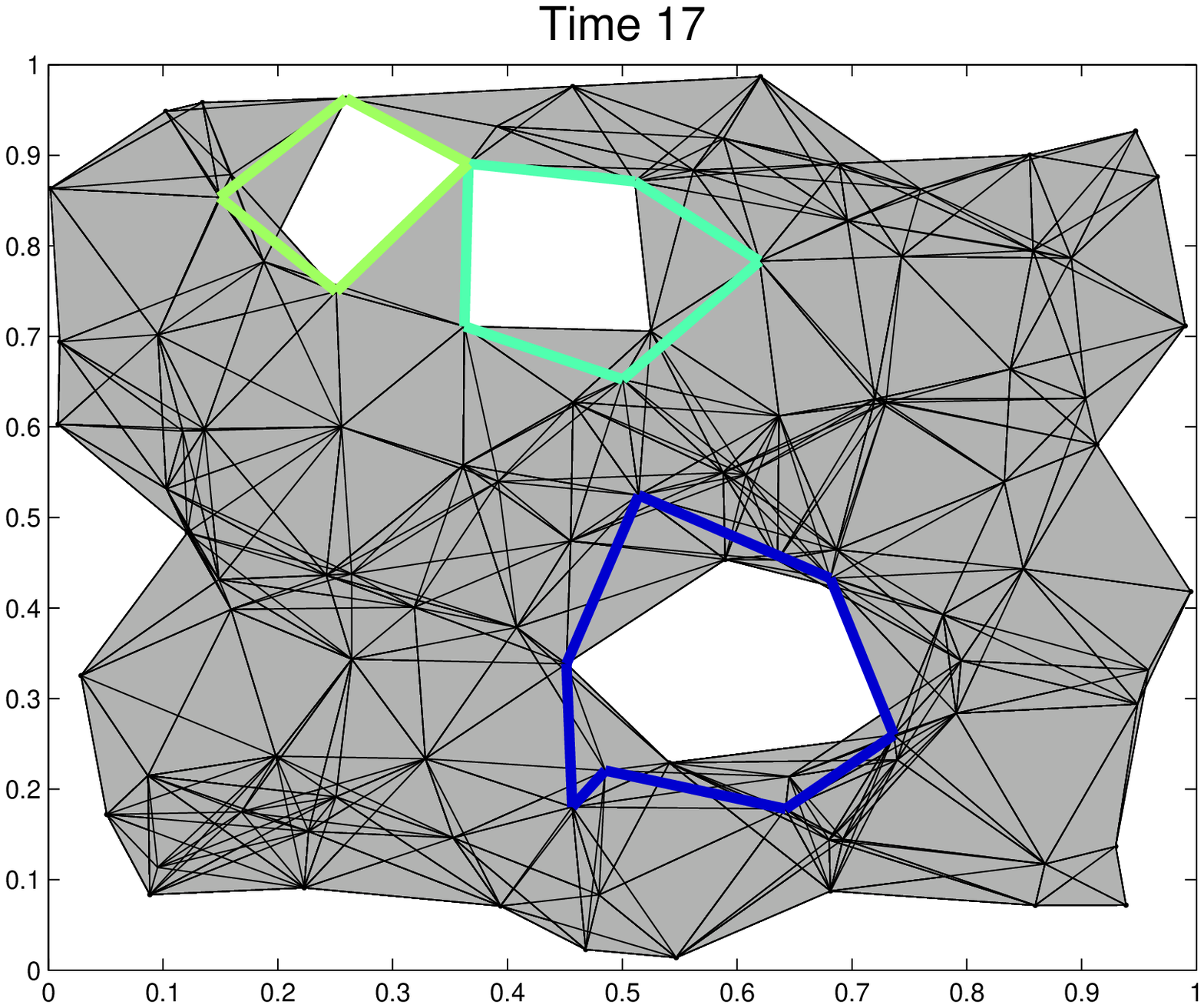} \\
\includegraphics[scale=0.22]{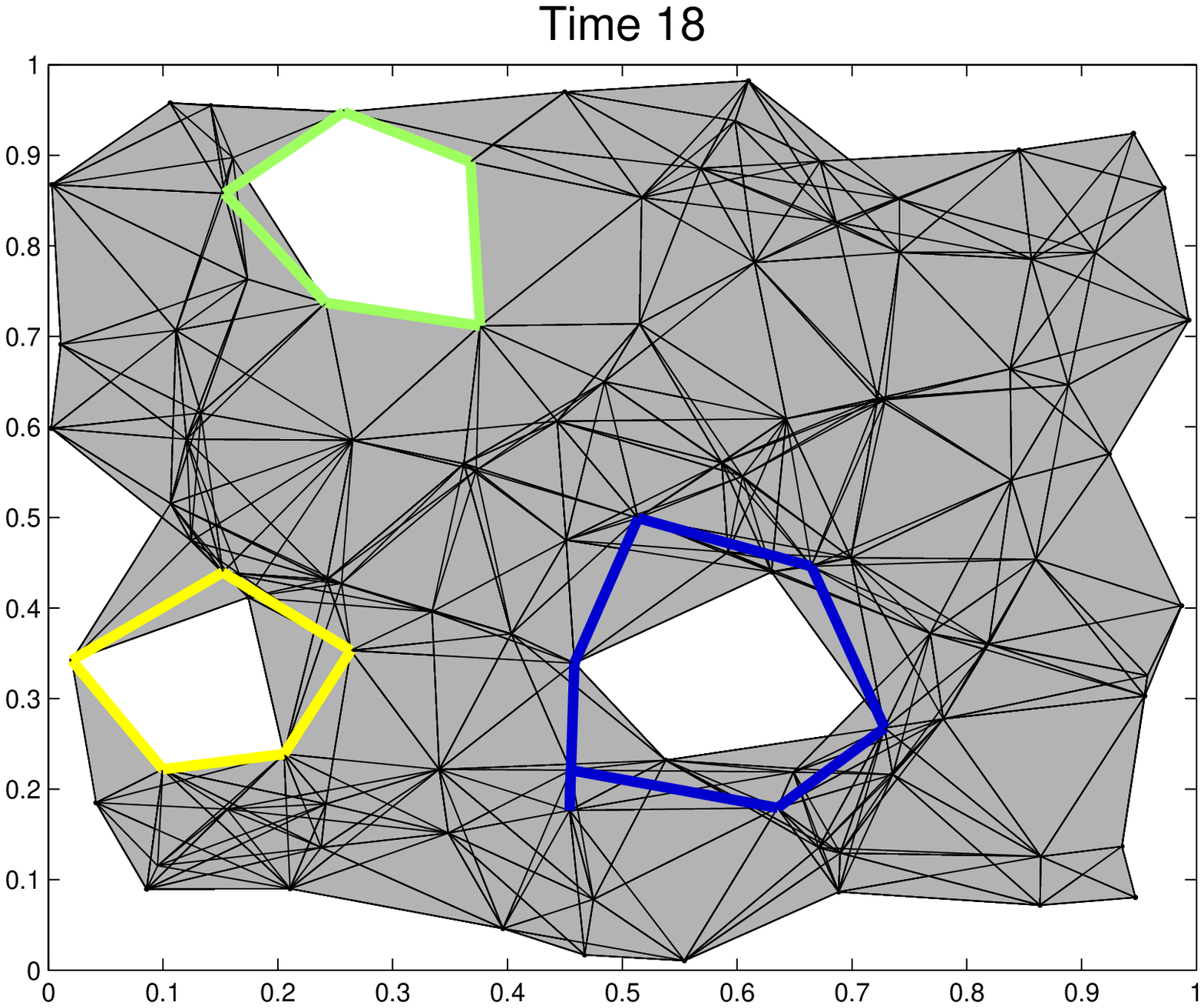} & \includegraphics[scale=0.22]{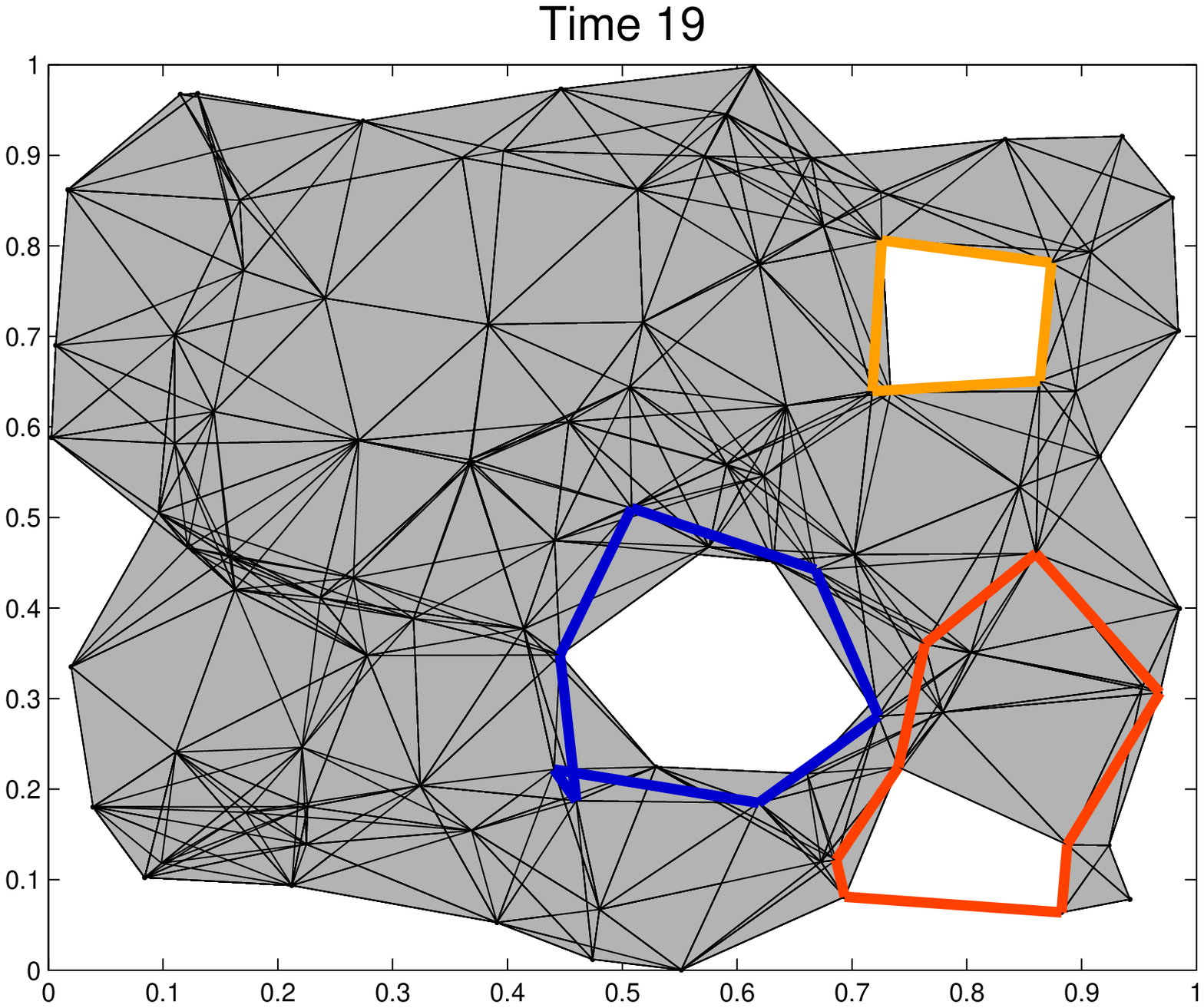} & \includegraphics[scale=0.22]{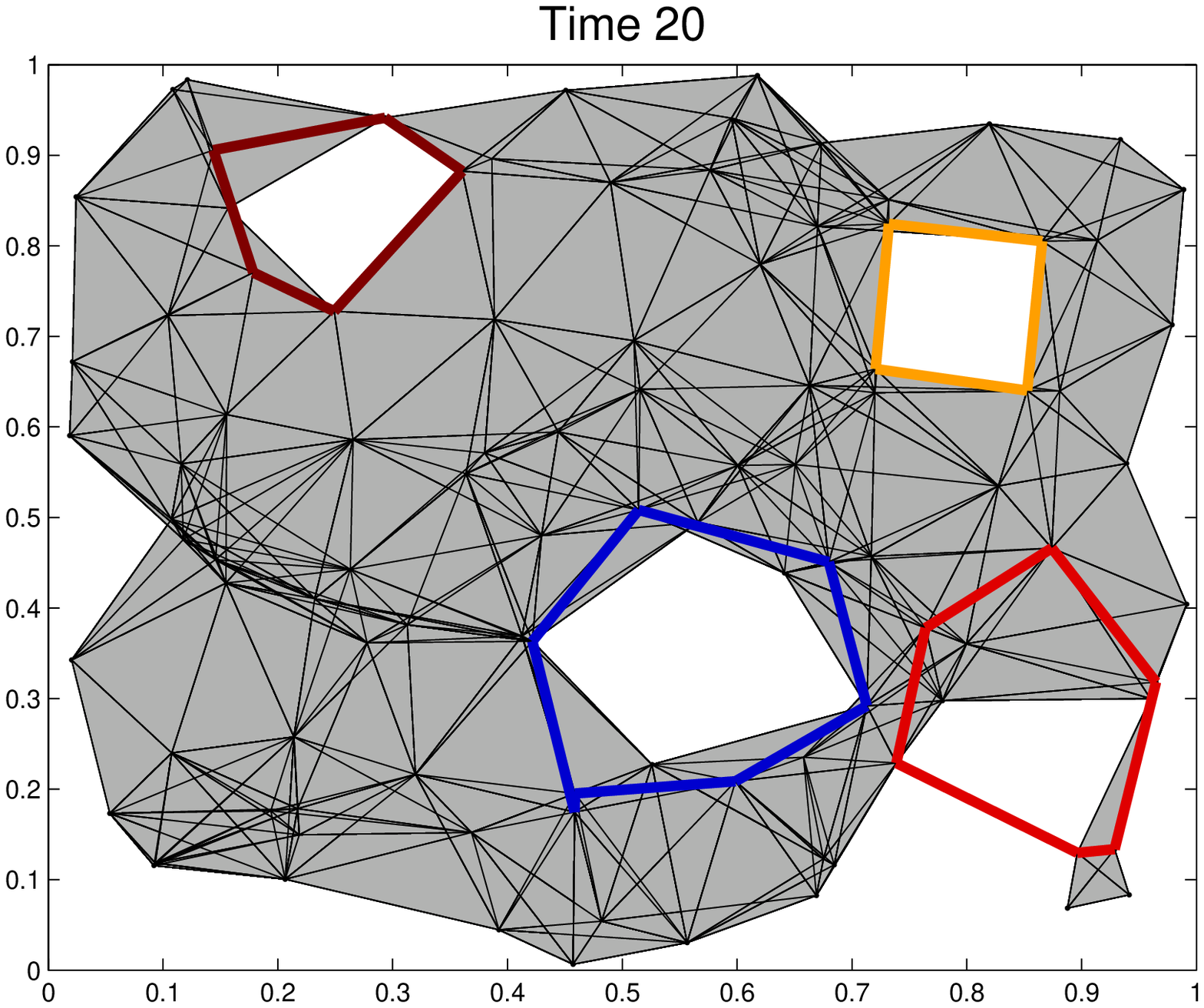} \\
\end{tabular}
\end{center}
\caption{Representative cycles for the intervals obtained using zigzag persistence in a dense network, with color-coding between the representative cycle at each time point and its corresponding interval (barcode - top left). \label{RepCycles}}
\end{figure}

Figure \ref{RepCycles} illustrates a network which is initially fully covered, and has a number of small coverage holes appearing over time, one of which is persistent. The barcode displaying lifetimes of homological features can be seen in the top left, with the bars color-coded to correspond to their associated representative cycles in the other figures. It can be seen that each representative cycle remains relatively tight around one coverage hole, and the set of cycles does correspond to a canonical basis at each time point. Overall, when a network is dense enough that its coverage holes appear and disappear in an isolated fashion (as opposed to splitting and merging with other holes), this method performs very well.

\subsubsection{Detecting and evaluating severity of expanding failure region}
\begin{figure}[htp]
\begin{center}
\begin{tabular}{ccc}
\includegraphics[scale=0.22]{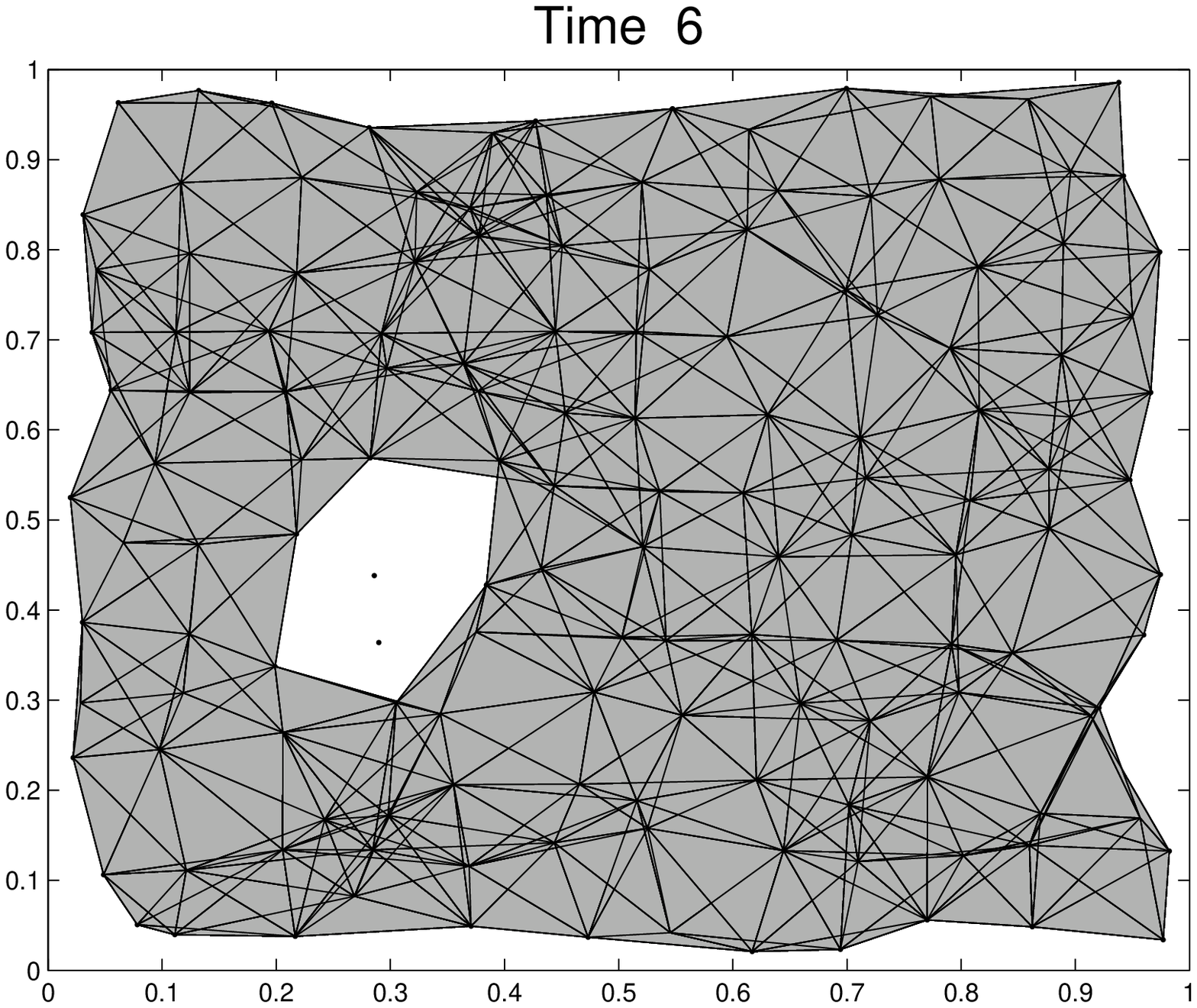} & \includegraphics[scale=0.22]{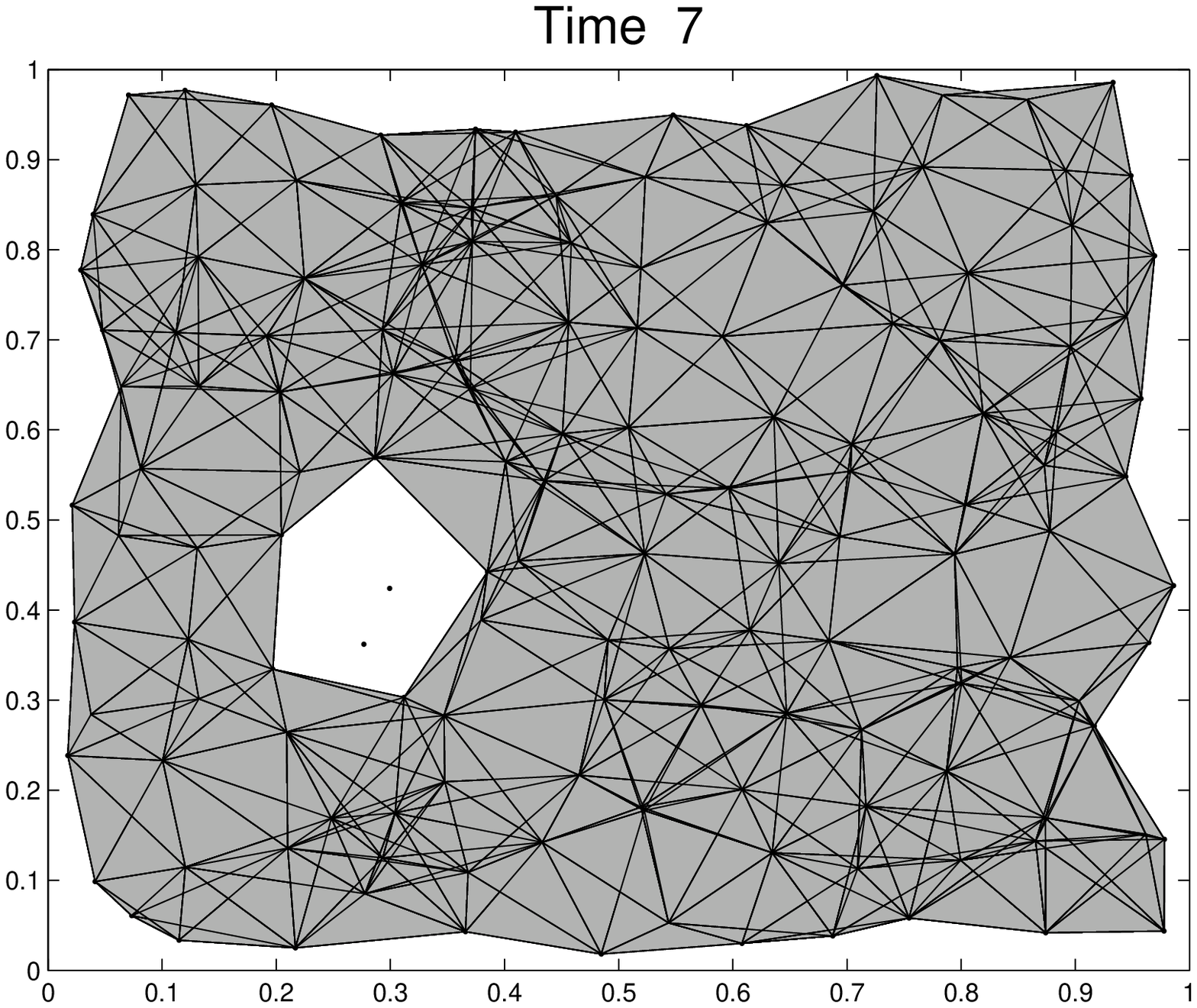} & \includegraphics[scale=0.22]{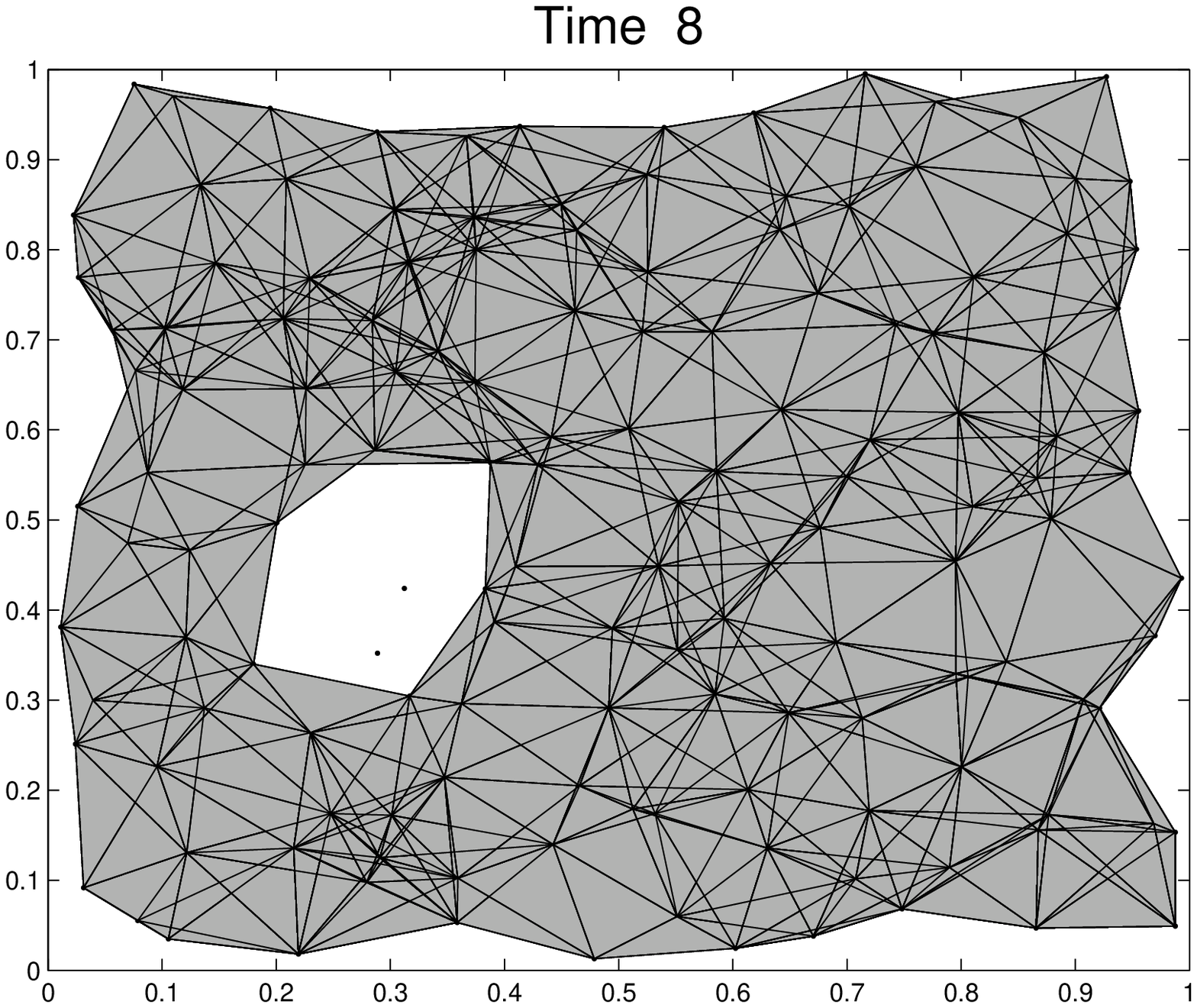} \\
\includegraphics[scale=0.22]{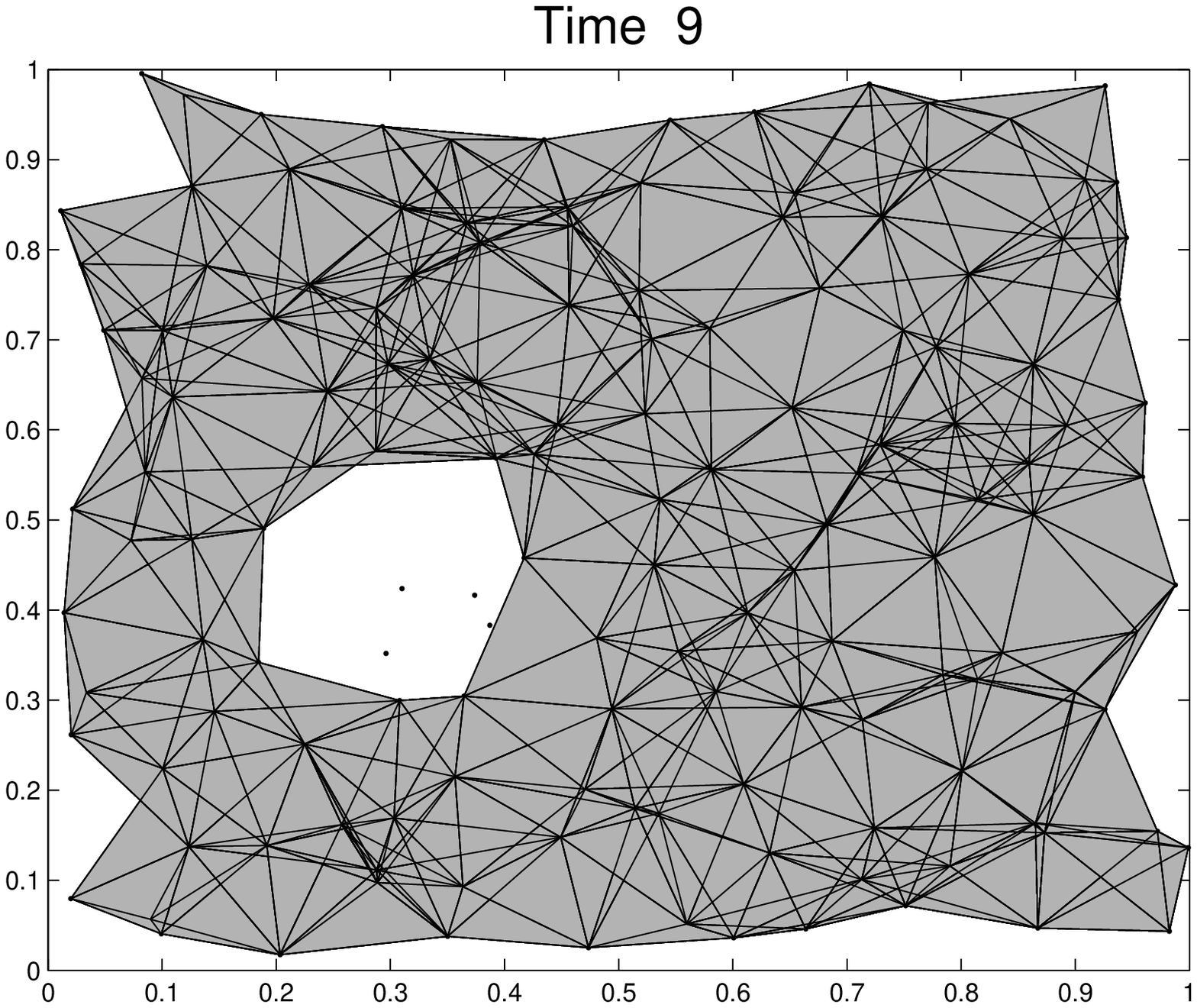} & \includegraphics[scale=0.22]{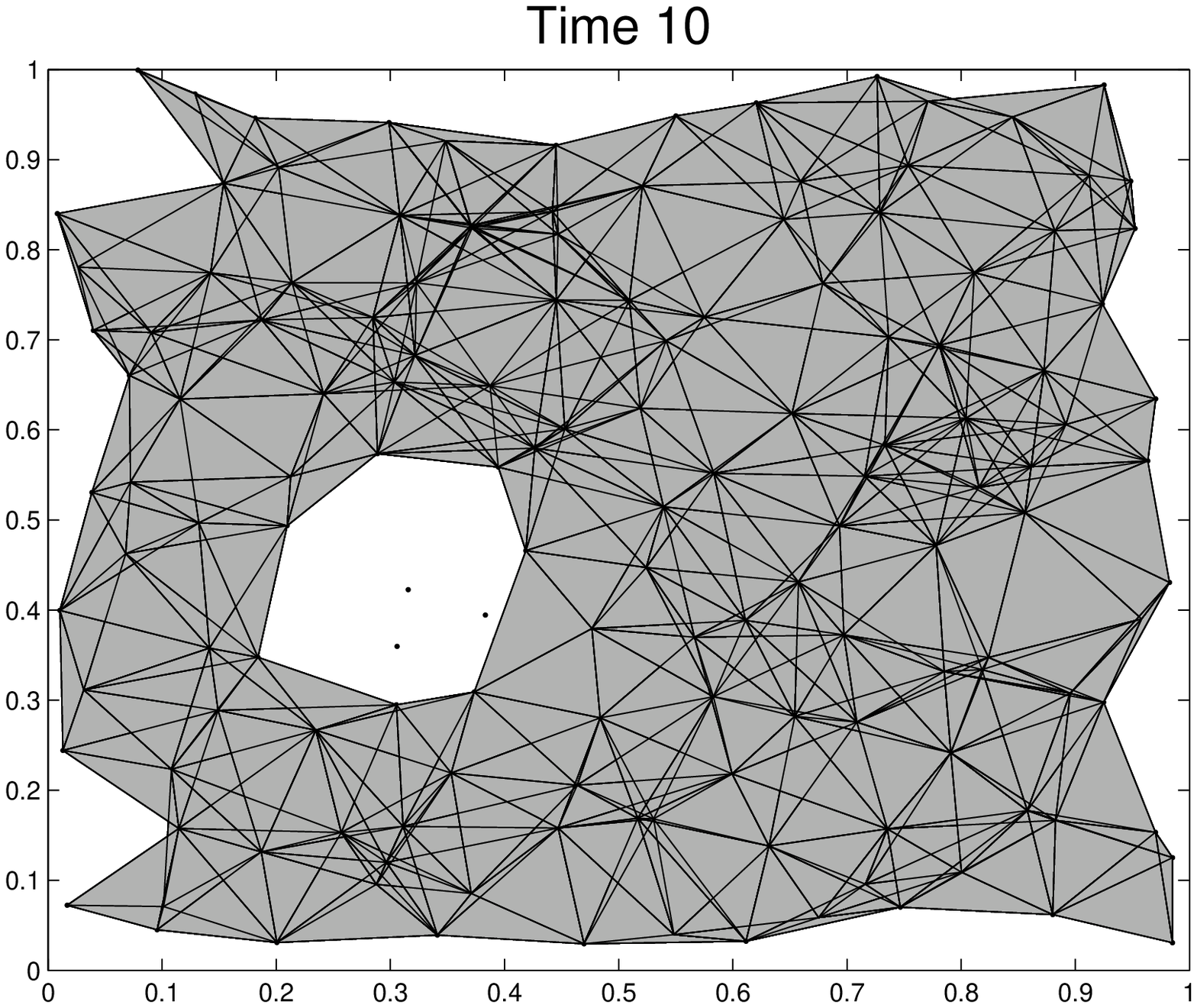} & \includegraphics[scale=0.22]{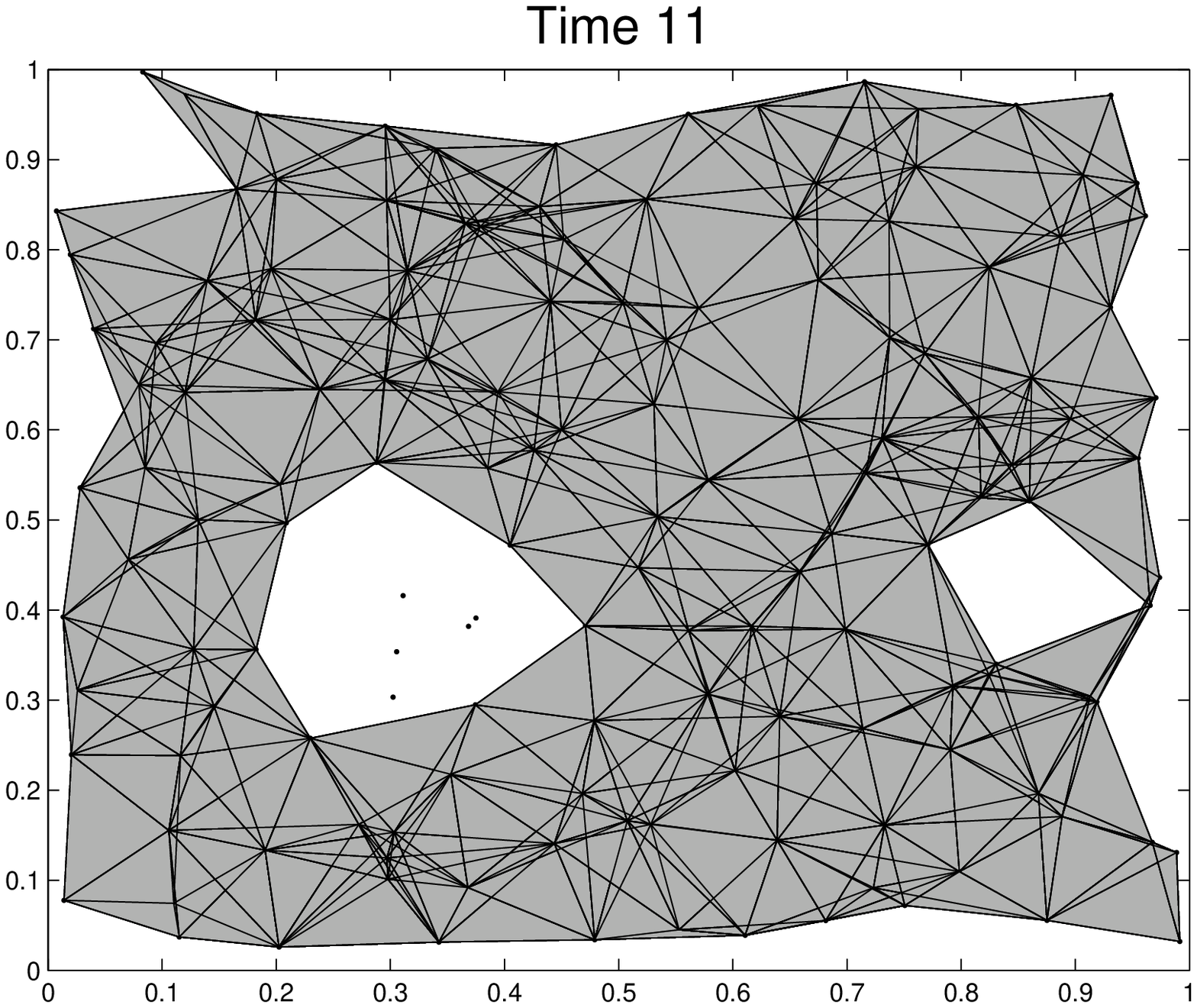}  \\
\includegraphics[scale=0.22]{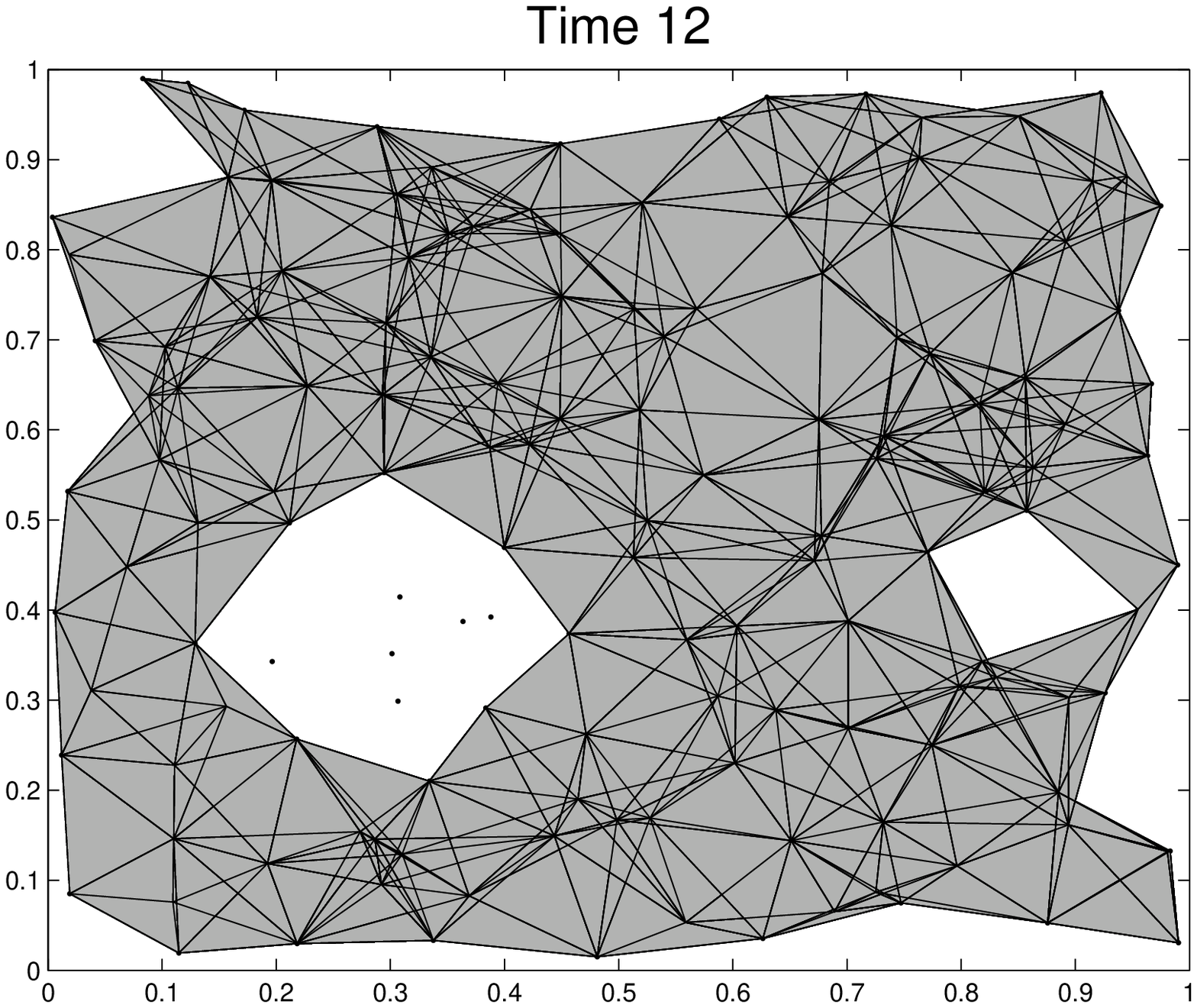} & \includegraphics[scale=0.22]{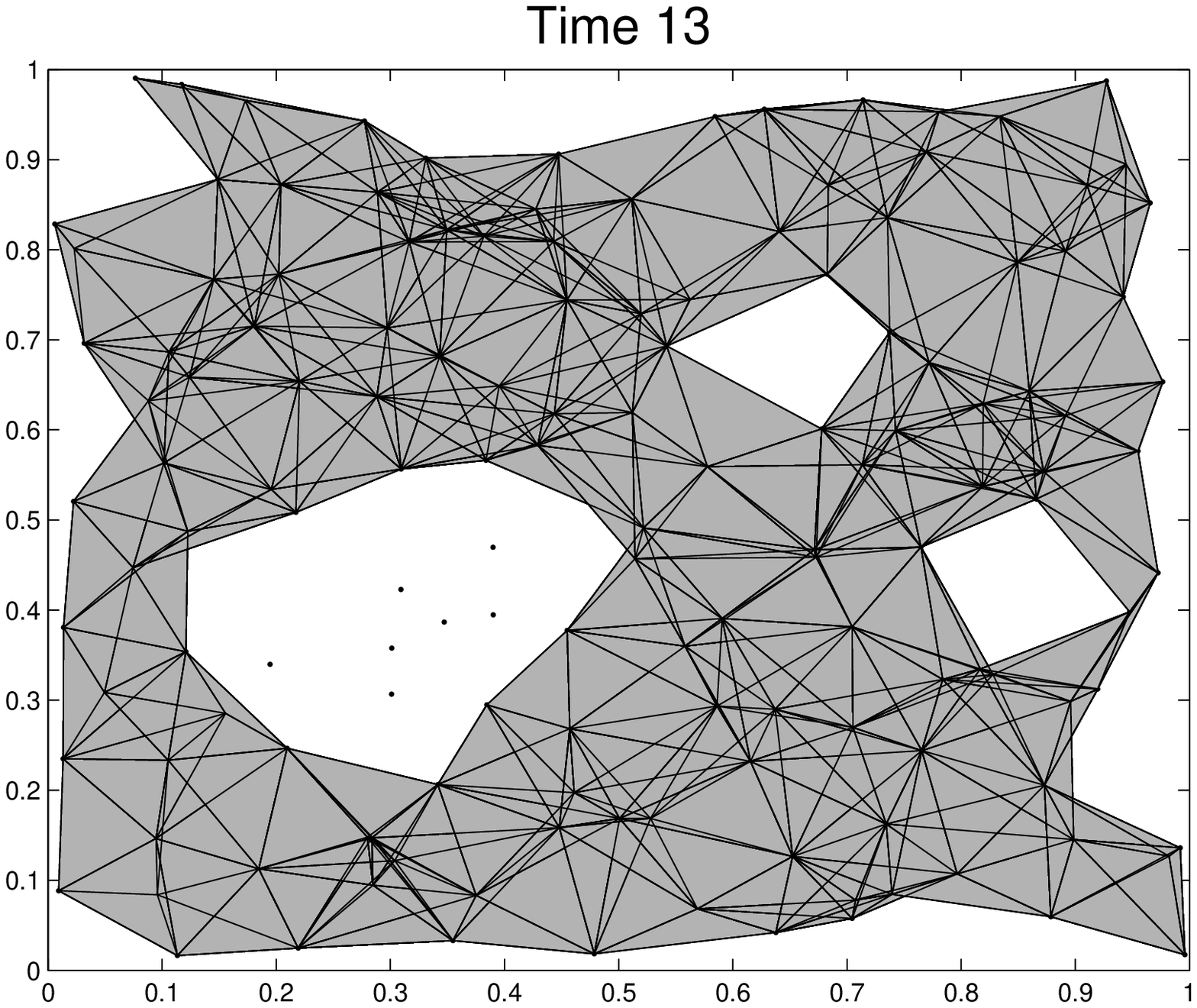} & \includegraphics[scale=0.22]{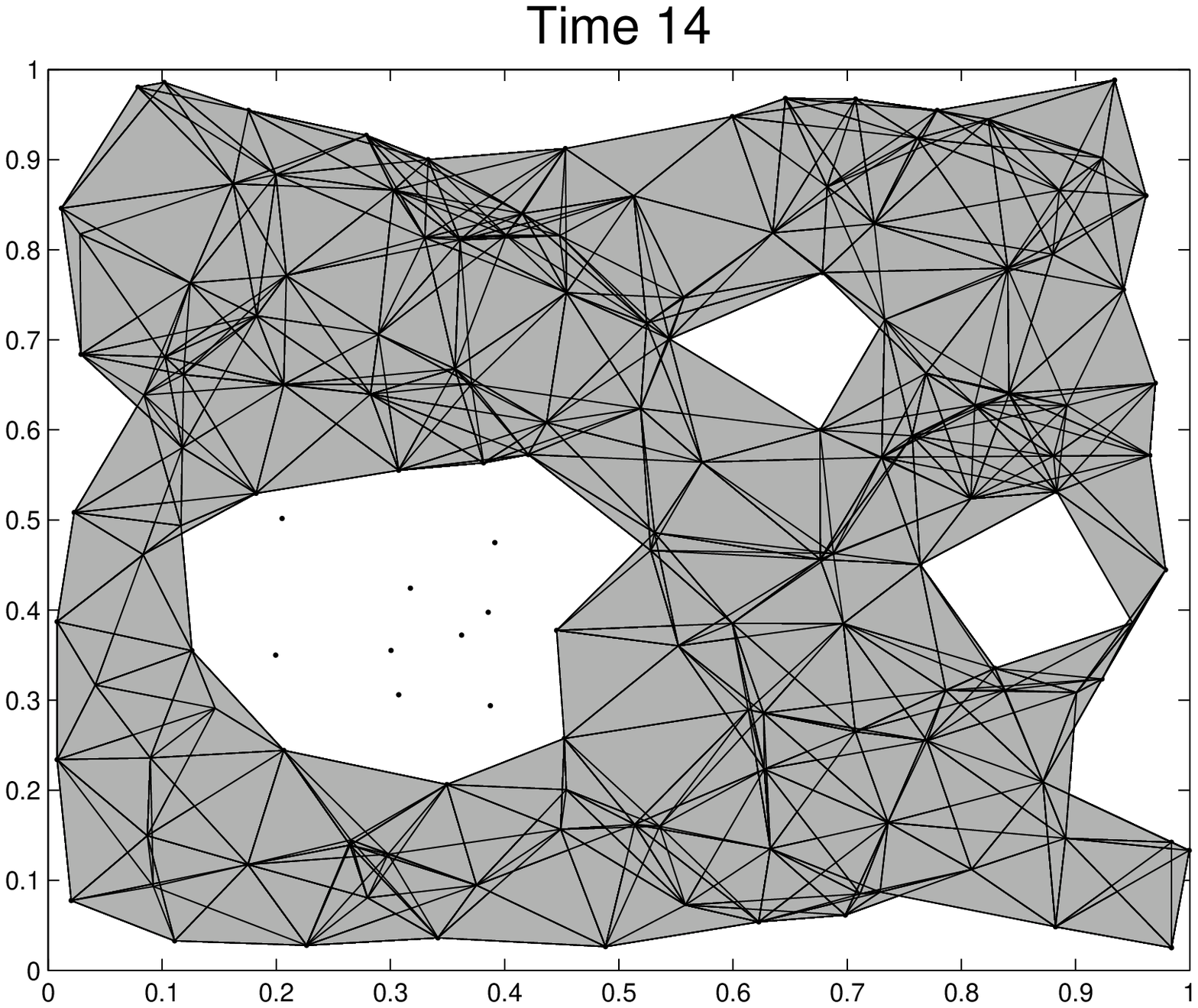} \\
\includegraphics[scale=0.22]{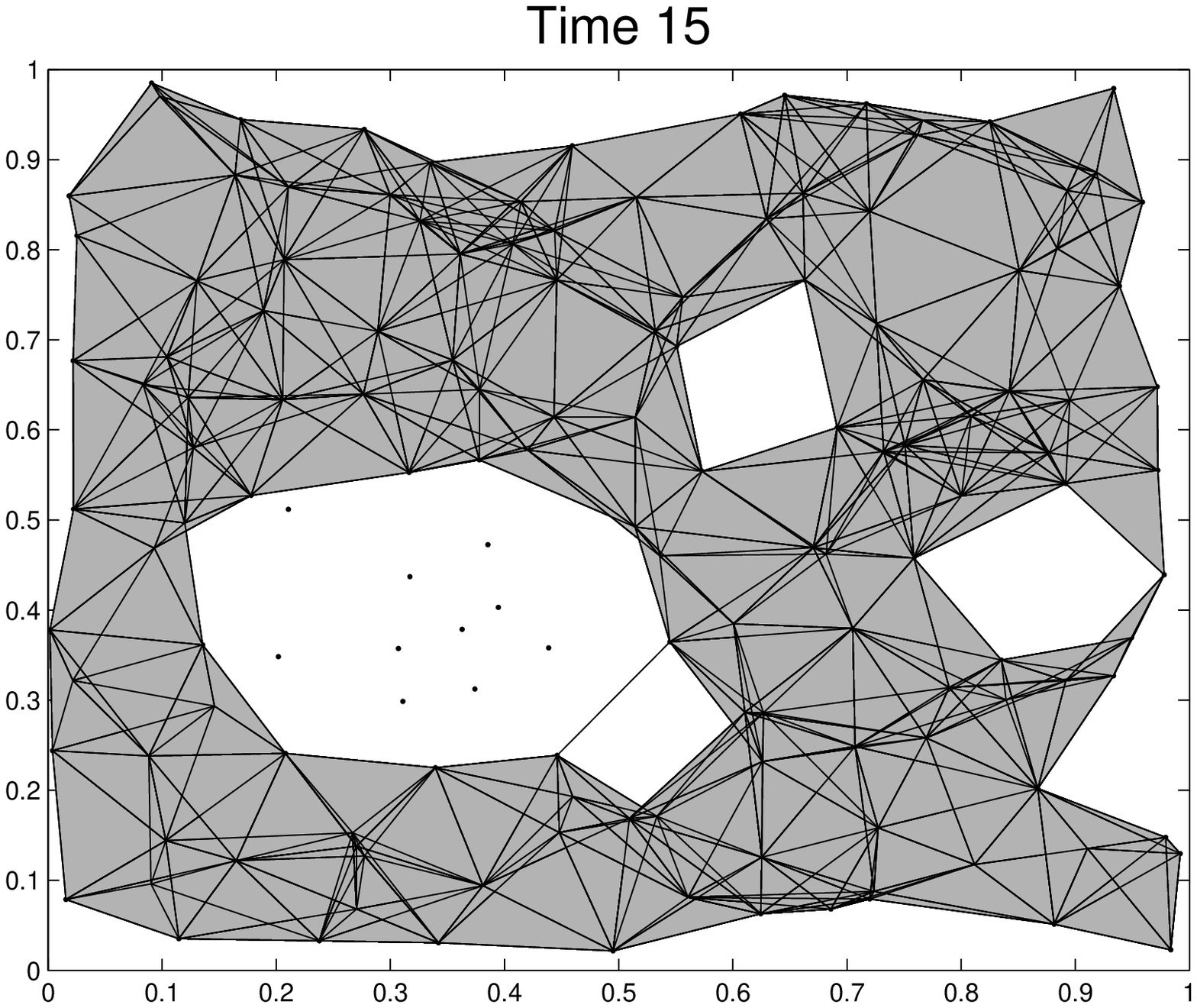} & \includegraphics[scale=0.22]{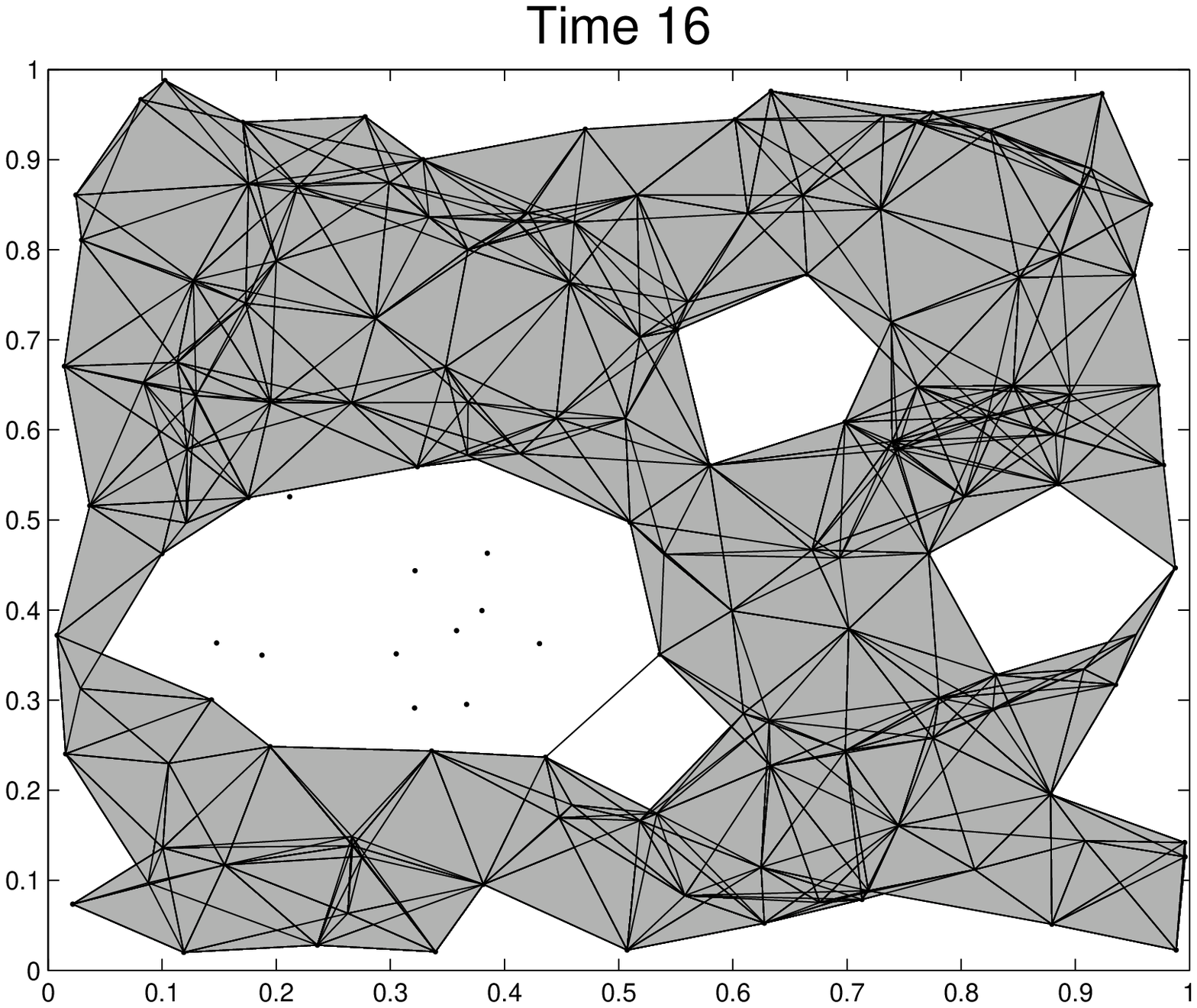} & \includegraphics[scale=0.22]{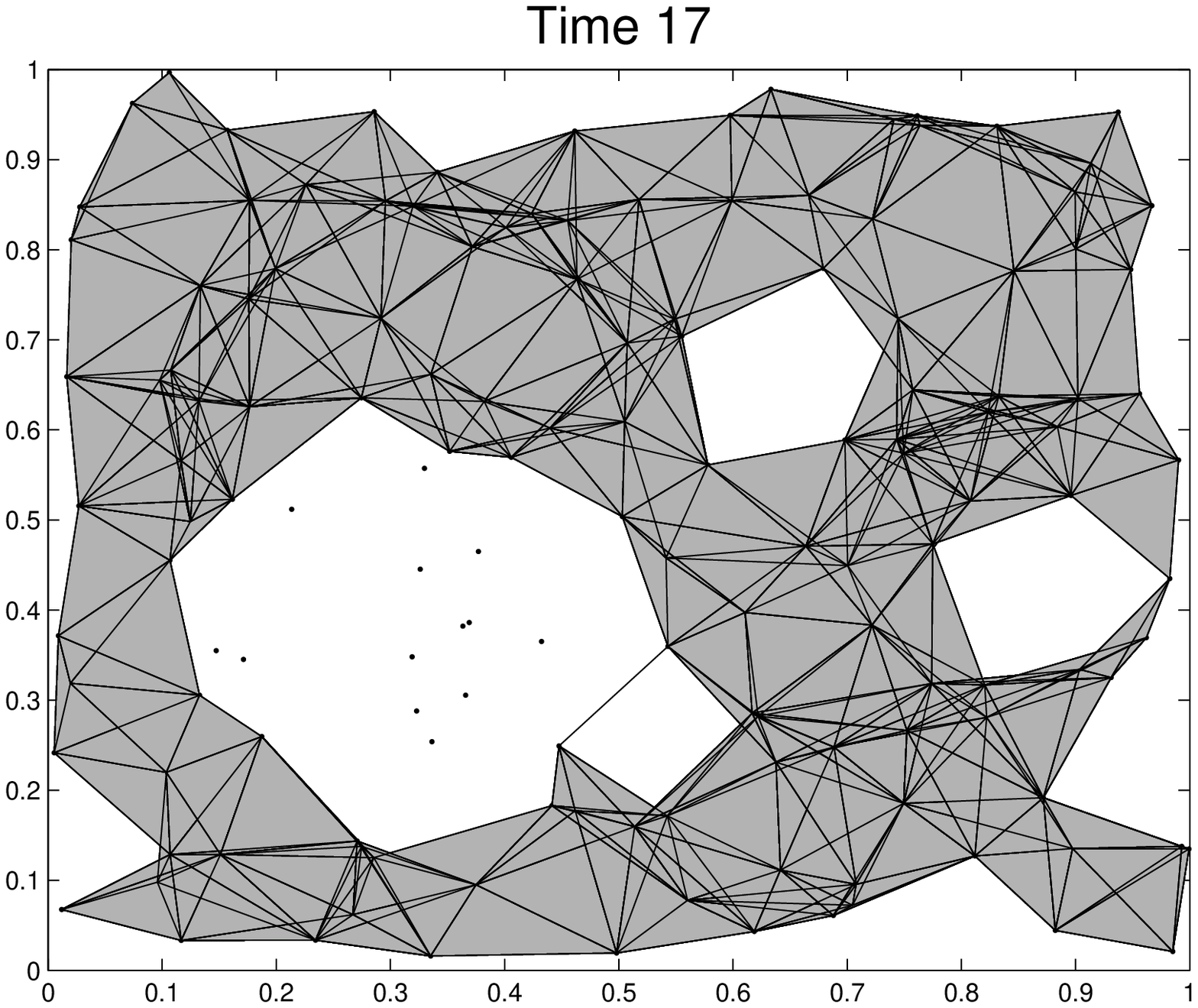} \\
\includegraphics[scale=0.22]{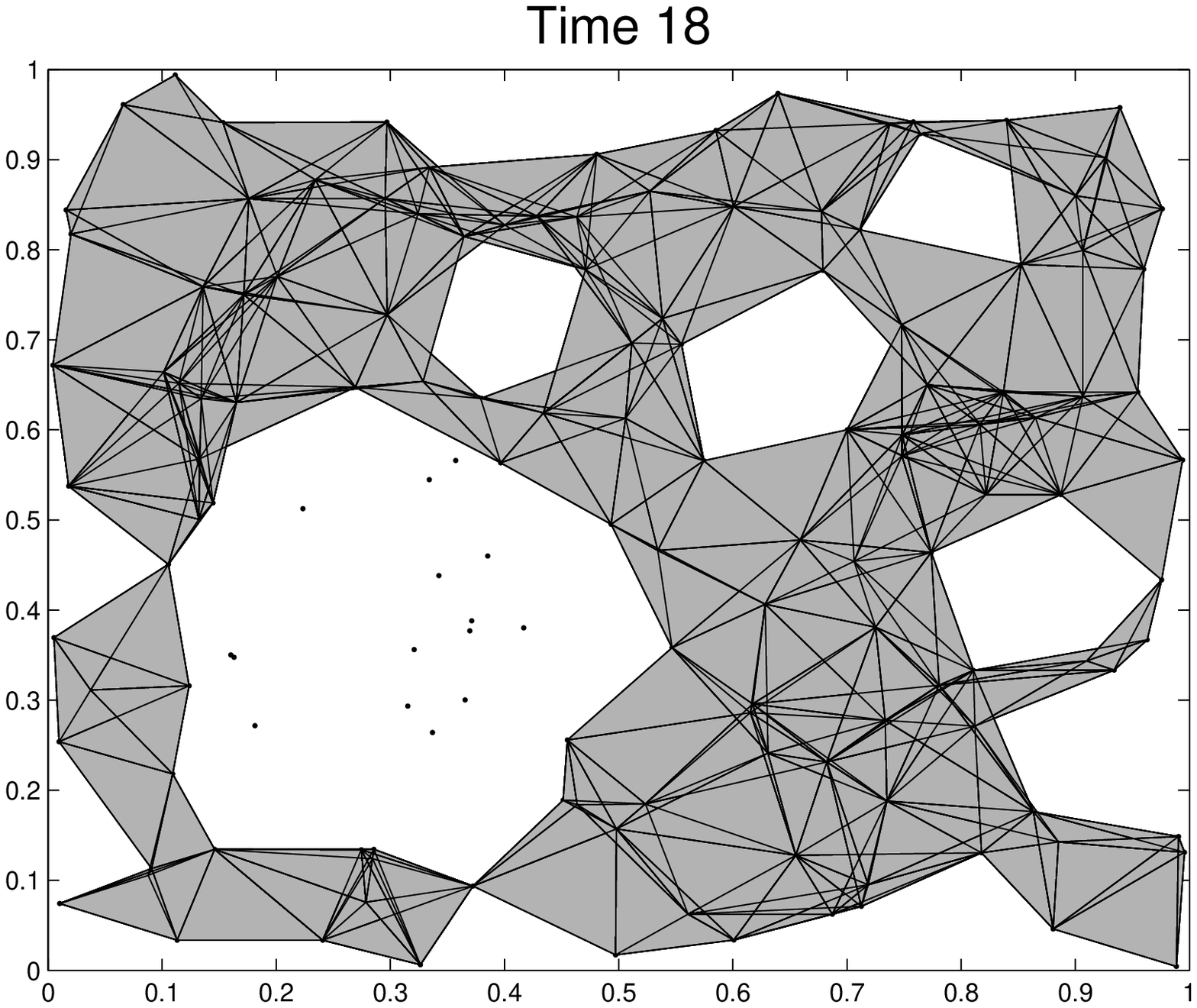} & \includegraphics[scale=0.22]{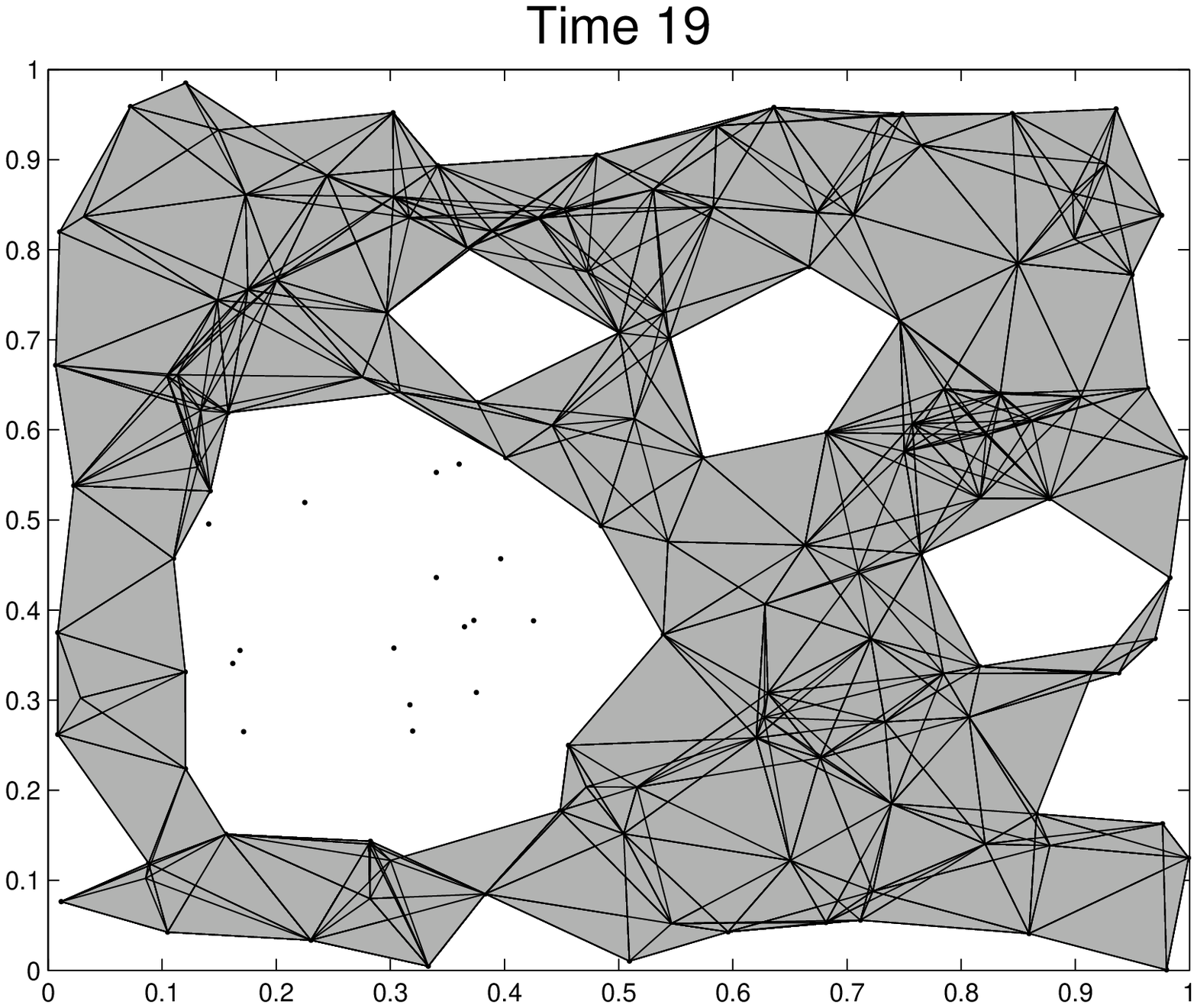} & \includegraphics[scale=0.22]{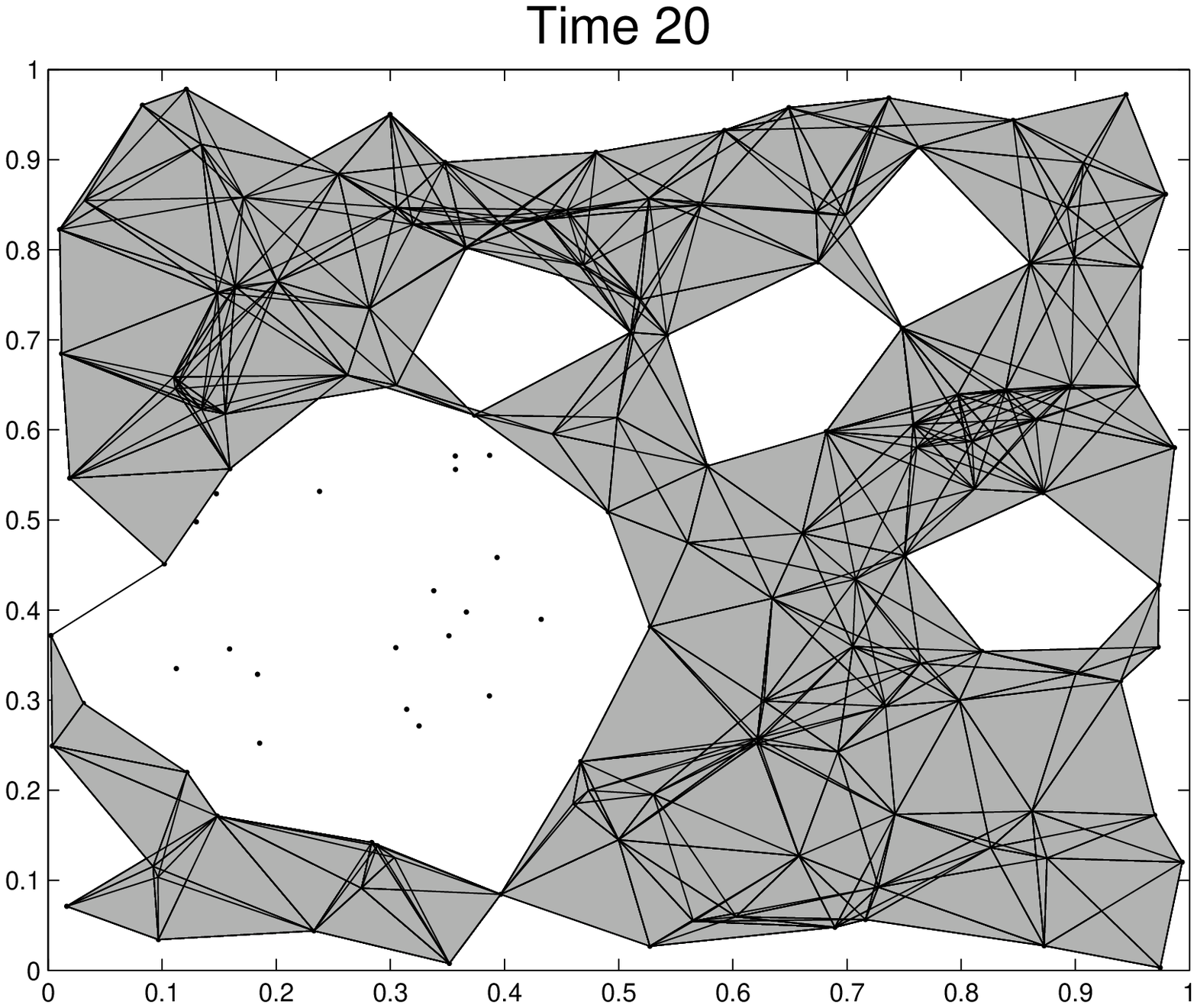} \\
\end{tabular}
\end{center}
\caption{A sequence of snapshots of a network with an expanding failure region. Nodes within the failure region can no longer sense or communicate.\label{ExpandingFailure}}
\end{figure}

\begin{figure}[htp]
\begin{center}
\begin{tabular}{l}
\includegraphics[scale=0.5]{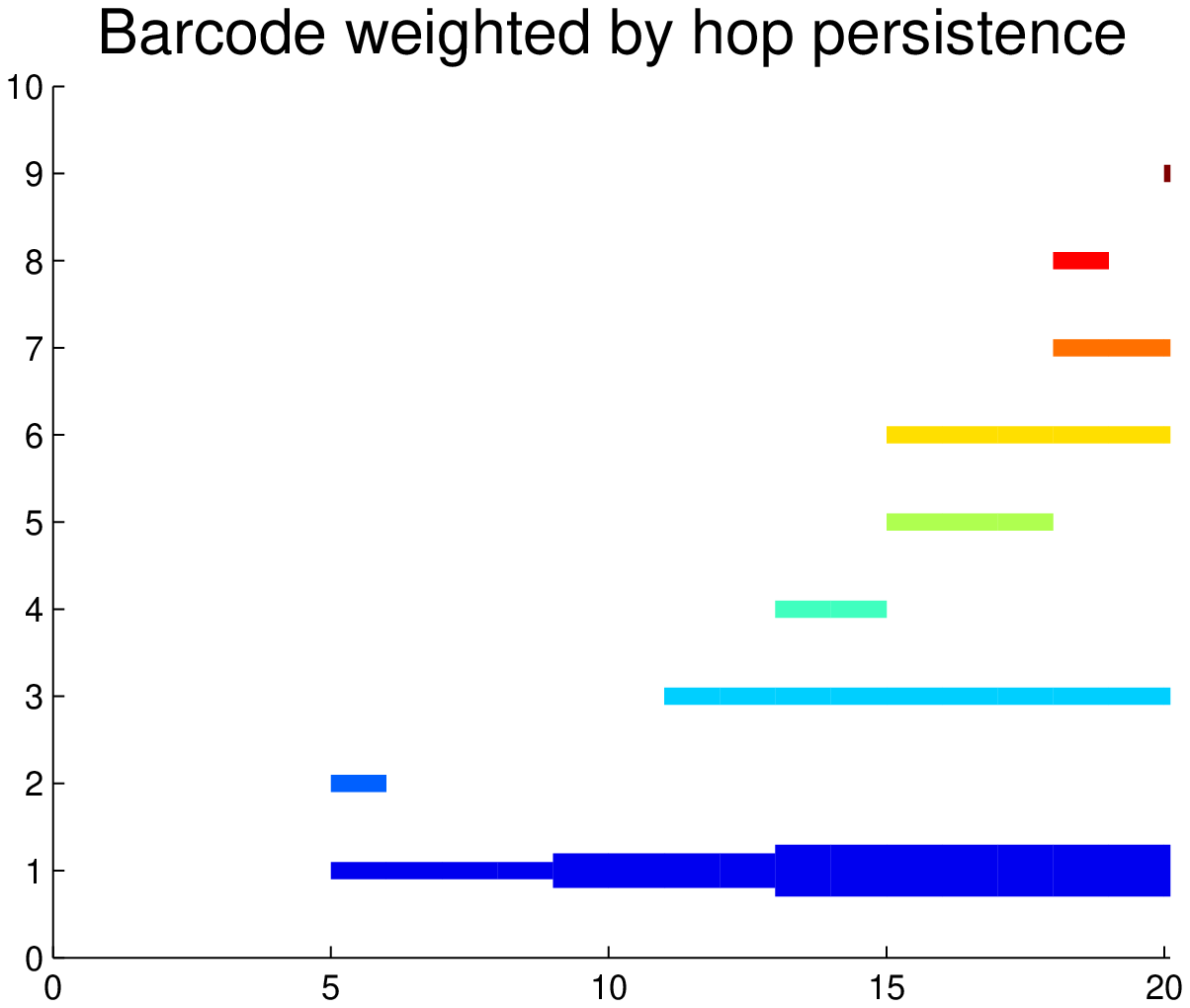} \\
\end{tabular}
\end{center}
\caption{Barcode from zigzag persistence on the network shown in Figure \ref{ExpandingFailure}. Each bar is weighted by the depth its adaptive representative cycle persists in the hop distance filtration at each time point. \label{WeightedBarcode}}
\end{figure}

In dense networks, the coverage holes are typically small and short-lived, so the representative cycles themselves provide fairly accurate tracking, but even in situations where the representative cycle itself does not surround a hole `tightly', its inclusion in the hop-distance filtration will still reflect the size of the hole accurately. This is especially useful for holes that are persistent over time, to better understand whether the hole is of increasing severity (perhaps due to a malicious attack or systematic failure). To compute dynamic size estimates for the hole(s) associated with each bar, the persistence in the hop-distance filtration for each representative cycle is attached to its corresponding bar at each time point. This is visualized in the barcode by thickening the bar by an amount proportional to the depth its representative cycle persists in the hop-distance filtration at that time. Figure \ref{ExpandingFailure} shows a time-varying network with an expanding failure region, and the associated thickened barcode is shown in Figure \ref{WeightedBarcode} (with hop-distance computed up to a maximum depth of 3). It can be seen that the hole which is growing in time is easily observed in the barcode as a bar which thickens over time.

\subsubsection{Maintaining perimeter around a guarded region}

\begin{figure}[htp]
\begin{center}
\begin{tabular}{ll}
\includegraphics[scale=0.35]{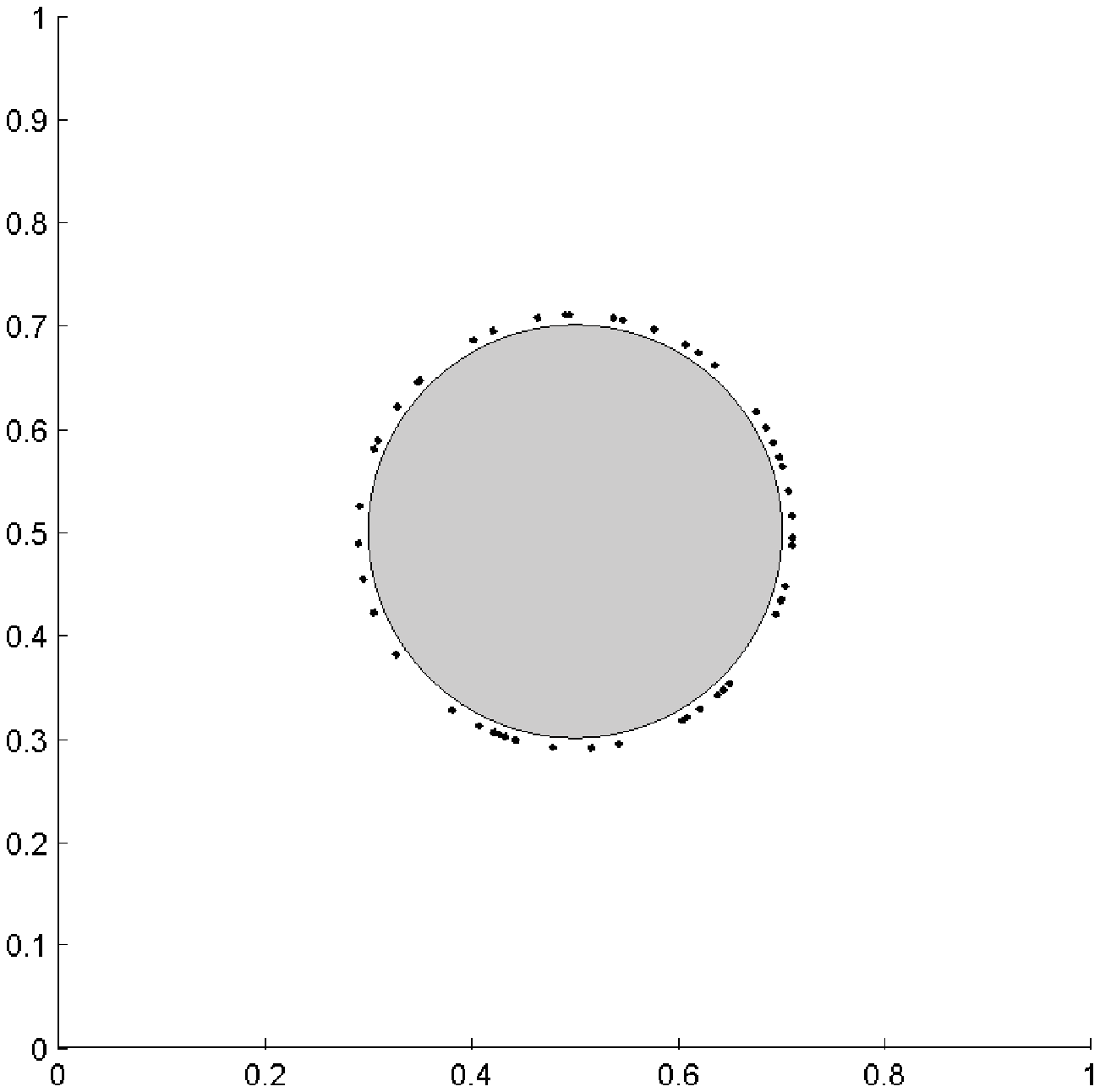} & \includegraphics[scale=0.35]{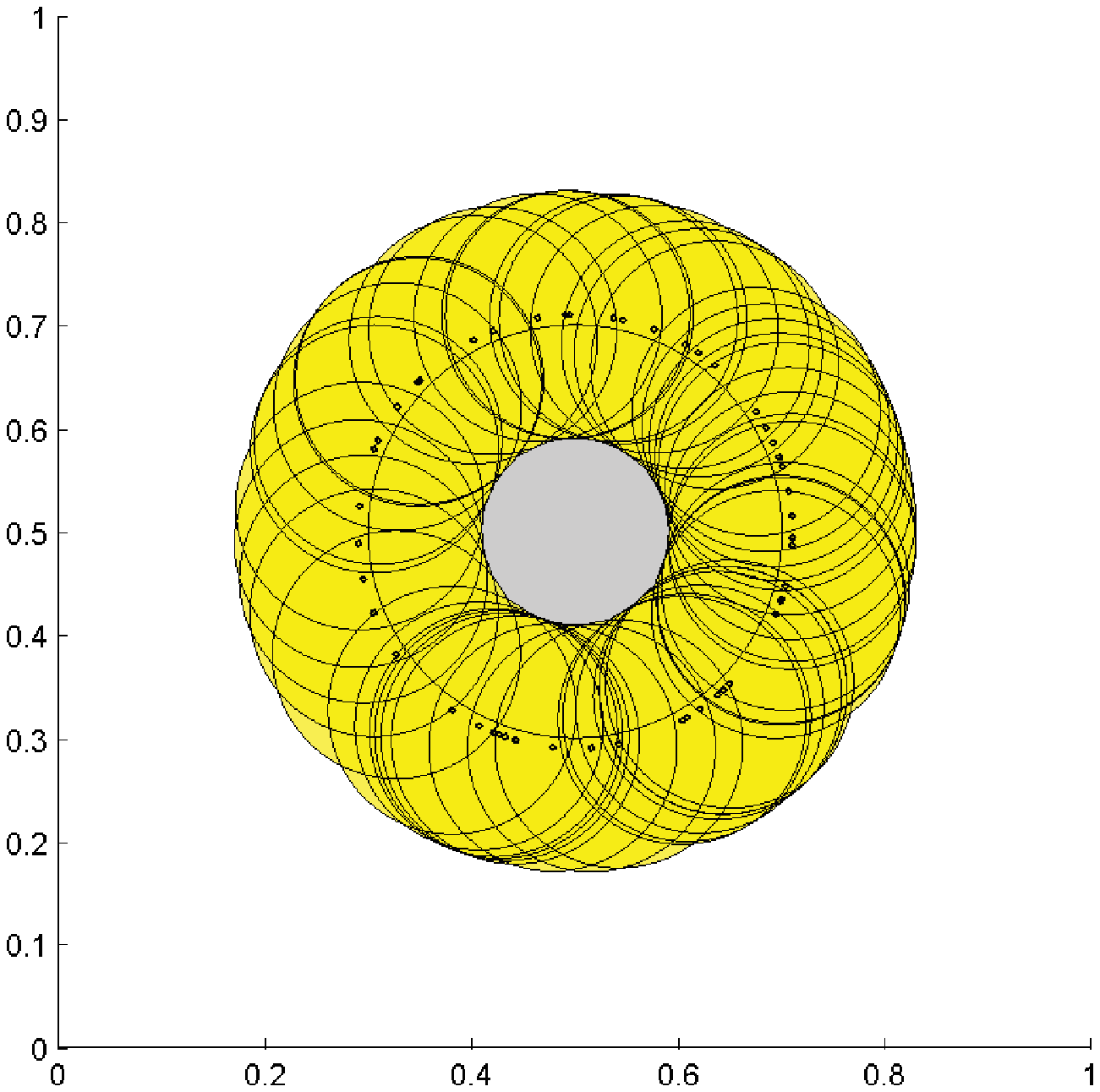} \\
\includegraphics[scale=0.35]{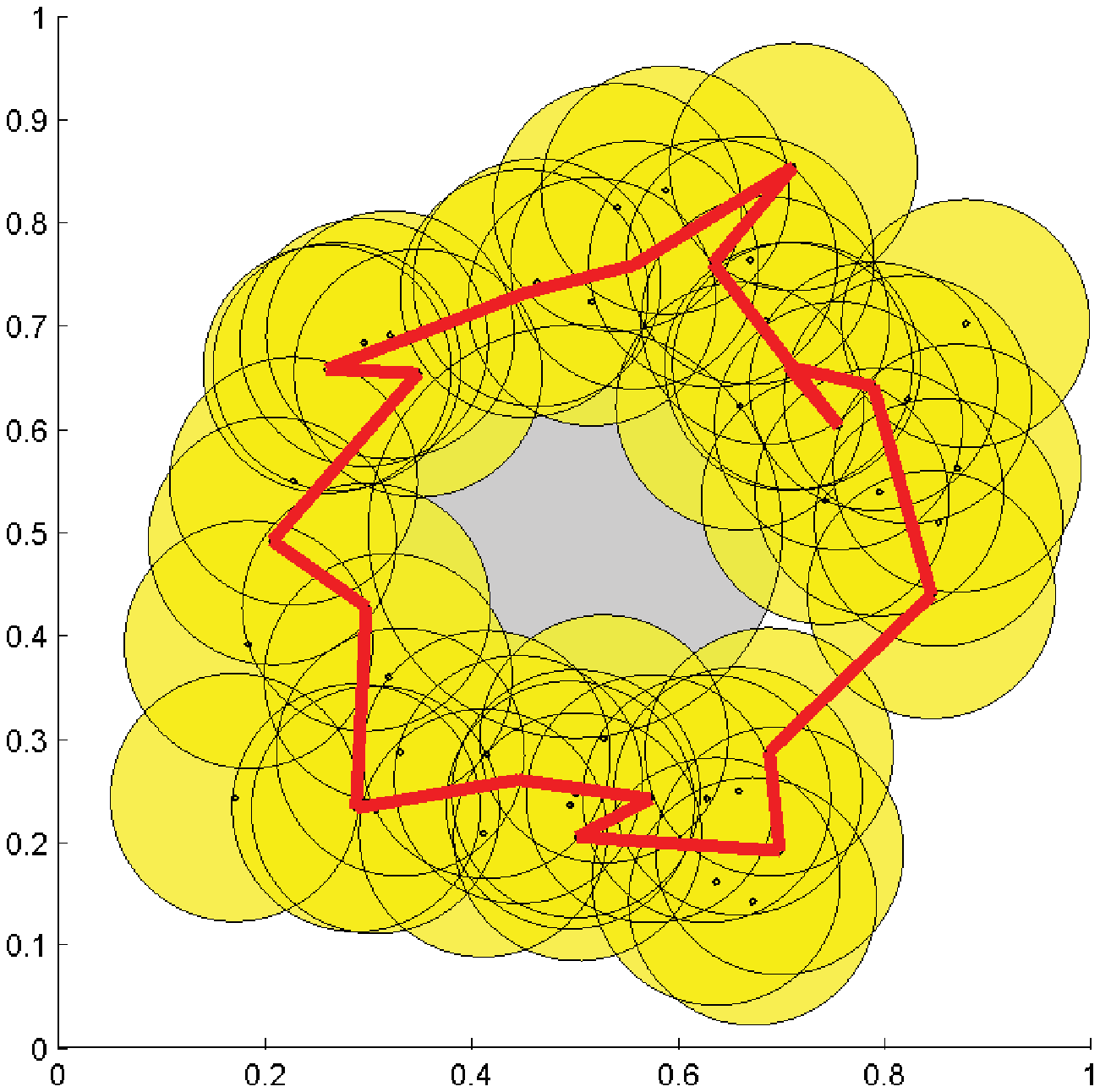} & \includegraphics[scale=0.35]{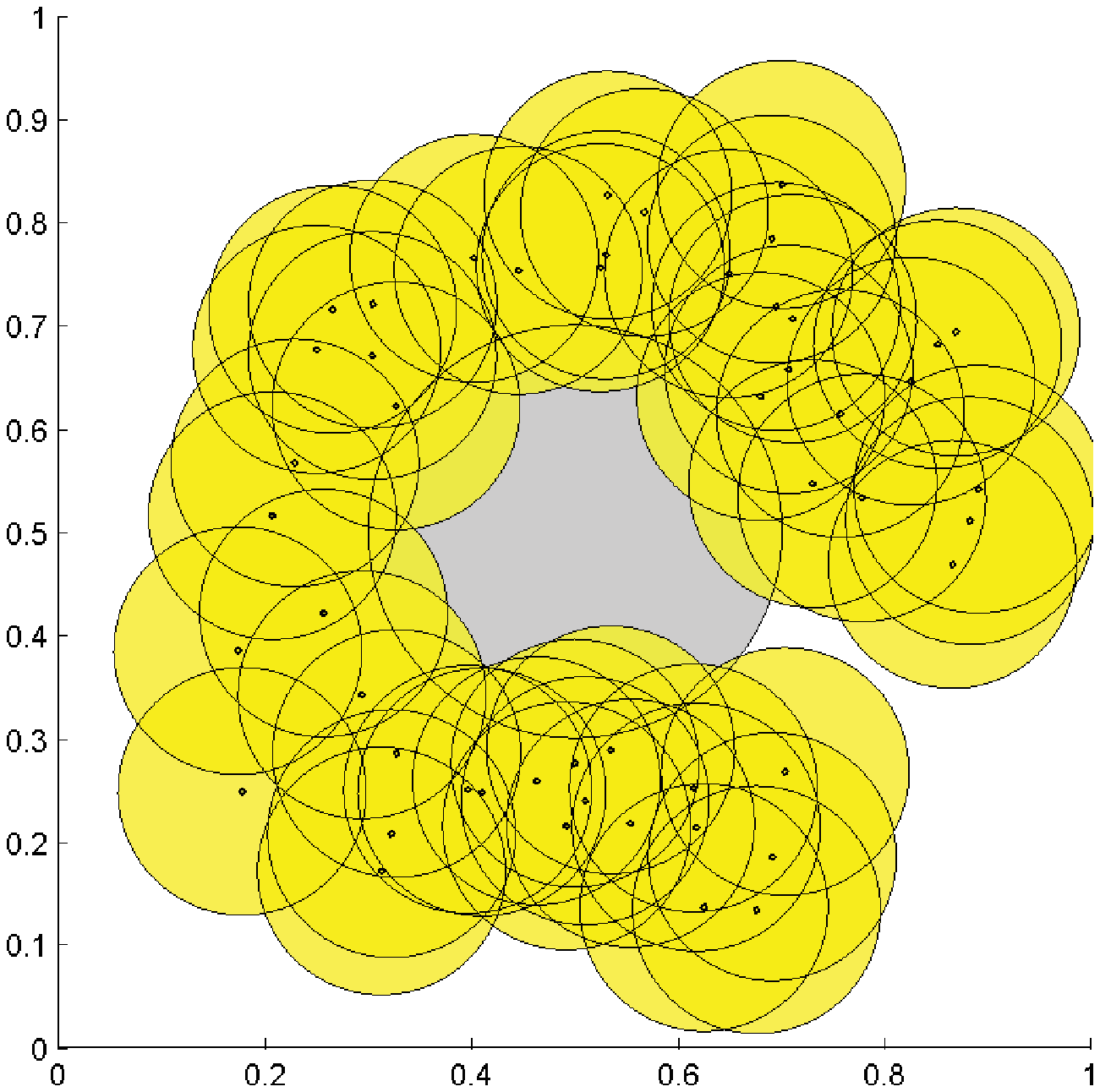}  \\
\end{tabular}
\end{center}
\caption{(Top left) A circular region to be protected, with guards stationed around perimeter. (Top right) Each guard can sense within a local disk. (Bottom left) While the guards are moving around the region, the presence of an unbroken cycle indicates separation of the protected area is maintained. (Bottom right) If the guards wander too far, we are alerted to the lapse in protection by observing the broken cycle. \label{Guarding}}
\end{figure}

Representative cycles can also be used to determine whether an existing cycle remains unbroken over time. This can be of particular use when there is an area the needs to remain isolated, while guards roam about the region surrounding it. Without requiring precise locations of the guards, we can determine whether there remains an unbroken cycle surrounding the protected area, by tracking the persistence of the cycle that is initially present. Figure \ref{Guarding} shows a set of sensors/guards which initially surround a protected area tightly (top row), and then begin randomly moving about the environment. After some time, the guards still form a cycle (drawn in red, bottom left) which has been continuously enclosing the protected area, but eventually when the guards wander too far apart we detect the breaking of the cycle (bottom right).

\section{Conclusions and Future Work}

We have presented here a set of methods which employ ideas from computational topology to describe time-varying coverage in a dynamic sensor network, while using only local information about which nodes neighbor each other at each time step.

Zigzag persistent homology takes the sequence of simplicial complexes (representing the dynamic network), and outputs a barcode of birth and death times of homological features in the sequence. We described the relationship between these birth-death intervals and the time-varying coverage holes in the network, and demonstrated how the barcode output is a useful quantitative descriptor to detect coverage differences when comparing sensor network mobility patterns.

We developed a method to obtain a specific set of geometrically-meaningful representative cycles for each birth-death interval, at each time point. This set of representative cycles is then used to track coverage holes over time, as well as to obtain size estimates (in conjunction with a hop-distance filtration) for the holes at each time point. This size information is then incorporated into the barcode, for a more complete description of the dynamic coverage of the network.

A surprising amount of information can be gleaned about the time-varying coverage of the network using homological methods, and all of this is achieved in a setting with no coordinate or edge-length information available, using only a binary adjacency matrix for the network at each time point.

\bibliographystyle{plain}
\bibliography{ZigzagCoverageBib}

\end{document}